\documentclass[12pt]{IEEEtran}
\usepackage{amssymb,stmaryrd,amsmath,amsfonts,rotating}
\usepackage{graphicx}
\usepackage{hyperref}
\usepackage{epic}
\RequirePackage{bbm}

\DeclareMathOperator{\sgn}{{\text sgn}}

\newtheorem{theorem}{Theorem}
\newtheorem{corollary}{Corollary}
\newtheorem{lemma}{Lemma}
\newtheorem{definition}{Definition}

\newtheorem{example}{Example}
\newtheorem{proposition}{Proposition}

\newtheorem{fact}{Fact}

\newcommand{\btheorem}{\begin{theorem}}
\newcommand{\etheorem}{\end{theorem}}
\newcommand{\bcorollary}{\begin{corollary}}
\newcommand{\ecorollary}{\end{corollary}}
\newcommand{\blemma}{\begin{lemma}}
\newcommand{\elemma}{\end{lemma}}
\newcommand{\bdefinition}{\begin{definition}}
\newcommand{\edefinition}{\end{definition}}
\newcommand{\bexample}{\begin{example}}
\newcommand{\eexample}{\end{example}}
\newcommand{\bproposition}{\begin{proposition}}
\newcommand{\eproposition}{\end{proposition}}
\newcommand{\bfact}{\begin{fact}}
\newcommand{\efact}{\end{fact}}

\newcommand{\ent}{\ensuremath{{\tt{h}}}} 
\newcommand{\BPsmall}{\ensuremath{\text{\tiny BP}}} 
\newcommand{\MAPsmall}{\ensuremath{\text{\tiny MAP}}} 
\newcommand{\GalAsmall}{\ensuremath{\text{\tiny Gal A}}} 
\newcommand{\GalBsmall}{\ensuremath{\text{\tiny Gal B}}} 
\newcommand{\EXITsmall}{\ensuremath{\text{\tiny EXIT}}} 
\newcommand{\MinSumsmall}{\ensuremath{\text{\tiny MinSum}}} 

\newcommand{\bproof}{\begin{proof}}
\newcommand{\eproof}{\end{proof}}
\newcommand{\bkeywords}{\begin{keywords}}
\newcommand{\ekeywords}{\end{keywords}}

\newcommand{\btable}{\begin{table}}
\newcommand{\etable}{\end{table}}
\newcommand{\bfigure}{\begin{figure}}
\newcommand{\efigure}{\end{figure}}

\newcommand{\reals}{\ensuremath{\mathbb{R}}}

\newcommand{\integers}{\ensuremath{\mathbb{Z}}}
\newcommand{\naturals}{\ensuremath{\mathbb{N}}}

\newcommand{\expectation}{\ensuremath{\mathbb{E}}}

\newcommand{\altPhi}{\phi}
\newcommand{\altPhiSI}{\xi_\phi}
\newcommand{\PhiSI}{\xi_\Phi}

\newcommand{\smthker}{\omega}
\newcommand{\intsmthker}{\Omega}
\newcommand{\discsmthker}{{\rm w}}

\newcommand{\fS}{f^{\smthker}}

\newcommand{\ftS}[1]{f^{#1,\smthker}}
\newcommand{\ftdS}[1]{f^{#1,\discsmthker}}

\newcommand{\gS}{g^{\smthker}}

\newcommand{\gtS}[1]{g^{#1,\smthker}}
\newcommand{\gtdS}[1]{g^{#1,\discsmthker}}

\newcommand{\fSi}{f^{\smthker_i}}
\newcommand{\gSi}{g^{\smthker_i}}

\newcommand{\gSx}{{g^{\smthker}_x}}
\newcommand{\hf}{h_{f}}
\newcommand{\hfinv}{h_{f}^{-1}}
\newcommand{\thf}{\widetilde{h}_{f}}
\newcommand{\thfinv}{\widetilde{h}_{f}^{-1}}
\newcommand{\hg}{h_{g}}
\newcommand{\thg}{\widetilde{h}_{g}}
\newcommand{\hginv}{h_{g}^{-1}}
\newcommand{\ashift}{{\frak s}}

\newcommand{\hfs}{{h_{f,\ashift}}}
\newcommand{\hfsinv}{{h^{-1}_{f,\ashift}}}
\newcommand{\hfzinv}{{h^{-1}_{f,0}}}
\newcommand{\hfz}{{h_{f,0}}}

\newcommand{\delhf}{\delta^{f}}
\newcommand{\delhg}{\delta^{g}}

\newcommand{\fSa}{f^{{\smthker,\ashift}}}

\newcommand{\gSa}{g^{{\smthker,\ashift}}}
\newcommand{\gSai}{g^{{\smthker,\ashift_i}}}
\newcommand{\gSiai}{g^{{\smthker_i,\ashift_i}}}

\newcommand{\zf}{{\rm z}^{f}}
\newcommand{\tzf}{\tilde{\rm z}^{f}}
\newcommand{\zg}{{\rm z}^{g}}

\newcommand{\uf}{{\rm u}^{f}}
\newcommand{\tuf}{\tilde{\rm u}^{f}}
\newcommand{\ug}{{\rm u}^{g}}
\newcommand{\Kf}{K^{f}}
\newcommand{\Kg}{K^{g}}

\newcommand{\Zg}{Z^{g}}
\newcommand{\Df}{D^{f}}
\newcommand{\Dg}{D^{g}}
\newcommand{\Bf}{N^{f}}
\newcommand{\Bg}{N^{g}}
\newcommand{\exitfns}{\Psi_{[0,1]}}
\newcommand{\sptfns}{\Psi_{(-\infty,+\infty)}}
\newcommand{\Dga}{D^{g,\ashift}}
\newcommand{\Bga}{N^{g,\ashift}}

\newcommand{\cross}{\chi}
\newcommand{\intcross}{\chi^{o}}
\newcommand{\intcrossing}{\chi^{o}}
\newcommand{\closure}[1]{{\overline{#1}}}

\newcommand{\xf}{v}
\newcommand{\xg}{u}
\newcommand{\ff}{{f}}
\newcommand{\tff}{\widetilde{f}}
\newcommand{\tfS}{\widetilde{f}^\smthker}
\newcommand{\fg}{{g}}
\newcommand{\tfg}{\widetilde{g}}
\newcommand{\tgS}{\widetilde{g}^\smthker}
\newcommand{\ffinv}{{f}^{-1}}
\newcommand{\fginv}{{g}^{-1}}

\newcommand{\minfty}{\text{\small{$-\infty$}}}
\newcommand{\pinfty}{\text{\small{$+\infty$}}}

\newcommand{\rarrowi}{\xrightarrow{i\rightarrow\infty}}

\newcommand{\neigh}[2]{\boldsymbol( #1 \boldsymbol)_{#2}}
\newcommand{\slanta}[1]{#1(;\tau)}
\newcommand{\half}{\textstyle{\frac{1}{2}}}

\newcommand{\dr}{{\rm d}_r}
\newcommand{\dl}{{\rm d}_l}

\newcommand{\ind}{\mathbbm{1}}
\newcommand{\indicator}[1]{\ind_{\{ #1 \}}}
\newcommand{\unitstep}{{\mathbf H}}
\newcommand{\veq}{\doteqdot}

\newcommand{\jump}[1]{{\cal J}_{#1}}
\newcommand{\hfiinv}{(\hf^i)^{-1}}
\newcommand{\hgiinv}{(\hg^i)^{-1}}

\newcommand{\flats}[1]{{\cal I}_{#1}}
\newcommand{\intflats}[1]{{\cal I}^{o}_{#1}}
\newcommand{\upperx}[1]{#1_{\!\rm{b}}}
\newcommand{\lowerx}[1]{#1_{\!\rm{a}}}
\newcommand{\tmplF}{\mathcal{F}}
\newcommand{\tmplG}{\mathcal{G}}
\newcommand{\dv}[1]{#1}
\newcommand{\gSdisc}{g^{\discsmthker}}

\newcommand{\defeq}{\stackrel{\text{def}}{=}}
\newcommand{\Avg}{{\cal A}}

\begin{document}
\title{Wave-Like Solutions of General One-Dimensional Spatially Coupled Systems}

\author{\authorblockN{Shrinivas
Kudekar\IEEEauthorrefmark{1}, Tom Richardson\IEEEauthorrefmark{1} and R{\"u}diger
Urbanke\IEEEauthorrefmark{2} \\ }
\authorblockA{\IEEEauthorrefmark{1}Qualcomm Inc., USA\\ Email: \{skudekar, tomr\}@qti.qualcomm.com} \\
\authorblockA{\IEEEauthorrefmark{2}School of Computer and Communication Sciences\\ EPFL, Lausanne, Switzerland\\ Email: ruediger.urbanke@epfl.ch}\\
 }
\date{\today}

\maketitle
\begin{abstract}
We establish the existence of wave-like solutions to
spatially coupled graphical models which, in the large size limit,
can be characterized by a one-dimensional real-valued state.
This is extended to a proof of the threshold saturation phenomenon for all
such models, which
includes spatially coupled irregular LDPC codes over the BEC, but
also addresses hard-decision decoding for transmission over general
channels, the CDMA multiple-access problem, compressed sensing, and some statistical
physics models.

For traditional uncoupled iterative coding systems with two components and
transmission over the BEC, the asymptotic convergence behavior is
completely characterized by the EXIT curves of the components. More
precisely, the system converges to the desired fixed point, which is the one corresponding
to perfect decoding, if and only if the two EXIT functions describing
the components do not cross.
For spatially coupled systems whose state is one-dimensional a
closely related graphical criterion applies. Now the curves are
allowed to cross, but not by too much.  More precisely, we show
that the threshold saturation phenomenon is related to the positivity
of the (signed) area enclosed by two EXIT-like functions associated to the
component systems, a very intuitive and easy-to-use graphical
characterization.

In the spirit of EXIT functions and Gaussian approximations, we
also show how to apply the technique to higher dimensional and even
infinite-dimensional cases. In these scenarios the method is no
longer rigorous, but it typically gives accurate predictions.  To
demonstrate this application, we discuss transmission over general
channels using both the belief-propagation as well as the min-sum
decoder.  \end{abstract}

\section{Introduction}\label{sec:introduction} The idea of spatial
coupling emerged in the coding context from the study of Low-Density
Parity-Check Convolutional (LDPCC) codes which were introduced
by Felstr{\"{o}}m and Zigangirov \cite{FeZ99}. We refer the reader
to \cite{EnZ99, ELZ99, LTZ01, TSSFC04} as well as to the introduction
in \cite{KRU10} which contains an extensive review.  
A critical discovey was the observation that 
LDPCC codes can outperform their block coding counterparts
\cite{SLCZ04, LSZC10, LSZC05}.  Subsequent work isolated
and identified the key system structure that is responsible for
this improvement.

It was conjectured in \cite{KRU10} that spatially
coupled systems exhibit BP threshold behavior corresponding to the MAP threshold behavior of
 uncoupled component system. This phenomenon was termed ``threshold saturation" 
and a
rigorous proof of the threshold saturation phenomenon over the BEC
and regular LDPC ensembles was given.  The proof was generalized
to all binary-input memoryless output-symmetric (BMS) channels in
\cite{KRU12}.  From these results it follows that universal
capacity-achieving codes for BMS channels can be constructed by
spatially coupling regular LDPC codes.  Spatial coupling has also
been successfully applied to the CDMA multiple-access channel
\cite{ScT11,TTK11}, to compressed sensing \cite{KP10,KMSSZ11,DMM,DJM11},
to the Slepian-Wolf coding problem \cite{YPN11}, to models in
statistical physics \cite{HMU11a,HMU11b}, and to many other problems in
communications and computer science, see
\cite{KRU12} for a review.

The purpose of this paper is two-fold. First, we establish the existence of wave-like solutions to
spatially coupled graphical models which, in the large size limit,
can be characterized by a one-dimensional real-valued state.
This is applied to give a rigorous
proof of the threshold saturation phenomenon for all such
models. This includes spatial
coupling of irregular LDPC codes over the BEC, but it also addresses
other cases like hard-decision decoding for transmission over general
channels, and the CDMA multiple-access problem \cite{ScT11,TTK11}
and compressed sensing \cite{DJM11}.
As mentioned above, transmission over the BEC using spatially-coupled
regular LDPC codes was already solved  in \cite{KRU10}, but our
current set-up is more general.  Whereas the proof in \cite{KRU10}
depends on specific features of the BEC, here we derive a graphical
characterization of the threshold saturation phenomena in terms of
EXIT-like functions for the underlying component system.  This broadens
the range of potential applications considerably.

Consider the example of coding over the BEC.  In
the traditional irregular LDPC EXIT chart setup the condition for successful decoding reduces
to the two EXIT charts not crossing. We will show that the EXIT
condition for good performance of the spatially-coupled system is
significantly relaxed and reduces to a balance condition on the area bounded
between the component EXIT functions.

\begin{figure}[htp]
{
\centering
\input{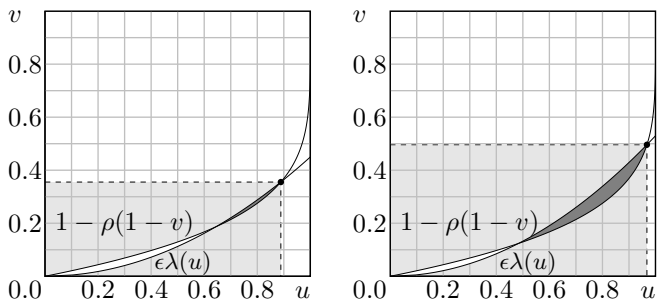}
}
\caption{\label{fig:positivegapbec36} Both pictures show the EXIT
curves for the $(3, 6)$ ensemble and transmission over the BEC.
Left: $\epsilon=0.45$. In this case $A=0.03125>0$, i.e., the white
area is larger than the dark gray area.  Right: $\epsilon=0.53$.
In this case $A=-0.0253749<0$, i.e., the white area is smaller than
the dark gray area.  } 
\end{figure}

The criteria is best demonstrated by a simple example. Consider transmission
over the BEC using the $(3, 6)$ ensemble.
Figure~\ref{fig:positivegapbec36} shows the corresponding EXIT
charts for $\epsilon=0.45$ and $\epsilon=0.53$. Note that both these
channel parameters are larger than the BP threshold which is
$\epsilon^{\text{\BPsmall}} \simeq 0.4294$.
If we consider the signed area bounded by the two
EXIT charts and integrate from $0$ to $u$ then on the left hand side, with $\epsilon=0.45,$ this
area is positive for all $u \in [0, 1]$.  This property guarantees
that the decoder for the spatially coupled system succeeds for this
case.  On the right-hand side with $\epsilon=0.53$, however, the area becomes
negative at some point (the total area in white is smaller than the
total area in dark gray) and by our condition this implies that the
decoder for the spatially coupled system does not succeed. The
threshold of the spatially coupled system is that channel
parameter such that the area in white and the area in dark gray are
exactly equal. 

This simple graphical condition is the essence of our result and
applies regardless whether we look at coding systems or other
graphical models. Given any system characterized by two EXIT functions,
we can plot these two functions and consider the signed area bound between them,
say for the first coordinate ranging from $0$ to a point $u$. As long as this area is positive for all
$u \in (0, 1]$ the iterative process succeeds, i.e., it converges to $0.$ Indeed, we will even
be able to make predictions on the speed of the process based on
the ``excess" area we have. 

A few conclusions can immediately be drawn from such a picture.
First, if the threshold of the uncoupled system is determined by
the so-called stability condition, i.e., the behavior of the EXIT
charts for $u$ around $0,$ then spatial coupling does not increase
the threshold. Indeed, if we increase the parameter beyond what is
allowed according to the stability condition, the area will become
negative around $0$.  Second, if the curves only have
one non-trivial crossing (besides the one at $0$ and at the right
end point) then the threshold is given by a balance of the two
enclosed areas.

For ``nice" EXIT charts (e.g., continuous, and with a finite number of
crossings) the above picture contains all that is needed.  But since
we develop the theory for the general case, some care is
needed when defining all relevant quantities. When reading the
technical parts below, it is probably a good idea to keep the above
simple picture in mind. For readers familiar with the so-called
Maxwell conjecture, it is worth pointing out that the above picture
shows that this conjecture is generically correct for coupled
systems.  To show that it is also correct for uncoupled systems one
needs to show in addition that under MAP decoding the coupled and
the uncoupled system behave identically. This can often be accomplished
by using the so-called interpolation method. For e.g., regular
ensembles with no odd check degrees this second step was shown to
be correct in \cite{GMU12}.

Let us point out a few differences to the set-up in \cite{KRU10}.
First, rather than analyzing directly the spatially-discrete system,
key results are established in the limit of continuum spatial
components. We will see that for such systems the solution for the
coupled system is characterized in terms of traveling waves, with special emphasis
on fixed (stationary) waves that we call {\em interpolating spatial fixed points}. The
spatially discrete version is then recovered as a sampling of the continuum
system.  The existence of traveling wave solutions and their
relationship to the EXIT charts of the underlying component systems
is the essential technical content of the analysis and does not depend
on information theoretic aspects of the coding case.

The second purpose of this paper is to show that the herein-developed
one-dimensional theory can model many higher-dimensional or even
infinite-dimensional systems to enable accurate prediction of their
performance. This is very much in the spirit of the use of EXIT
charts and Gaussian approximations for the the design of iterative
systems.  
Using this interpretation,
we apply our method to channel coding over general channels.  Even
though the method is no longer rigorous in these cases, we show
that our graphical characterization gives very good predictions of
system performance is therefore a convenient design
tool.

Recently several alternative approaches to the analysis of
spatially-coupled systems have been developed by
various authors \cite{TTK11b, DJM11, YJNP12a, YJNP12b}.  These
approaches share some important aspects with our work but there are
also some important differences. Let us quickly discuss this. 

In \cite{DJM11} a proof was given showing that spatially coupled measurement
matrices, together with a suitable iterative decoding algorithm,
the so-called {\em approximate message-passing} (AMP) algorithm,
allows to achieve the information theoretic limits of compressive sensing
in a wide array of settings. The key technical idea is to show that
the iterative system is characterized in the limit of large block
sizes by a one-dimensional parameter (which in this case represents
per-coordinate mean square error) and which can be tracked faithfully by
{\em state evolution equations}.  To ease the analysis the authors
consider the continuum limit of the state evolution equations for the coupled system. The spatially discrete state
evolution is then obtained by sampling the continuous state evolution
equations.  This is a strategy we also adopt.
The most important ingredient of the proof is a construction
of an appropriate free energy or potential function for the system such
that the spatial fixed points of the coupled state evolution are the stationary
points of the potential.  It is shown that if the compressed sensing under-sampling ratio is greater
than the information dimension, then the solution of the state evolution
becomes arbitrarily small.  The proof is by contradiction where one supposes
 that instead
the system converges to an interpolating spatial fixed point. By perturbing slightly the interpolating fixed point (the solution
is ``moved'' inside) one can show that the potential strictly decreases to first order with the perturbation.
Since a spatial fixed point is necesarily a stationary point of
the potential function, this gives a contradiction. 
The analysis in the present paper gives a sharp condition for the existence of an
interpolating spatial fixed point and when the condition is not met the system is shown
to exhibit travelling wave solutions that imply convergence of the system to the desired
state evolution fixed point.

In \cite{YJNP12a, YJNP12b} the two main ingredients are also the
characterization of the iterative system by a one-dimensional (or
finite-dimensional) parameter and the construction of a suitable
potential function whose stationary points are the spatial fixed points of
coupled system density evolution.  A significant innovation introduced in
\cite{YJNP12a, YJNP12b} is that it is shown how to {\em systematically}
construct such a potential function in a very general setting.  This
makes it possible to apply the analysis to a wide array of settings
and provides a systematic framework for the proof. In addition,
this framework allows not only to attack the scalar case but can
be carried over to vector-valued states.

Our starting point is the set of EXIT functions, a familiar tool
in the setting of iterative systems.
We also use a type of potential function but ours applies to the underlying component system.
Unlike the works mentioned above we retain the symmetry of the iterative system rather than
collapsing one of the equations.
The form of the potential function is such that each step of the uncoupled DE iteration 
minimizes the potential function over the variable being updated.
The potential function can be lifted to the spatially coupled system but that is not the approach we take in this paper.
Rather, we find an intimate connection between spatial fixed points of the coupled system and the potential of the
underlying component system.
Spatial fixed point solutions that interpolate between fixed points of the component system are of particular importance in our analysis.
Surprisingly, if we take as arguments of the component potential function values appropriately
sampled from the spatial fixed point then the component potential function value 
is determined by a portion of the spatial fixed point solution that is local to the evaluation points.
This basic result yields structural information on the structure of interpolating fixed point solutions and
it is used as a foundation to characterize and construct wave-like solutions for 
spatially coupled systems.
Perhaps one of the strong points of the current paper is
that it gives a fairly detailed and complete picture of the system
behavior.  I.e., we not only characterize the threshold(s) but
we also are able to characterize {\em how} the system converges to
the various FPs (these are the wave solutions) and {\em how fast}
it does so.

The outline of the paper is as follows. In Section~\ref{sec:main}
we consider an abstract system, characterized by two EXIT-like
functions.  In terms of these functions we state a graphical criterion
for the occurrence of threshold saturation.  In
Section~\ref{sec:applications} we then apply the method to several
one-dimensional systems. We will see that in each case the analysis
is accomplished in just a few paragraphs by applying the general
framework to the specific setting.  In Section~\ref{sec:gaussapprox}
we develop a framework that can be used to analyze higher-dimensional
systems in a manner analogous to the way the Gaussian approximation
is used together with EXIT charts in iterative system design.  We
also show by means of several examples that this approach typically
gives accurate predictions. In Section~\ref{sec:proof} we give a
proof of the main results. The proof includes ana anlysis of spatial 
fixed points and the construction of wave-like solutions.
Many of the supporting lemmas and bounds
are in the appendices.

\section{Threshold Saturation in One-Dimensional Systems}\label{sec:main}
In
this section we develop and state the main ingredients which we
will later use to analyze various spatially coupled systems.  Although
in most cases we are ultimately interested in ``spatially discrete'' and
``finite-length'' coupled systems, i.e., systems where we have a
finite number of the underlying ``component'' systems that are spatially coupled along
a line, it turns out that the theory is more elegant and simpler
to derive if we start with spatially continuous and unterminated systems, i.e., stretching from
$-\infty$ to $\infty$. Once a suitably defined spatially continuous system
is understood, one can make contact with the actual system at hand
by spatially discretizing it and by imposing specific boundary
conditions.

Throughout this section we use the example of the spatially-coupled
$(\dl, \dr)$-regular LDPC ensemble.  \bexample[$(\dl, \dr, w, L)$
Ensemble]\label{def:ensemble} The $(\dl, \dr, w, L)$ random ensemble
is defined as follows, see \cite{KRU10}.  In the ensuing paragraphs
we use $[a, a+b]\Delta$, for integers $a$ and $b$, $b\geq0$, and
the real non-negative number $\Delta$, to denote the set of points
$a\Delta, (a+1)\Delta, \dots, (a+b)\Delta$.

We assume that the variable nodes are located at positions $[0,
L] \Delta$, where $L \in \naturals$ and $\Delta >0$. At each position
there are $M$ variable nodes, $M \in \naturals$. Conceptually we
think of the check nodes as located at all positions $[- \infty,
\infty] \Delta$.  Only some of these positions contain check nodes
that are actually connected to
variable nodes.  
At each position there are $\frac{\dl}{\dr}
M$ check nodes. Unconnected check nodes are not used.
It remains to describe how the connections are
chosen.

We assume that each of the $\dl$ neighbors of a variable node at
position $i \Delta$ is uniformly and independently chosen from the
range $[i-w, \dots, i+w] \Delta$, where $w$ is a ``smoothing''
parameter.\footnote{Full independence is not possible while satisfying the degree constraints.
This does not affect the analysis since we only need the independence to hold asymptotically in large block size over finite neighborhoods in the graph.}
 In the same way, we assume that each of the $\dr$
connections of a check node at position $i$ is independently chosen
from the range $[i-w, \dots, i+w] \Delta$.  Note that this deviates
from the definition in \cite{KRU10} where the ranges were $[i,
\dots, i+w-1] \Delta$ and $[i-w+1, \dots, i] \Delta$ respectively.
In our current setting the symmetry of the current definition
simplifies the presentation.  The present definition is equivalent
to the previous one with $w$ replaced by $2w+1.$

This ensemble is spatially discrete. As we mentioned earlier, it
is somewhat simpler to start with a system which is spatially
continuous.  We will discuss later on in detail how to connect these
two points of view. Just to get started -- how might one go from a
spatially discrete system as the $(\dl, \dr, w, L)$ ensemble to a
spatially continuous system?  Assume that we let $\Delta$ tend to
$0$ while $L$ and $w$ tend to infinity so that $L \Delta$ tends to $\infty$ and
$W=w\Delta$ is held constant. In
this case we can imagine that in the limit there is a component
code at each location $x \in (-\infty, +\infty)$ in space and that
a component at position $x$ ``interacts'' with all components in a
particular ``neighborhood'' of $x$ of width $2W.$ {\hfill $\ensuremath{\Box}$}
\eexample

Consider a system on $(-\infty, +\infty)$ (the spatial component)
whose ``state'' at each point (in space) is described by a scalar
(more precisely an element of $[0, 1]$). This means, the state of
the system at iteration $t$, $t \in \naturals$, is described by a function
$\ff^t$, where $\ff^t(x) \in [0, 1]$, $x \in (-\infty, \infty)$.

\bexample[Coding for the BEC]
Consider transmission over a binary erasure channel (BEC) using the
$(\dl, \dr, w, L)$ ensemble described in Definition~\ref{def:ensemble}.
Then the ``state'' of each component code at a particular iteration
 is the fraction of erasure messages that are
emitted by variable nodes at this iteration.  Hence the state of
each component is indeed an element of $[0, 1]$.  {\hfill
$\ensuremath{\Box}$} \eexample

\bdefinition
We denote the space of non-decreasing functions $[0,1] \rightarrow [0,1]$  by $\exitfns.$
A function $h\in\exitfns$ has right limits $h(x+)$ for $x\in [0,1)$ and left limits $h(x-)$ for $x\in(0,1].$
To simplify some notation we define $h(0-)=0$ and $h(1+)=1.$
The function $h$ is continuous at $x$ if $h(x-) = h(x+).$ 

Similarly, let $\sptfns$ denote the space
of non-decreasing functions on $(\minfty, \pinfty)$
taking values in $[0,1].$ 
We denote $\lim_{x\rightarrow -\infty} f(x)$ as $f(\minfty)$
$\lim_{x\rightarrow +\infty} f(x)$ as 
$f(\pinfty).$  
We call a function $f \in \sptfns$ {\em$(a,b)$-interpolating} if
$f(\minfty) = a$
and
$f(\pinfty) = b.$
We will generally use the term ``interpolating" with the understanding that $b>a.$
The canonical case will be $(0,1)$-interpolating functions and we will also use the term
``$(0,1)$-interpolating spatial fixed point'' to refer to a pair of $(0,1)$-interpolating functions.

In general we work with potentially discontinuous functions.  Because of this we
occasionally need to distinguish between functions in $\exitfns$ or in $\sptfns$
that differ only on a set of measure $0.$ 
We say $h_1 \equiv h_2$ if $h_1$ and $h_2$ differ on a set of measure $0.$
These functions are equivalent in the $L_1$ sense.
We still enforce monotonicity so equivalent functions can differ only
at points of discontinuity.
 \edefinition

We think of  $\hf$ and $\hg$ as EXIT-like functions describing the
evolution of the underlying component system under an iterative
operation.  Usually, we will have $(0,0)$ and $(1,1)$ as key fixed points.

We say that a sequence $h_i \rightarrow h$ in $\exitfns$ if $h_i(u)
\rightarrow h(u)$ for all points of continuity of $h.$  We use a
similar definition of convergence in $\sptfns.$  In general only the
equivalence class of the limit is determined. I.e., if the limit $h$ is
discontinuous then it is not uniquely determined.

Any function $h \in \exitfns$ has a unique equivalence class of inverse functions in $\exitfns$. 
For $h \in \exitfns$ we will use $h^{-1}$ to denote any member of the equivalence class.
Formally, we can set $h^{-1}(v)$ to any value $u$ such that
$v \in [  h(u-), h(u+) ].$ 
Note that $h^{-1}(v-)$ and $h^{-1}(v+)$ are uniquely determined
for each $v\in [0,1].$ 
Thus, we see that the function $h^{-1}$ is uniquely determined at all of its points of continuity and
it is not uniquely determined at points of discontinuity.
Similarly, any function $f \in \sptfns$ has
a well defined monotonically non-decreasing inverse equivalence class
and we use $f^{-1}:[f(-\infty),f(+\infty)]\rightarrow [-\infty,\infty]$ to denote any member.

We assume that the dynamics of the underlying component system is
described by iterative updates according to the two functions $\hf,\hg \in \exitfns.$
In deference to standard nomenclature in coding, we refer to these
iterative updates as the {\em density evolution} (DE) equations.
If we assume that $\xf$ and $\xg$ are scalars describing the component
system state then these update equations are given by
\begin{equation}\label{eqn:DE}
\begin{split}
{\xg}^{t} & =  \hg (\xf^t), \\
{\xf}^{t+1} & = \hf (\xg^{t})\,.
\end{split}
\end{equation}

\bexample[DE for the BEC]
Consider a $(\dl, \dr)$-regular ensemble.
Let $\lambda(u)=u^{\dl-1}$ and $\rho(v)=v^{\dr-1}$.  Let $\xf^t$ be
the fraction of erasure messages emitted at variable nodes at iteration $t$
and let $\xg^t$ be the fraction of erasure messages emitted at
check nodes at iteration $t$.\footnote{Conventionally, in iterative coding these quantities are denoted by $x$ and $y$.
But since we soon will introduce a continuous spatial dimension, which naturally is denoted by $x$, we prefer
to stick with this new notation to minimize confusion.}
Let $\epsilon$ be the
channel parameter.  Then we have
\begin{equation}\label{eqn:DEBEC}
\begin{split}
\xg^{t} & =  1-\rho(1-\xf^t), \\
\xf^{t+1} & = \epsilon \lambda(\xg^{t})\,.
\end{split}
\end{equation}
In words, we have the correspondences $\hg(\xf)=1-\rho(1-\xf)$, and
$\hf(\xg)=\epsilon \lambda(\xg)$.  As written, the function $\hf(\xg)$
is not continuous at $\xg=1.$ More explicitly, $\hf(1) = \epsilon <1$, whereas
we defined the right limit at $1$ to be generically equal to $1$.
We will see shortly how to deal with this.  {\hfill $\ensuremath{\Box}$}
\eexample

Let us now discuss DE for the spatial continuum version. 
Letting $x$ denote the spatial variable, e.g. $x \in \reals,$ the spatially-coupled
system has the following update equations:
\begin{equation}\label{eqn:gfrecursion}
\begin{split}
\fg^{t}(x) & =  \hg ((\ff^t \otimes \smthker) (x)), \\
\ff^{t+1}(x) & = \hf ( (\fg^{t} \otimes \smthker) (x) )\,.
\end{split}
\end{equation}
Here, $\otimes$ denotes the standard convolution operator on $\reals$
and $\smthker$ is an {\em averaging kernel.}
\begin{definition}[Averaging Kernel]
An averaging kernel $\smthker$ is a non-negative even function,
$\smthker(x)=\smthker(-x)$, of bounded variation that integrates to $1,$
i.e., $\int \smthker(x) \text{d} x =1.$
We call $\smthker$ {\em regular} if there exists $W \in (0,\pinfty]$  such that
$\smthker(x) = 0$  for $x \not\in [-W,W]$ and 
$\smthker(x) > 0$ for $x \in (-W,W).$  Note that we do not require $W$ to be finite,
we may have $W=\infty.$ We also introduce the notation
\[
\Omega(x) \defeq\int_{-\infty}^x \smthker(z)\,dz\,.
\]
\end{definition}
%

\bexample[Continuous Version of DE for the BEC]
If we specialize the maps to the case of transmission over the BEC we get
the  update equations:
\begin{equation}\label{eqn:gfrecursionBECcont}
\begin{split}
\fg^{t}(x) & =  1-\rho(1-(\ff^t \otimes \smthker) (x)), \\
\ff^{t+1}(x) & = \epsilon \lambda( (\fg^{t} \otimes \smthker) (x) )\,.
\end{split}
\end{equation}
{\hfill $\ensuremath{\Box}$}
\eexample
For compactness we will often use the shorthand notation
$\fS$ to denote $\ff \otimes \smthker.$

\begin{figure}[htp]
{
\centering
\input{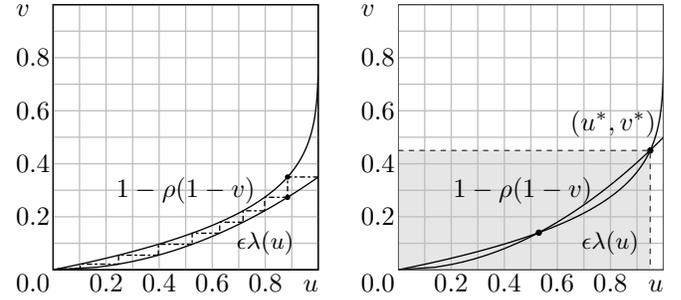}
}
\caption{\label{fig:exitbec36} Left: The figure shows the EXIT
functions $\hf(\xg)=\epsilon \lambda(\xg)$ and $\hg(\xf)=1-\rho(1-\xf)$ for the $(3, 6)$-regular
ensemble and $\epsilon=0.35$.  Note that the horizontal axis is $\xg$
and the vertical axis is $\xf$ so that we effectively plot the inverse
of the function $1-\rho(1-\xf)$. Since $0.35=\epsilon < \epsilon^{\BPsmall}
\approx 0.4292$, the two curves do not cross. The dashed ``staircase''
shaped curve indicates how DE proceeds. Right: In this figure the
channel parameter is $\epsilon=0.5 > \epsilon^{\BPsmall}$.  Hence,
the two EXIT curves cross. In fact, they cross exactly twice (besides
the trivial FP at $(0, 0)$), the first point corresponds to an
unstable FP of DE, whereas the second one is a stable FP. }
\end{figure}

In the usual manner of EXIT chart analysis, it is convenient to
consider simultaneously the plots of $\hf$ and the reflected plot
of $\hg.$ More precisely, in the unit square $[0,1]^2,$ we consider
the monotonic curves\footnote{If $\hf$ or $\hg$ is discontinuous then the curve interpolates
the jump with a line segment.} $(\xg, \hf(\xg))$ and $(\hg(\xf), \xf)$ for $\xf, \xg \in [0,1].$ 
Density evolution (DE)
of the underlying (uncoupled) iterative system can then be viewed as a path drawn out by
moving alternately between these two curves (see Fig. \ref{fig:exitbec36}).
This path has the characteristic ``staircase'' shape.  
We will sometimes refer to the system being defined on $[0,1]\times[0,1]$ with this picture in mind.
The fixed
points of DE of the uncoupled system correspond to the
points where these two curves meet or cross. Assuming continuity of $\hf$ and $\hg,$ they are the
points $(\xg, \xf)$ such that $(\xg, \hf(\xg)) = (\hg(\xf), \xf).$

To help with analysis in the potentially discontinuous case we introduce the following notation.
For any $h \in \exitfns$ we write
\[
u \veq h(v)
\]
to mean $u \in [h(v-),h(v+)].$
\begin{definition}[Crossing Points]
Given $(\hf,\hg)\in\exitfns^2$ and we say that $(\xg, \xf)$ is a crossing point
if 
\[ 
\xg \veq \hg(\xf),\text{ and } \xf \veq \hf(\xg)\,.
\]
The following are three equivalent characterizations of crossing points.
\begin{itemize}
\item $u \veq \hfinv(v)$ and $v \veq \hginv(u),$
\item $u \veq \hg(v)$ and $u \veq \hfinv(v),$
\item $v \veq \hf(u)$ and $v \veq \hginv(u).$
\end{itemize}

The set of all crossing points will be denoted $\cross(\hf,\hg).$ 
It is easy to see that $\cross(\hf,\hg)$ is closed as a subset of $[0,1]^2.$
By definition of $\exitfns$,
we have $(0,0) \in \cross(\hf,\hf)$ and $(1,1) \in \cross(\hf,\hg).$ 
We term $(0,0)$ and $(1,1)$
the {\em trivial} crossing points and denote the non-trivial crossing points by
\[
\intcross (\hf,\hg) = \cross(\hf,\hg)\backslash \{(0,0),(1,1)\}.
\]
\end{definition}

If $(u,v)\in\cross(\hf,\hf)$ and $\hf$ and $\hg$ are continuous at
$u$ and $v$ respectively then $(u,v)$ is a fixed point of density
evolution.  In general, if $(u,v)\in\cross(\hf,\hf)$ then $(u,v)$
is a fixed point of density evolution for a pair of EXIT functions
equivalent to the pair $(\hf,\hg).$ 

\begin{lemma}\label{lem:crossorder}
For any $\hf,\hg \in \exitfns$ the set $\cross(\hf,\hg)$ is component-wise ordered,
i.e., given $(u_1,v_1),(u_2,v_2) \in \cross(\hf,\hg)$ we have  
$(u_2-u_1)(v_2-v_1)\ge 0.$
\end{lemma}
\begin{IEEEproof}
Let $(u_1,v_1),(u_2,v_2) \in \cross(\hf,\hg).$
If $u_1<u_2$ then $\hf(u_1+) \le \hf(u_2-)$ 
and, since  $v_1 \le \hf(u_1+)$ and $v_2 \ge \hf(u_2-),$
 we obtain $v_1 \le v_2.$
All other cases can be shown similarly.
\end{IEEEproof}

For a set $S,$ which may be a subset of  $[0,1]^2$ or a subset of $\reals,$
we use $\neigh{S}{\epsilon}$ to denote the $\epsilon$ neighborhood of $S:$
\begin{equation}\label{eqn:neighdef}
\neigh{S}{\epsilon} = \{ x: \exists y\in S, |y-x|<\epsilon\}\,.
\end{equation}

\begin{lemma}\label{lem:crosspointlimit}
If $(\hf^i,\hg^i)\rightarrow(\hf,\hg)$ then, for any $\delta>0,$ we have
$\cross(\hf^i,\hg^i) \subset \neigh{\cross(\hf,\hg)}{\delta}$ for
all $i$ sufficiently large.
\end{lemma}
\begin{IEEEproof}
Assume  $(u^i,v^i)\in\cross(\hf^i,\hg^i)$ converges in $i$ to a limit point $(u,v).$ 
Since $\hf^i(u) \rightarrow \hf(u)$ at points of continuity of $\hf$
it is easy to see that
\(
\liminf_{i\rightarrow \infty} \hf^i(u^i-)\ge \hf(u-)
\)
and that
\(
\limsup_{i\rightarrow \infty} \hf^i(u^i+)\le \hf(u+)
\)
and it follows that $v \veq \hf(u).$
Similarly, $u \veq\hg(v).$
Hence, $(u,v) \in \cross(\hf,\hg).$
 
Since $[0,1]^2 \backslash \neigh{\cross(\hf,\hg)}{\delta}$ is compact
and the same argument applies to subsequences
the lemma follows.
\end{IEEEproof}

\begin{lemma}
Consider initialization of system \eqref{eqn:DE} with an arbitrary choice of $u^0.$
Then the sequence $(u_1,v_1),(u_2,v_2),\ldots$ is monotonic (either non-increasing or non-decreasing)
in both coordinates.
\end{lemma}
\begin{IEEEproof}  
If $v^{t+1} = \hf(u^t) \ge v^t$ then $u^{t+1}=\hg(v^{t+1}) \ge \hg(v^t) = u^t.$
And if $u^{t+1} \ge   u^t$ then  $v^{t+2}=\hg(u^{t+1}) \ge \hg(u^t) = v^{t+1}.$
\end{IEEEproof}
It follows that the sequence $(u_i,v_i)$ converges and the limit point is clearly a crossing point of $(\hf,\hg).$
Thus, the limiting behavior of the scalar component system is governed by crossing points.
In the spatially coupled system the behavior often involves a pair
of crossing points from the underlying component system. 

Suppose $(u_1,v_1) < (u_2,v_2)$ are fixed points of the component DE.
If $\ff^0(x)\in[v_1,v_2]$ for all $x$
then $\ff^t(x)\in[v_1,v_2]$ and
then $\fg^t(x)\in[u_1,u_2]$ 
for all $x$ and $t.$ 
Thus, in this situation the system is effectively confined to $[u_1,u_2]\times [v_1,v_2].$
This circumstance occurs frequently but we can easily transform this into our canonical form.
We can introduce new coordinates $\tilde{u},\tilde{v}$ an affine transformation 
of $(u,v),$ characterized by its inverse affine map
\begin{align*}
u &   =  a \tilde{u} + b \\
v &   =  c \tilde{v} + d\,.
\end{align*}
By choosing $(b,d) = (u_1,v_1)$ and $(a,c) = (u_2-u_1,v_2-v_1)$
we map $(u_1,v_1)\rightarrow (0,0)$ and $(u_2,u_2)\rightarrow (1,1).$
Similarly, by choosing $(b,d) = (u_2,v_2)$ and $(a,c) = (u_1-u_2,v_1-v_2)$
we map $(u_2,v_2) \rightarrow (0,0)$ and $(u_1,v_1)\rightarrow(1,1).$
(This shows the symmetry that allows us to occasionally interchange $(0,0)$ and $(1,1)$
in the analysis.)
By rescaling the update functions appropriately we can thereby redefine the system on $[0,1]\times[0,1].$
In particular we can define ${\thf}(\tilde{u}) = \frac{1}{c}(\hf( a \tilde{u} + b) - d)$
and ${\thg}(\tilde{v}) = \frac{1}{a}(\hg( c \tilde{v} + d) - b).$

\bexample[EXIT Chart Analysis for the BEC] Figure~\ref{fig:exitbec36}
shows the EXIT chart analysis for the $(3, 6)$-regular ensemble
when transmission takes place over the BEC. The left picture shows
the situation when the channel parameter is below the BP threshold.
In this case we only have the trivial FP at $(0, 0)$.  According to our definition we
have $(1,1)$ as a crossing point, but it is not a fixed point because $\hf(1)=\epsilon<1.$ The right
picture shows a situation when we transmit above
the BP threshold. We now see two further crossings of the EXIT curves
and so $\cross(\hf,\hg)$ is non-trivial.

In this case we can renormalize the system according to our prescription as follows.
Consider the DE
equations stated in (\ref{eqn:DEBEC}).  If $(\xg^*, \xf^*)$ is the
largest (in both components) FP of the corresponding
DE and if we set $\hf(\xg)=\epsilon\lambda(\xg \xg^*)/\xf^*$
and $\hg(\xf)=(1-\rho(1 - \xf \xf^*))/\xg^*$ then system \eqref{eqn:DE}
is again equivalent to \eqref{eqn:DEBEC} on the restricted domain but, in addition, the component
functions are continuous at $0$ and $1$ and $(0,0)$ and $(1,1)$ are the relevant fixed points.  This rescaling is indicated
in the right picture of Figure~\ref{fig:exitbec36} through the
dashed gray box.  Since the standard (unscaled) EXIT chart picture
is very familiar in the coding context, we will continue to plot
the unscaled picture.  But we will always indicate the scaled version
by drawing a gray box as in the right picture of
Figure~\ref{fig:exitbec36}. This hopefully will not cause any
confusion. There is perhaps only one point of caution. The behavior
of the coupled system depends on certain areas in this EXIT chart.
These areas are defined in the scaled version and are different by
a factor $\xg^* \xf^*$ in the unscaled version.
{\hfill $\ensuremath{\Box}$} \eexample

So far we have considered the uncoupled system and seen that its
behavior can be characterized in terms of fixed points, or more generally crossing points. 
The behavior of the
spatially coupled system can also be characterized by its FPs.  For the
spatially coupled system a FP is not a pair of scalars, but a pair
of functions $(\tmplF(x), \tmplG(x))$ such that if we set $\ff^t(x)=\tmplF(x)$
and $\fg^t(x)=\tmplG(x)$, $t \geq 0$, then these functions fulfill
(\ref{eqn:gfrecursion}). One set of FPs are the constant functions
corresponding to the fixed points of the underlying component system.
The crucial phenomena in spatial coupling is the emergence of interpolating spatial
fixed points, i.e., non-constant monotonic fixed point solutions.
For the coupled system it is
fruitful not only to look at interpolating FPs but slightly more general objects,
namely interpolating {\em waves}. Here a wave is like a FP, except that it
{\em shifts}. I.e., for $(\tmplF(x), \tmplG(x))$ fixed and for some real
value $\ashift$, if we set $\ff^t(x)=\tmplF(x-\ashift t)$ and $\fg^t(x)=\tmplG(x-\ashift
t)$, $t \geq 0$, then these functions fulfill (\ref{eqn:gfrecursion}).
We will see that the behavior of coupled systems is governed
by the (non)existence of such waves and this
(non)existence has a simple graphical characterization in terms of
the component-wise EXIT functions and their associated FPs.  This
is the main technical content of this paper.  In fact, the {\em
direction} of travel of the wave
depends in a simple way on the EXIT functions and the area bound
by them.  
The extremal values of spatial wave solutions (the limit values at $\minfty$ and $\pinfty$)
are generally crossing points of the
underlying component system.  One important aspect of the analysis involves
determining the pairs of crossing points that can appear as such extremal values.
The answer is formulated in terms of the following definition.

\begin{definition}[Component Potential Functions]
For any pair $(\hf,\hg) \in \exitfns^2$ and point $(u,v) \in [0,1]^2,$ we define
\[
\altPhi(\hf,\hg;u,v) =  \int_0^u \hginv(u')\text{d}u' +  \int_0^v \hfinv(v')\text{d}v' \, - uv\,.
\]
\end{definition}

{\em Discussion:} The functional  $\altPhi$ serves as a potential function
for the scalar system. 
Assuming continuity of $\hfinv$ at $v$ and continuity of $\hginv$ at $u$ we have
$\nabla \altPhi(\hf,\hg; u,v) = (\hginv(u)-v,\hfinv(v)-u).$
Thus, under some regularity conditions
a crossing point $(\xg,\xf)$ is a stationary point of  $\altPhi.$
i.e., $ \nabla \altPhi(\hf,\hg; u,v) =0$
for $(u,v)\in \cross(\hf,\hg).$

In the definition of $\altPhi$ we have used
$(0,0)$ as an originating point.  We can choose the origin arbitrarily and it
is the differences in potential that matter most.
We have
\begin{align}\begin{split}\label{eqn:potdiff}
&\altPhi(\hf,\hg;u,v) - \altPhi(\hf,\hg;u_1,v_1)\\
= &
\int_{u_1}^u \hginv(u')\text{d}u' +  \int_{v_1}^v \hfinv(v')\text{d}v' \, - uv + u_1 v_1
\\= &
\int_{u_1}^u (\hginv(u')-v_1)\text{d}u' +  \int_{v_1}^v (\hfinv(v')-u_1)\text{d}v' \,
\\&\qquad - (u-u_1)(v-v_1)
\end{split}
\end{align}
and we see that taking differences as above
is equivalent to placing the origin at $(u_1,v_1).$ 

A straightforward calculation, noting that
\(
u\hf(u) = \int_0^u \hf(u')\, du' + \int_0^{\hf(u)} \hfinv(v)\, dv
\)
and
\(
v\hg(v) = \int_0^v \hg(v')\, dv + \int_0^{\hg(v)} \hginv(u)\, du\,,
\)
shows that for all $(\hf,\hg)$ we have
\begin{align}
\begin{split}\label{eqn:altPhiPhi}
&\altPhi(\hf,\hg;\hg(v),\hf(u))\\
=&
uv - (u-\hg(v))(v-\hf(u)) \\
& -  \int_0^u \hf(u')\text{d}u' - \int_0^v \hg(v')\text{d}v'\, .
\end{split}
\end{align}
A similar potential function form, along the lines of \eqref{eqn:altPhiPhi}, is
\begin{align*}
\Phi(\hf,\hg;u,v) = uv -  \int_0^u \hf(u')\text{d}u' - \int_0^v \hg(v')\text{d}v'\, .
\end{align*}
This functional is also stationary on the FPs of the component density evolution and is equal
to $\altPhi(\hf,\hg;\cdot,\cdot)$ on the crossing points points of $\hf,\hg.$
This form underlies the work in \cite{DJM11, TTK11b,  YJNP12a, YJNP12b}.
We prefer $\altPhi$ because of various properties developed below.  
A particularly useful fact is that density evolution is equivalent to coordinatewise successive minimization of $\altPhi$ (see Lemma \ref{lem:monotonic}).
The two forms are
related through Legendre transforms, e.g., $\int_0^u \hf(u')\text{d}u'$ is the Legendre transform
of $\int_0^v \hfinv(v')\text{d}v'.$ 
Evaluating $\altPhi$ at points on the graph of $\hf$ is equivalent,
up to reparametrization, to evaluating $\Phi$ on the graph of $\hg$
and vice-versa.
More specifically, we have 
\begin{align}
\label{eqn:altPhiuhfu}
\altPhi(\hf,\hg;& u,\hf(u)) = \Phi(\hf,\hg;u,\hginv(u))  \nonumber \\
&=  \int_0^u (\hginv(u')-\hf(u'))\,du'\\
\intertext{and}
\altPhi(\hf,\hg;& \hg(v),v) = \Phi(\hf,\hg;\hfinv(v),v)  \nonumber \\
&= \int_0^v (\hfinv(v')-\hg(v'))\,dv' \label{eqn:altPhivhgv}
\end{align}

\begin{lemma}\label{lem:monotonic}
The function $\altPhi(\hf,\hg;u,v)$ is convex in $u$ for fixed $v$ and convex in $v$ for fixed $u.$
In addition, for all $(u,v) \in[0,1]^2$ we have
\begin{align*}
\altPhi(\hf,\hg;u,v) &\ge \altPhi(\hf,\hg;u,\hf(u)) \\
\altPhi(\hf,\hg;u,v) &\ge \altPhi(\hf,\hg;\hg(v),v) 
\end{align*}
with equality holding in the first case if and only if $v\veq\hf(u)$
and
 in the second case if and only if $u\veq\hg(v).$
\end{lemma}
\begin{IEEEproof}
It is easy to check that $\altPhi(\hf,\hg;u,v)$ is Lipschitz (hence absolutely) continuous   
and we have almost everywhere
\begin{align}
\begin{split}\label{eqn:altPhiderivatives}
\frac{\partial}{\partial u} \altPhi(\hf,\hg;u,v)  &= \hginv(u) - v, \\
\frac{\partial}{\partial v} \altPhi(\hf,\hg;u,v)  &= \hfinv(v) - u \,.
\end{split}
\end{align}
The lemma now follows immediately from the monotonicity (non-decreasing) of $\hginv$ and $\hfinv.$
\end{IEEEproof}
We have immediately the following two results.
\begin{corollary}\label{cor:monotonic}
If $(u^0,v^0) \in [0,1]^2$ and we define $(u^t,v^t)$ for $t \ge 1$ via
\eqref{eqn:DE} then $\altPhi(\hf,\hg;u^t,v^t)$ is a non-increasing sequence in $t.$
\end{corollary}
\begin{corollary}\label{lem:miniscross}
We have $(u,v) \in\cross(\hf,\hg)$ if and only if
$\altPhi(\hf,\hg;u',v)$ is minimized at $u'=u$ and
$\altPhi(\hf,\hg;u,v')$ is minimized at $v'=v$ in some neighborhood of $(u,v).$
\end{corollary}



One of the key results on the existence of wave solutions, and especially spatial fixed points,
is that the crossing points associated to the extremal values of the solution are extreme
(minimizing) values of $\altPhi$ over the range spanned by the solution.  The  following definition characterizes this.

\begin{definition}[Strictly Positive Gap Condition]
We say that the pair of functions $(\hf,\hg)$ satisfies the {\em
strictly positive gap condition} if $\cross(\hf,\hg)$ is {\em
non-trivial} and if 
\[
(u,v) \in \intcross(\hf,\hg) \Rightarrow \altPhi(\hf,\hg;u,v) > \max\{0,A(\hf,\hg)\}
\] 
where we define the {\em total gap} $A(\hf,\hg) \defeq \altPhi(\hf,\hg;1,1).$
We say that the pair of functions $(\hf,\hg)$ satisfies the {\em
positive gap condition} (no longer strict) if $\cross(\hf,\hg)$ is {\em
non-trivial} and 
\[
(u,v) \in \intcross(\hf,\hg) \Rightarrow \altPhi(\hf,\hg;u,v) \ge \max\{0,A(\hf,\hg)\}\,.
\] 
 {\hfill $\ensuremath{\Box}$}
\end{definition}
For systems that are not normalized to $[0,1]^2$ we may
say a system satisfies the strictly positive gap condition over $[u,u']\times[v,v']$
where $(u,u')$ and $(v,v')$ are component fixed points.

{\em Discussion:} The (strictly) positive gap condition is related to the existence of interpolating spatial fixed point solutions.
In particular, we will see that systems possessing $(0,1)$-interpolating fixed point solutions must satisfy the positive gap condition
and have $A(\hf,\hg)=0.$  Systems satisfying the strictly positive gap condition with $A(\hf,\hg)=0$ will be proven to
possess $(0,1)$-interpolating spatial fixed point solutions.  The cases where $A(\hf,\hg) \neq 0$ correspond to $(0,1)$-interpolating traveling wave solutions.
In this case we show that the strictly positive gap condition is sufficient
for the existence of a wave-like solution but the positive gap condition
is not known to be necessary. We conjecture that it is not in fact necessary.

\blemma[Trivial Behavior]\label{lem:trivialbehavior}
If $\cross(\hf,\hg)$ is trivial then the system behavior is simplified
and under DE, i.e., under $\eqref{eqn:gfrecursion}$, the only spatial fixed points are with
$\ff^t$ and $\fg^t$ set to either the constant $0$ or the constant
$1,$ one of which is stable and one of which is unstable.
The system converges for all initial values, other than the unstable spatial fixed point itself,
to the stable spatial fixed point.
\elemma

Now that we have covered the ``trivial'' cases, let us consider
the system behavior when $\cross(\hf,\hg)$ is non-trivial. As we
will see, it is qualitatively different.  
The value of the total gap $A(\hf,\hg)$
plays an important role
in the behavior of the system. This is why we 
introduced a special notation for it.  The strictly positive gap condition
implies that the value of $\altPhi(\hf,\hg;\xg,\xf)$ for $(\xg,\xf)\in \intcross(\hf,\hg)$
is strictly larger than
the values $0$ and $A(\hf,\hg)$ found at the two trivial fixed points.  We will
see that this condition is related to the existence of wave-like
solutions that interpolate between the two trivial fixed
points.

\bexample[Positive Gap Condition for the BEC]
Figure~\ref{fig:positivegapbec36} illustrates the (strictly) positive
gap condition for the $(3, 6)$-regular ensemble when transmission
takes place over the BEC. The left picture shows the situation when
the channel parameter is between the BP and the MAP threshold of
the underlying ensemble. The right picture shows the situation when
the channel parameter is above the MAP threshold of the underlying
ensemble.  In both cases $\cross(\hf,\hg)$ contains one non-trivial
FP $(\xg, \xf)$ and for this FP $\altPhi(\hf,\hg;\xg,\xf) > \max\{0,A\}$, i.e.,
both cases fulfill the strictly positive gap condition.
In the first case $A>0$, whereas in the second case $A<0$.  We
will see in Theorem~\ref{thm:mainexist} below that this
change in the sign of $A$ leads to a reversal of direction of a wave-like
solution to the system and hence to fundamentally different asymptotic behavior.
Both pictures show the unscaled curve and the lightly shaded box
shows what the picture would look like if we rescaled it so that
the largest FP appears at $(1, 1)$.

It is not hard to see that the strictly positive gap
condition is necessarily satisfied for any $\hf,\hg$ for which $\cross(\hf,\hg)$
has a single non-trivial crossing point, and for which $(0,0)$ and
$(1,1)$ are stable fixed points under the DE equations \eqref{eqn:DE}.
{\hfill $\ensuremath{\Box}$}
\eexample

We are now ready to state the main result concerning the existence of interpolating wave solutions.

\begin{theorem}[Existence of Continuum Spatial Waves]\label{thm:mainexist}
Assume that $\smthker$ is a regular averaging kernel.
Let $(\hf,\hg)$ be a pair of functions in $\exitfns$ satisfying the strictly positive
gap condition.

%
Then there exist $(0,1)$-interpolating functions $\tmplF,\tmplG \in\sptfns$ and a
real-valued constant $\ashift,$ satisfying $\sgn(\ashift) = \sgn(A(\hf,\hg))$
and $|\ashift| \ge |A(\hf,\hg)|/\|\smthker\|_\infty$, such that setting
$f^t(x) = \tmplF(x - \ashift t)$ and $g^t(x) = \tmplG(x - \ashift t)$ for
$t=0,1,\ldots$ solves \eqref{eqn:gfrecursion}\,.
\end{theorem}

We remark that we can relax the regularity condition on $\smthker$ if
$\hf$ and $\hg$ are continuous; cf. Lemma \ref{lem:pathology}.

\begin{figure}[htp]
{
\centering
\input{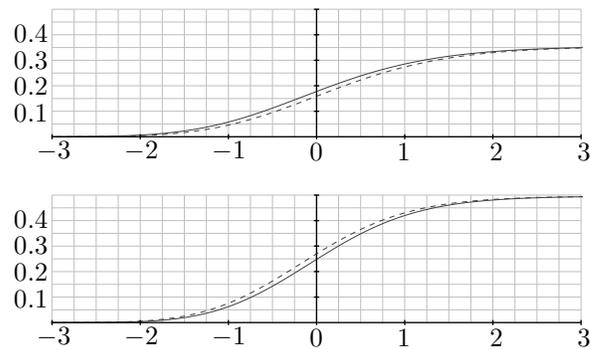}
}
\caption{\label{fig:spatialfpbec36}
FPs whose existence is guaranteed by Theorem~\ref{thm:mainexist}
for the $(3, 6)$ ensemble and transmission over the BEC. The top
picture corresponds to the cases $\epsilon=0.45$ and the bottom
picture to the case $\epsilon=0.53$.  In both cases we  used the
smoothing kernel $\omega(x)=\frac 12 \mathbbm{1}_{\{|x|\leq 1\}}$.  The dashed curve
is the result of applying one step of DE to the solid curve. As
predicted, in the top picture the curve moves to the right (the
corresponding gap $A$ in Figure~\ref{fig:positivegapbec36} is
positive) whereas in bottom picture the curve moves to the left
(the corresponding gap $A$ is negative).  The shifts are $0.142$ and
$-0.102$, respectively.  } \end{figure}

\bexample[Spatial wave for the BEC]
Figure~\ref{fig:spatialfpbec36} shows the spatial waves whose existence
is guaranteed by Theorem~\ref{thm:mainexist} for the $(3, 6)$
ensemble and transmission over the BEC. The top picture corresponds
to the cases $\epsilon=0.45$ and the bottom picture to the case
$\epsilon=0.53$.  In both cases we  used the smoothing kernel
$\omega(x)=\frac12 \mathbbm{1}_{\{|x|\leq 1\}}$.  As predicted, in
the first case the curve moves to the right by a value of
$0.142 \geq |A|/\|\smthker\|_\infty = 0.03125 \times 2 =0.06245$
and in the second case the curve moves to the left by an amount
of $0.101 \geq |A|/\|\smthker\|_\infty = 0.0253740 \times 2 = 0.0507498$.
 {\hfill $\ensuremath{\Box}$}
\eexample

One consequence of Theorem \ref{thm:mainexist} is that the existence of a
$(0,1)$-interpolating fixed point implies $A(\hf,\hg) = 0.$  This is true even without
regularity assumptions.

\begin{theorem}[Continuum Fixed Point Positivity]\label{thm:FPAzero}
Let  $\smthker$  be an averaging kernel (not necessarily regular) and
assume that there exists a $(0,1)$-interpolating fixed point solution to \eqref{eqn:gfrecursion}.
Then $(\hf,\hg)$ satisfies the positive gap condition and $A(\hf,\hg) = 0.$
\end{theorem}
A more general version of this result appears as Lemma \ref{lem:FPAzero}, for which
a proof is given.

Theorem \ref{thm:mainexist} is our most fundamental result concerning the spatially coupled system.
One limitation of the result arises in cases with infinitely many crossing points.  In such a case it
can be difficult to extract asymptotic behavior since there may exist many wave-like solutions
and the strictly positive gap condition may not hold globally.
For such cases we develop the following altered analysis.

Let $\hf$ and $\hg$ be given and define
\[
m(\hf,\hg)\defeq\min_{(u,v) \in [0,1]^2} \altPhi(\hf,\hg;u,v)
\]
and
\[
\cross_m(\hf,\hg) \defeq \{(u,v)\in\cross(\hf,\hg) : \altPhi(\hf,\hg;u,v) = m\}\,.
\]
Since $\altPhi(\hf,\hg;\cdot,\cdot)$ is continuous it follows that
$\cross_m(\hf,\hg)$ is closed.  Since $\cross(\hf,\hg)$ is component-wise linearly
ordered we can define
\begin{align*}
(u',v') & = \min \cross_m(\hf,\hg)\\
\intertext{and}
(u'',v'')& = \max \cross_m(\hf,\hg)\\
\end{align*}
where $\min$ and $\max$ are taken component-wise.

\begin{theorem}[General Continuum Convergence]\label{thm:globalconv}
Let $(\hf,\hg)$ be given as above, let $\smthker$ be regular,
and assume $\ff^0 \in \sptfns$ is given
with $\ff^0(\minfty)\le v''$ and 
$\ff^0(\pinfty)\ge v'.$
Then in system \eqref{eqn:gfrecursion} we have for all $x\in \reals$
\begin{align*}
\liminf_{t\rightarrow \infty} f^t(x) &\ge v' \quad  \liminf_{t\rightarrow \infty} g^t(x) \ge u' \\
\limsup_{t\rightarrow \infty} f^t(x) &\le v'' \quad \limsup_{t\rightarrow \infty} g^t(x) \le u'' \,.
\end{align*}
\end{theorem}
The proof may be found in appendix \ref{app:E}.

Note, in particular, that if $\altPhi$ is uniquely minimized at some point $(u,v),$ then
this point is a fixed point of the component system and if the spatial system is initialized (either $f$ or $g$) with this point (the appropriate coordinate)
 in the closed range spanned by the initial condition, 
i.e. $(u,v) \in [g(\minfty),g(\pinfty)]\times[f(\minfty),f(\pinfty)],$
then the coupled system
globally converges to the constant function associated to this fixed point.

\subsection{Discrete Spatial Sampling}

In many applications the setup is spatially discrete and finite length.
The analysis can be applied to these cases with suitable adjustments.  As a first step we state a
result analogous to Theorem \ref{thm:mainexist}
for a spatially discrete system.  The DE equations
for the spatially discrete version can be written as in
\eqref{eqn:gfrecursion} with the following modifications:
the variable $x$ is discrete, the averaging kernel is a discrete sequence, and
the convolution
operation is convolution of discrete sequences.  The analysis views the
spatially discrete problem as a sampled version of the continuum
version. In the limit of infinitely fine sampling the discrete
version converges to the continuum version.

Let $x_i = i \Delta$ and let $\discsmthker$ be a non-negative
function over $\integers$ that is even, $\discsmthker_i =
\discsmthker_{-i},$ and sums to $1,$ $\sum_i \discsmthker_i = 1.$
It is convenient to interpret $\discsmthker$ as a discretization of
$\smthker,$ i.e.,
\begin{equation}\label{eqn:kerdiscretetosmth}
\discsmthker_i = \int_{(i-\frac{1}{2})\Delta}^{(i+\frac{1}{2})\Delta} \smthker(z)\text{d}z.
\end{equation}
This relationship then makes it clear that the discrete ``width'' of
spatial averaging is inversely proportional to $\Delta.$
A good example is the smoothing kernel
$\smthker(x) = \frac{1}{2} \mathbbm{1}_{\{|x|\le 1\}}.$
If we set $\Delta = \frac{2}{2W+1}$ then $\discsmthker_i =
\frac{1}{2W+1} \mathbbm{1}_{\{|i|\le W\}}.$
Given a real-valued function $\fg$ defined on $\Delta\integers$ we will call 
the function $\tfg \in \sptfns,$  defined as 
$\tfg(x)=\fg(x_i)$ for $x\in[x_i-\Delta/2,x_i+\Delta/2),$
the {\em piecewise constant extension} of $\fg.$ 
Note that by this definition, we have
\begin{align*}
\tfg^{\smthker}(x_i)&=\int_{-\infty}^\infty \smthker(x_i-y)\tfg(y)dy
\\&=
\sum_{j=-\infty}^\infty
\int_{x_j-\Delta/2}^{x_j+\Delta/2} \smthker(x_i-y)\tfg(y)dy
\\
&=
\sum_{j=-\infty}^\infty \discsmthker_{i-j}\fg(x_j)\,
\\&=
\fg^{\discsmthker}(x_i)
\end{align*}

With this framework in mind, we can write the spatially discrete DE equations as follows.
\begin{equation}\label{eqn:discretegfrecursion}
\begin{split}
\fg^{t}(x_i) & =  \hg ((\ff^t \otimes \discsmthker) (x_i)) \\
\ff^{t+1}(x_i) & = \hf ( (\fg^{t} \otimes \discsmthker) (x_i) )\,.
\end{split}
\end{equation}

\bexample[Spatially Discrete DE for the BEC]
\begin{equation}\label{eqn:gfrecursionBEC}
\begin{split}
\fg^{t}(x_i) & =  1-\rho((\ff^t \otimes \discsmthker) (x_i)), \\
\ff^{t+1}(x_i) & = \epsilon \lambda( (\fg^{t} \otimes \discsmthker) (x_i) )\,.
\end{split}
\end{equation}
{\hfill $\ensuremath{\Box}$}
\eexample

An elementary but critical result relating the spatially continuous case to the
discrete case is the following.
\begin{lemma}\label{lem:disccontbnd}
Let $\tmplF \in \sptfns$ and let $f$ be a real valued function defined on $\Delta \integers.$
Then, if for all $i$ we have
\(
f(x_i) \le \tmplF(x_i)
\)
then 
\(
f^{\discsmthker}(x_i) \le \tmplF^{\smthker}(x_i+\half\Delta)
\)
and if
\(
f(x_i) \ge \tmplF(x_i)
\)
then 
\(
f^{\discsmthker}(x_i) \ge \tmplF^{\smthker}(x_i-\half\Delta)
\)
\end{lemma}
\begin{IEEEproof}
Assume
$\ff(x_i) \le \tmplF(x_i)$ (for all $i$).
Consider the piecewise constant extension $\tilde{\ff}.$ 
It follows that $\tilde{\ff}(x) \le  \tmplF(x+\half\Delta)$ for all $x$
and so $\ff^{\discsmthker}(x_i) = \tilde{\ff}^{\smthker}(x_i) \le  \tmplF^{\smthker}(x_i+\half\Delta)$
for each $i.$

The opposite inequality is handled similarly.
\end{IEEEproof}

Applying the lemma to system \eqref{eqn:discretegfrecursion} we obtain the following.
\begin{theorem}[Continuum-Discrete Bounds] \label{thm:mainquantize}
Assume that $\discsmthker$ is a discrete sequence
related to a regular smoothing kernel $\smthker$
as indicated in \eqref{eqn:kerdiscretetosmth}.
Let $\ff^t_c,\fg^t_c \in \sptfns,\, t=0,1,2,\ldots$ denote spatially continuous functions determined
according to \eqref{eqn:gfrecursion} and let
$\ff^t,\fg^t$ denote spatially discrete functions determined
according to \eqref{eqn:discretegfrecursion}.
If
$\ff^0(x_i) \le \ff^0_c (x_i)$ (for all $i$) then $\ff^t(x_i) \le \ff^t_c(x_i + t\Delta )$ and
$\fg^t(x_i) \le \fg_c^t(x_i + (t+\half) \Delta)$ for all $t.$
Similarly, if
$\ff^0(x_i) \ge \ff^0_c (x_i)$ (for all $i$) then $\ff^t(x_i) \ge \ff^t_c(x_i - t\Delta )$ and
if $\fg^t(x_i) \ge \fg_c^t(x_i - (t+\half) \Delta)$ for all $t.$
\end{theorem}
\begin{IEEEproof}
Assume
$\ff^0(x_i) \le \ff_c^0(x_i)$ (for all $i$).
By Lemma \ref{lem:disccontbnd} $\ftdS{0}(x_i) \le \ff_c^{0,\smthker}(x_i+\half\Delta)$ for each $i.$
By monotoniciy of $\hg$ we have
\[
\fg^0(x_i) = \hg(\ff^0(x_i))\le \hg(\ff_c^{0,\smthker}(x_i+\half\Delta)) = \fg_c^0(x_i+\half\Delta).
\]
By the same argument we obtain
\(
\gtdS{0}(x_i) \le \fg_c^{0,\smthker}(x_i+\Delta)\,
\)
and
\(
\ff^1(x_i)\le \ff_c^1(x_i+\Delta).
\)
The general result now follows by induction.

The opposite inequality can be handled similarly.
\end{IEEEproof}

This result is convenient to apply when there exist wave-like solutions.
For example, if $\ff_c^t(x) = \tmplF(x-\ashift t)$ with $\ashift>0$  and
$\tmplF$ is a $(0,1)$-interpolating function, then
$\ff^0(x_i) \le \ff_c^0(x_i),$ implies
$\ff^t(x_i) \le \tmplF(x_i-(\ashift-\Delta) t).$
Thus, if $\ashift > \Delta$ then we obtain asymptotic convergence for the
spatially discrete case.

\begin{theorem}[Discrete Spatial Convergence]\label{thm:mainqqqqq}
Assume that $\smthker$ is a regular averaging kernel.
Let $(\hf,\hg)$ be a pair of functions in $\exitfns$ satisfying the strictly positive
gap condition.
Assume $\Delta < |A(\hf,\hg)|/\|\smthker\|_\infty$
and initialize system \eqref{eqn:discretegfrecursion} with
any $(0,1)$ interpolating $\ff^0 \in \sptfns.$
If $A(\hf,\hg)>0$ then $\ff^t(x_i) \rightarrow 0$ and
if $A(\hf,\hg)<0$ then $\ff^t(x_i) \rightarrow 1$ 
for all $x_i$ 
\end{theorem}

This result gives order ${\Delta}$ convergence of the spatially discrete system to the continuum one (under positive gap assumptions).
Much faster convergence is observed in many situations. 
In \cite{HMU11b} a particular example is presented with a compelling heuristic argument for exponential convergence.
In general the rate of convergence appears to depend on the regularity of $\hf$ and $\hg$ and
$\smthker.$
A $(0,1)$-interpolating spatial fixed point does not sample $\hf$ and $\hg$ at every value, so one cannot
conclude that $A(\hf,\hg) = 0$ and,  indeed, this generally does not hold.  One can construct fixed point examples where
$|A(\hf,\hg)|$ is of order $\Delta.$
As a general result we have the following.
\begin{theorem}\label{thm:discreteFPDelta}
Assume $\hf$ and $\hg$ have a $(0,1)$-interpolating fixed point
for the spatially discrete system.  Then,
\[
|A(\hf,\hg) | \le {\Delta}\|\smthker\|_\infty\,.
\]
\end{theorem}
As indicated, regularity assumptions on $\hf,\hg$ can lead to stronger results. 
In this direction we have the following.
\begin{theorem}[$C^2$ Discrete Fixed Point Bound]\label{thm:discreteFPsum}
Assume $\hf$ and $\hg$ are $C^2$ and there exists an $(0,1)$-interpolating
spatial fixed point for the spatially discrete system.  Then
\[
|A(\hf,\hg) | \le 
\frac{1}{2} (\|\hf''\|_\infty+\|\hg''\|_\infty)\|\smthker\|_\infty^2{\Delta^2}
\]
\end{theorem}
Proofs for the above are presented in appendix \ref{app:discretecont}.
Note that they do not require regularity on $\smthker.$

For discrete systems where gap conditions may be difficult to verify we may require more general
results.
Especially challenging are cases with an infinite number of crossing points
clustering near the extremal ones. For such generic situations we have
the following spatially discrete version of Theorem \ref{thm:globalconv}.
\begin{theorem}[General Discrete Convergence]\label{thm:discreteglobalconv}
Let $(\hf,\hg)$ be given as in Theorem \ref{thm:globalconv}, let $\smthker$ be regular,
and assume $\ff^0 \in \sptfns$ is given
with $\ff^0(\minfty)\le v''$ and 
$\ff^0(\pinfty)\ge v'.$
Then, for any $\epsilon>0,$ in system \eqref{eqn:gfrecursion} 
with $\Delta$ sufficiently small we have for all $x\in \reals$
\begin{align*}
\liminf_{t\rightarrow \infty} f^t(x) &\ge v'-\epsilon \quad  \liminf_{t\rightarrow \infty} g^t(x) \ge u'-\epsilon \\
\limsup_{t\rightarrow \infty} f^t(x) &\le v''+\epsilon \quad \limsup_{t\rightarrow \infty} g^t(x) \le u''+\epsilon \,.
\end{align*}
\end{theorem}
The proof may be found in appendix \ref{app:E}.

\subsection{Termination}

Finite length systems can be modeled by introducing spatial dependence
into the definition $\hf$ and/or $\hg.$ For example, in the LDPC-BEC case termination corresponds
to setting $\hf = 0$ outside some finite region.
When $A(\hf,\hg)>0$ and the strictly positive gap condition holds we can apply Theorem \ref{thm:mainexist}
to conclude that the infinite length unterminated system has a wave-like solution
that converges point-wise to $0.$
Such a solution can often be used to bound from above the
solutions for terminated cases to show that
their solutions also tend to $0.$
Alternatively, we can apply Theorem \ref{thm:globalconv} to conclude that even if we remove the termination
after initialization the system will converge to $0.$ 

Setting $\hf=0$ over some region reduces $f$ relative to the
unterminated case making it more difficult to obtain lower bounds for
the terminated case.  It turns out for one-sided termination,
however, that an analogy can be drawn between the spatial variation
in $\hf$ and a global perturbation in $\hf$ that is spatially invariant
and which then allows application of Theorem \ref{thm:mainexist}.
Here we see a useful application of discontinuous $\hf.$

\subsubsection{One-sided Termination}

Let us formally define the one-sided termination version of
\eqref{eqn:gfrecursion} to be the system that follows
\eqref{eqn:gfrecursion}
except that when $x<0$ we set $f^t(x)=0$ regardless of $g^{t-1}.$
This is equivalent to redefining $\hf = \hf(u;x)$ to have spatial dependence so that
when $x<0$ we have $\hf(u;x)=0$  and for $x \ge 0$ we have $\hf(u;x) =\hf(u)$ as before.

Since this system is not translation invariant,
it does not admit interpolating traveling wave-like solutions.  It does, however,
admit interpolating spatial fixed points.

Let us introduce the notation
\[
\unitstep_a (x) \defeq \begin{cases}
0 & x  <0 \\
a & x = 0 \\
1 & x > 0
\end{cases}
\]
In some cases the value of $a$ is immaterial and we may drop the subscript from the 
notation. 

\begin{theorem}[Continuum Terminated Fixed Point]\label{thm:terminatedexist}
Assume $\smthker$ is regular.
Let $(\hf,\hg)\in \exitfns^2$ and assume that $\hg$ is continuous at $0$ and that
$\altPhi(\hf,\hg;\cdot,\cdot)$ is uniquely minimized at $(1,1)$
(hence $A(\hf,\hg) < 0$  but we do not assume that the strictly positive gap condition holds).
Then there exists $(0,1)$-interpolating $f,g \in \sptfns$
that form a fixed point of the one-sided termination of
\eqref{eqn:gfrecursion}.
\end{theorem}
\begin{IEEEproof}
Define ${\hf}(u;z) = \hf(u)\wedge \unitstep(u-z)$  (where $a \wedge b = \min \{a,b\}$) and choose $z\in (0,1)$ so that
$A(\hf(\cdot;z),\hg) =0.$
We claim that $(\hf(\cdot;z),\hg)$ satisfies the strictly positive gap condition.

Since $A(\hf(\cdot;z),\hg)=0$ we see that $\intcross(\hf(\cdot;z),\hg)$ cannot
be empty.
Let $(u,v) \in \intcross(\hf(\cdot;z),\hg)$ then, since $\hg$ is continuous at $0,$ we have
$u \ge z.$  If $u=z$ then clearly $\altPhi(\hf(\cdot;z),\hg;u,v) = \int_0^z \hginv(u)du>0,$ since $\hg$ is continuous at $0.$
If $u>z$ then $(u,v) \in \intcross(\hf,\hg)$ and
it now follows from \eqref{eqn:altPhiuhfu} that
\begin{align*}
\altPhi(\hf(\cdot;z),\hg;u,v) 
&= \altPhi(\hf(\cdot;z),\hg;u,v) - A(\hf(\cdot;z),\hg) 
\\&= \altPhi(\hf,\hg;u,v) - A(\hf,\hg) 
\\&> 0.
\end{align*}

By Theorem \ref{thm:mainexist} there exists $f,g \in \Psi_{[-\infty,\infty]}$
that form a $(0,1)$-interpolating spatial fixed point ($\ashift=0$) for
\eqref{eqn:gfrecursion} with $\hf(\cdot;z)$ replacing $\hf.$
It is easy to see that there is some finite maximal $y$ such that
$f(x)=0$ for $x < y.$  Translate $f$ and $g$ so that $y=0$ and
it follows that the resulting $f,g$ pair is a fixed point of the one-sided termination version of
\eqref{eqn:gfrecursion}.
\end{IEEEproof}
It is interesting to note in the above construction that the fixed point solution
has $g^{\smthker}(0)=z$ and $f(0+) = \hf(z+).$  Hence the value of 
the discontinuity at the boundary of the termination is determined by the condition
$A=0.$
In the case where $\hg$ is not continuous at $0,$ i.e., $\hg(0+)>0$ we can construct a fixed point solution
as above with $\tmplG(\minfty) =\hg(0+).$

For the case $A(\hf,\hg) \ge 0$ we have the following.
\begin{theorem}[Continuum Terminated Convergence]\label{thm:terminatedzero}
Assume $\smthker$ is regular.
Let $(\hf,\hg) \in \exitfns^2$ and assume that
$\altPhi(\hf,\hg;u,v) > 0$ for $(u,v) \neq (0,0).$
Then $\ff^t \rightarrow 0$ for the
one-sided termination of
\eqref{eqn:gfrecursion}
for any choice of $\ff^0.$
If $\hf(x) > 0$ and $\hg(x)>0$ on $(0,1]$ then
$\ff^t \rightarrow 0$ also when $\altPhi(\hf,\hg;\cdot,\cdot) \ge 0$
and $A(\hf,\hg) = 0.$
\end{theorem}
The proof is presented in Appendix \ref{app:E}.

We can, of course, also terminate the spatially
discrete versions of the system.
Thus, consider the one sided termination of
\eqref{eqn:discretegfrecursion}
in which the equations are modified so that
we set $f^t(x_i)=0$ if $x_i < 0,$
which is equivalent to redefining $\hf$ to have spatial dependence so that
$\hf = 0$ if $x_i < 0.$
We assume that $\discsmthker$ is related to $\smthker$ (for a continuum version)
as indicated in \eqref{eqn:kerdiscretetosmth}.
For this case we have the following quantitative result.
\begin{theorem}[Discrete Fixed Point Positive Gap]\label{thm:discreteterminatedexist}
Assume $\smthker$ is regular.
Let $(\hf,\hg)$ satisfy the strictly positive gap condition and assume that
$A(\hf,\hg) < -\Delta\|\smthker\|_\infty.$
Then there exists $(0,1)$ interpolating $f,g \in \sptfns$
that form a spatial fixed point of the one-sided termination of
\eqref{eqn:discretegfrecursion}.
\end{theorem}
\begin{IEEEproof}
Define $\hf(u;z) = \hf(u) \wedge \unitstep_1(u-z)$
 with $z\in (0,1)$ chosen sufficiently small
so that $A(\hf(\cdot;z),\hg)  \le -\Delta\|\smthker\|_\infty.$
By Theorem \ref{thm:mainexist}
there exists $\tmplF,\tmplG \in \Psi_{[-\infty,\infty]}$
that form a spatial wave solution for
\eqref{eqn:gfrecursion} with $\hf(\cdot;z)$ replacing $\hf$
and $\ashift \le -\Delta.$
By Theorem \ref{thm:mainquantize} we see that by setting
$\ff^0(x_i) = {\tmplF}(x_i)$ in
\eqref{eqn:discretegfrecursion} (the non-terminated case)
we have $\ff^1(x_i) \ge {\tmplF}(x_i - (s+\Delta) ) \ge {\tmplF}(x_i ).$
By translation, we can assume that ${\tmplF}(x_i)=0$ for $x_i<0.$
Now, the inequality $\ff^1(x_i) \ge {\tmplF}(x_i)$ also holds in the one sided termination case
since the values of $\ff^1(x_i)$ are unchanged from the unterminated
case for $x_i\ge 0.$
Thus, in the one-sided termination case the sequence $\ff^t$ is monotonically non-decreasing
for each $x_i$ and must therefore have a limit $\ff^{\infty}.$
If $\hf$ and $\hg$ are continuous then the pair $\ff^{\infty},\fg^{\infty}$  constitute a fixed point of the one-sided termination case.
If $\hf$ and $\hg$ are not continuous then it is possible that the pair $\ff^{\infty},\fg^{\infty}$ does not constitute a fixed point and that initializing with
$\ff^{\infty}$ we obtain another non-decreasing sequence.
In general we can use transfinite recursion together with monotonicity in $x$
to conclude the existence of
 a fixed point at least as large point-wise as $ (\ff^{\infty},\fg^{\infty}).$ 
\end{IEEEproof}

The previous result gives quantitative information on the discrete approximation but it requires the strictly positive gap assumption.
The following result, whose proof is in Appendix \ref{app:E},
removes that requirement at the cost of the quantitative bound.

\begin{theorem}[Discrete Fixed Point General]\label{thm:discreteterminatedexistB}
Assume $\smthker$ is regular.
Assume that
$\altPhi(\hf,\hg;\cdot,\cdot)$ is uniquely minimized at $(1,1)$
with $A(\hf,\hg) < 0.$ Then
for all $\Delta$ sufficiently small
there exists $f,g\in \sptfns$
that form a spatial fixed point of the one-sided termination of
\eqref{eqn:discretegfrecursion} with $\lim_{\Delta\rightarrow 0}f(\pinfty) =1.$
\end{theorem}

For the case $A(\hf,\hg) \ge 0$ we have the following quantitative result.
\begin{theorem}[Discrete Terminated Convergence]\label{thm:discreteterminatedzero}
Assume that $\smthker$ is regular.
Let $(\hf,\hg)$ satisfy the strictly positive gap condition and assume that
$A(\hf,\hg) > \Delta\|\smthker\|_\infty.$
Then $\ff^t \rightarrow 0$ for the
one-sided termination of
\eqref{eqn:discretegfrecursion}
for any choice of $\ff^0.$
\end{theorem}
\begin{IEEEproof}
Theorem \ref{thm:mainqqqqq} gives $\ff^t \rightarrow 0$ in the unterminated case
which clearly implies the same for the terminated case.
%
%
\end{IEEEproof}

%
%

\subsubsection{Two-sided Termination}
The two-sided termination of system \eqref{eqn:gfrecursion}
is defined by setting $\ff^t(x) = 0$ for
all $x$ outside some finite region, say $[0,Z]$ for all $t.$
This can be understood as a spatial dependence of $\hf =\hf(u;x)$
where $\hf(u;x) =0$ for $x \not\in [0,Z]$ and $\hf(u;x) =\hf(u)$ as before otherwise.
This system can be bounded from above by the one-sided termination case.
Thus, Theorem \ref{thm:terminatedzero}
and Theorem \ref{thm:discreteterminatedzero}
(convergence to $0$)
apply equally well to the two-sided terminated case.
Theorem \ref{thm:discreteterminatedexist} 
(interpolating fixed point existence)
on the other hand does not
immediately generalize, but a similar statement holds.

\begin{theorem}[Two Sided Continuum Fixed Point]\label{thm:twoterminatedexist}
Assume that $\smthker$ is regular.
Let $(\hf,\hg)$ satisfy the strictly positive gap condition and let
$A(\hf,\hg) < 0.$
Then, for any $\epsilon > 0,$ and for all $Z$ sufficiently large,
there exists $\ff,\fg$
that form a fixed point of the two-sided termination of
\eqref{eqn:gfrecursion} such that
$\ff$ and $\fg$ are symmetric about $\frac{Z}{2},$ monotonically non-decreasing on $(-\infty,\frac{Z}{2}]$ and have left and right limits at least
$1-\epsilon$ at $\frac{Z}{2}.$
\end{theorem}
The proof is presented in appendix \ref{app:C}.

We have also the following spatially discrete version of the above,
whose proof is also in appendix \ref{app:C}.
In the discrete case the termination is taken to hold for $x_i < 0$ and
$x_i > Z = L\Delta$ where $L$ is an integer.
Symmetry in the spatial dimension then takes the form
$\ff(x_i) = \ff(x_{L-i}).$ 
\begin{theorem}[Two Sided Discrete Fixed Point with Gap]\label{thm:discretetwoterminatedexist}
Assume that $\smthker$ is regular.
Let $(\hf,\hg)$ satisfy the strictly positive gap condition and assume that
$A(\hf,\hg) < -\Delta\|\smthker\|_\infty.$
Then, for any $\epsilon > 0,$ and for all $Z$ sufficiently large,
there exists $\tmplF,\tmplG$
that form a fixed point of the two-sided termination of
\eqref{eqn:discretegfrecursion}
such that
$\tmplF$ and $\tmplG$ are spatially symmetric, monotonically non-decreasing on $(-\infty,\half Z]$ and satisfy $\max_i \{\tmplF(x_i)\} \ge 1-\epsilon$
and $\max \{\tmplG(x_i)\} \ge 1-\epsilon.$ 
\end{theorem}

We have also the following qualitative version that relaxes the strictly positive gap condition
and
whose proof is in appendix \ref{app:E}.
\begin{theorem}[Two Sided Discrete Fixed Point]\label{thm:discretetwoterminatedexistGB}
Assume that $\smthker$ is regular.
Let $(\hf,\hg)$ be given such that $\altPhi(\hf,\hg;\cdot,\cdot)$ is
uniquely minimized at $(1,1)$ and therefore 
$A(\hf,\hg) < 0.$
Then, for any $\epsilon > 0,$ and for all $Z=L\Delta$ sufficiently large and $\Delta$ sufficiently small,
there exists $\tmplF,\tmplG$
that form a fixed point of the two-sided termination of
\eqref{eqn:discretegfrecursion}
such that
$\tmplF$ and $\tmplG$ are spatially symmetric, monotonically non-decreasing on $(-\infty,\half Z)$ and satisfy $\max_i \{\tmplF(x_i)\} \ge 1-\epsilon$
and $\max_i \{\tmplG(x_i)\} \ge 1-\epsilon.$ 
\end{theorem}

\subsection{Sensitivity to Irregular Smoothing and other Pathologies}\label{sec:pathology}

In this section we illustrate by example some of the subtlety that
can arise with non-regular smoothing kernels.  We also show how
non-uniqueness of fixed point solutions can occur when the positive
gap condition is satisfied but the strictly positive gap condition is not satisfied.

The following example shows that changing $\hf$ or $\hg$ on a set of measure $0$
can, for some choices  of $\smthker,$ have a dramatic effect on the solution
to \eqref{eqn:gfrecursion}.
Assume an averaging kernel $\smthker$ that is positive everywhere on $\reals$ except on
$[-2,2],$ where it equals $0.$
Consider
\[
\hf(u) = \unitstep_a(u-\frac{1}{2})
\,\,\text{ and }\,\,
\hg(u) = \unitstep_b(u-\frac{1}{2})
\]
where  $a$ and $b$ are specified below.
Let $\ff(x) = \unitstep(x),$ then we have 
we have 
$\fS(x)<\frac{1}{2}$ for $x \in (-\infty,-2),$
$\fS(x)=\frac{1}{2}$ for $x \in [-2,2],$ and
$\fS(x)>\frac{1}{2}$ for $x \in (2,\infty)\,.$
Consider initializing system \eqref{eqn:gfrecursion} with $\ff^0(x)=\unitstep(x).$ 
If $a=b=\frac{1}{2}$ then the solution is the fixed point
\[
\ff^t(x)=\fg^t(x) =\frac{1}{2}(\unitstep_1(x+2)+\unitstep_0(x-2))\,.
\]
If $a=b=1$ then the solution is 
\begin{align*}
\ff^t(x)&= \unitstep_1(x+4t)
\\
\fg^t(x)&=\unitstep_1(x+4t+2)\,,
\end{align*}
and $\ff^t(x) \rightarrow 1.$
If $a=b=0$ then the solution is 
\begin{align*}
\ff^t(x)&= \unitstep_0(x-4t)
\\
\fg^t(x)&=\unitstep_0(x-4t-2)\,,
\end{align*}
and $\ff^t(x) \rightarrow 0.$
If $a=0$ and $b=1$ then the solution is 
\begin{align*}
\ff^t(x)&= \unitstep_0(x)
\\
\fg^t(x)&=\unitstep_1(x-2)\,,
\end{align*}
another fixed point.

To give a more general example, let $\ff$ and $\fg$ be any functions in $\sptfns$ that
equal $0$ on $(\minfty,-1)$ and $1$ on $(1,\pinfty)$
then we have 
$\fg \veq \hg\circ\fS$ 
and
$\ff \veq \hf\circ\gS.$ 
It follows that for all such $f,g$ we have 
$h_{[\ff,\gS]} \equiv \hf$ and
$h_{[\fg,\fS]} \equiv \hg.$
(For an explanation of notation please see section \ref{sec:notation}.)
Hence, it is possible for some $\hf,\hg$ to have many distinct interpolating solutions that 
satisfy
$\fg \veq \hg\circ\fS$ 
and
$\ff \veq \hf\circ\gS.$ 
\subsubsection{Non-Unique Solutions}\label{sec:nonunique}
Let $\smthker = \frac{1}{2}\indicator{|x|<1}.$
and let $\tilde{f}$ and $\tilde{g}$ be any functions in $\sptfns$ that
equal $0$ on $(\minfty,-1)$ and $1$ on $(1,\pinfty)$ and take values in
$(0,1)$ on $(-1,1).$
Now consider
\begin{align*}
\ff_a(x) &= \frac{1}{2} \bigl(\tilde{f}(x+a)+\tilde{f}(x-a)\bigr) \\
\fg_a(x) &= \frac{1}{2} \bigl(\tilde{g}(x+a)+\tilde{g}(x-a)\bigr) 
\end{align*}
For all  $a>3$ we see that
$(h_{[\ff_a,\gS_a]},h_{[\fg_a,\fS_a]})$
does not depend on $a$ and the given functions form a family
of spatial fixed points for the system.
This gives an example where system \eqref{eqn:DE} exhibits multiple
spatial fixed point solutions.
Note that $(h_{[\ff_a,\gS_a]},h_{[\fg_a,\fS_a]})$ does not satisfy the strictly positive
gap condition since $\altPhi(h_{[\ff_a,\gS_a]},h_{[\fg_a,\fS_a]};\frac{1}{2},\frac{1}{2})=0.$

%


\section{Examples of 1-D Systems}\label{sec:applications}

\subsection{Binary Erasure Channel}
Let us start by re-deriving a proof that for transmission over the
BEC regular spatially-coupled ensembles achieve the MAP threshold
of the underlying ensemble. By keeping the rate fixed and by
increasing the degrees it then follows that one can achieve capacity
this way.  This was first shown in \cite{KRU10}.  Given the current
framework, this can be accomplished in a few lines. Before we prove this
let us see a few more examples.
\begin{figure}[htp]
{
\centering
\input{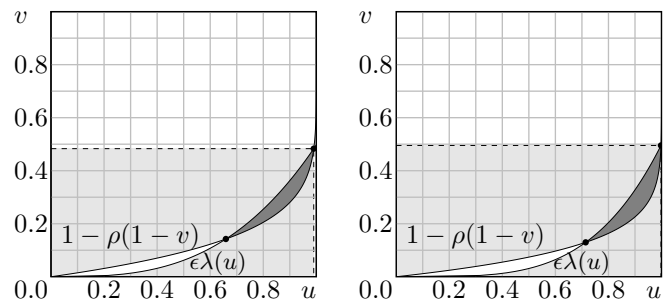}
}
\caption{\label{fig:capacitybec}
EXIT charts for the $(4, 8)$-regular (left) and the $(5, 10)$-regular (right)
degree distributions and transmission over the BEC. The respective coupled thresholds
are
$\epsilon^{\BPsmall}_{\text{\tiny coupled}}(4, 8)=0.497741$, and
$\epsilon^{\BPsmall}_{\text{\tiny coupled}}(5, 10)=0.499486$.
}
\end{figure}
We have already seen the corresponding EXIT charts for the $(3,
6)$-regular case in Figure~\ref{fig:exitbec36}. Figure~\ref{fig:capacitybec}
shows two more examples, namely the $(4, 8)$-regular as well as the
$(5, 10)$-regular case. Numerically, the thresholds are
$\epsilon^{\BPsmall}_{\text{\tiny coupled}}(3, 6)=0.48814$,
$\epsilon^{\BPsmall}_{\text{\tiny coupled}}(4, 8)=0.497741$, and
$\epsilon^{\BPsmall}_{\text{\tiny coupled}}(5, 10)=0.499486$. As
we see these thresholds quickly approach the Shannon limit of
one-half.

Consider now a degree distribution pair $(\lambda, \rho)$. The BP
threshold of the uncoupled system is determined by the maximum
channel parameter $\epsilon$ so that $\epsilon \lambda(x) \leq
1-\rho^{-1}(1-x)$ for all $x \in (0, 1]$. Therefore, dividing both
sides by $\lambda(x)$ we get for each $x \in (0, 1]$ an upper bound
on the BP threshold. In other words, the BP threshold of the uncoupled
ensemble can be characterized as
\begin{align*}
\epsilon^{\BPsmall}_{\text{\tiny uncoupled}} =
\inf_{x\in(0,1]}\frac{1-\rho^{-1}(1-x)}{\lambda(x)}\,.
\end{align*}
The limiting spatially coupled threshold (when $L$ and $w$ tend to infinity)
can be characterized in a similar way. In this case the determining quantity is
the area enclosed by the curves. Therefore,
\begin{align*}
\epsilon^{\BPsmall}_{\text{\tiny coupled}} =
\inf_{x\in(0,1]}\frac{\int_0^x 1-\rho^{-1}(1-u)\,\text{d}u}{\int_0^x \lambda(u)\text{d}u}\,.
\end{align*}
In the case where the BP threshold equals $\frac{1}{\lambda'(0)\rho'(1)},$
i.e., when the threshold equals the stability threshold, then the
spatially coupled threshold equals the BP threshold.

In the regular case and in many other cases
\begin{align*}
\epsilon^{\BPsmall}_{\text{\tiny coupled}} =
\frac{\int_0^{x^*} 1-\rho^{-1}(1-u)\,\text{d}u}{\int_0^{x^*} \lambda(u)\text{d}u}\,
\end{align*}
where $x^*$ corresponds to the forward BP fixed point with channel
parameter $\epsilon^{\small}_{\text{\tiny coupled}}.$ In this case
one can check that the threshold is exactly equal to the area
threshold. Further, we already know that the area threshold is an
upper bound on the MAP threshold of the underlying ensemble and we
know that the MAP threshold of the underlying system is equal to
the MAP threshold of the coupled system when $L$ tends to infinity.
We therefore conclude that for all such underlying ensembles where
the area threshold satisfies the strictly positive gap condition, the area
threshold equals the MAP threshold.

\begin{figure}[htp]
\centering
\input{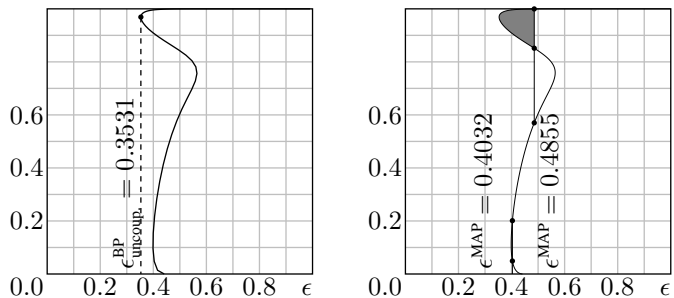}
\caption{\label{fig:complicated}  BP EXIT curves for the ensemble
$(\lambda(x)=\frac{3 x+3 x^2+14 x^{50}}{20}, \rho(x)=x^{15})$ and
transmission over the BEC. Left: Determination of the BP threshold.
Right: Determination of MAP behavior as conjectured by the Maxwell construction.}
\end{figure}

Our current framework can also be adapted to more complicated
cases.  The following example is from \cite[Fig. 4.15]{Mea06}.  Consider the
degree distribution $(\lambda(x)=\frac{3 x+3 x^2+14 x^{50}}{20},
\rho(x)=x^{15})$.  The left picture in Figure~\ref{fig:complicated}
shows the BP EXIT curve of the whole code.
As one can see, the BP threshold of the uncoupled ensemble in this
case is $\epsilon^{\BPsmall}_{\text{\tiny uncoup.}} = 0.3531$ and
the BP EXIT curve has a single jump.

The right picture shows the MAP EXIT curve according to the Maxwell
construction, see \cite[Section 3.20]{RiU08}. According to this
construction, the MAP EXIT curve has two jumps, namely at
$\epsilon=0.403174$, the conjectured MAP threshold, and at
$\epsilon=0.4855$. These two thresholds are determined by local
balances of areas. This is in particular easy to see for the threshold
at $\epsilon=0.4855$, where the two areas are quite large.

Let us now show that for the coupled ensemble the Maxwell conjecture
is indeed correct, i.e., we show that the asymptotic (in the coupling
length $L$) BP EXIT curve for the spatially-coupled ensemble indeed
looks as shown in the right-hand side of Figure~\ref{fig:complicated}.
To show that the Maxwell conjecture is also correct for the uncoupled
system requires a second step which we do not address here. This
second step consists in showing that the MAP behavior of the uncoupled
and coupled system is identical and is typically accomplished by using
the so called ``interpolation'' technique.

\begin{figure}[htp]
{
\centering
\input{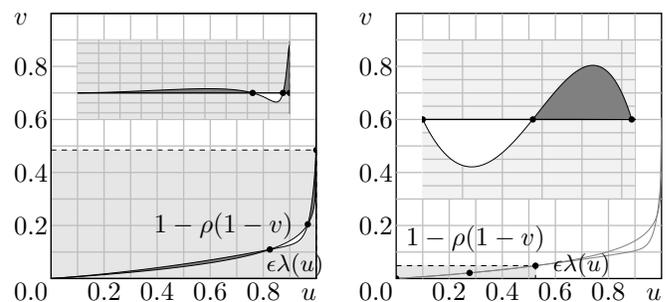}
}
\caption{\label{fig:complicatedgap}
Confirmation of the Maxwell conjecture using the one-dimensional framework of
spatial coupling for the ensemble $(\lambda(x)=\frac{3 x+3 x^2+14 x^{50}}{20}, \rho(x)=x^{15})$
and transmission over the BEC.
The two inlets show in a magnified way the behavior of the curves
inside the two gray boxes.
}
\end{figure}

The left picture in Figure~\ref{fig:complicatedgap} shows the
individual EXIT curves according to our framework for $\epsilon=0.4855$.
For this channel parameter the two EXIT curves cross four times,
namely for $u=0$, $u=0.824784$, $u=0.967733$, and $u=0.999952$.
Note that for this channel parameter the curves do not fulfill the
positive gap condition since initially the curve $\epsilon \lambda(\xg)$
is above the curve $1-\rho(1-\xf)$. Nevertheless we can use our
formalism. Let us explain the idea informally.  Let us first check
the behavior of the system for $\epsilon=0.4855$. Let us shift both
curves and renormalize them in such a way that first (from the left)
non-trivial FP is mapped to zero and the last FP (on the right) is
mapped to one. Then these curves {\em do} fulfill the our conditions
and our theory applies. This shows that once the channel parameter
has reached slightly below $0.4855$, the EXIT function drops as
indicated in the righ-hand side of Figure~\ref{fig:complicated}.

Now where we know what the curve looks like above $\epsilon=0.4855$
we can look at the remaining part.  The right picture in
Figure~\ref{fig:complicatedgap} shows the individual EXIT curves
according to our framework for $\epsilon=0.4032$. Again, we can
redefine our curves above this parameter and reparametrize and then
they do fulfill the positive gap condition. So this marks the second
threshold. The inlet shows the curve magnified by 1.5 and 15 respectively.
From this we see that the curves are quite well matched, so the areas
are not so easy to see.

\subsection{Hard-Decision Decoding}
Low-dimensional descriptions appear naturally when we investigate
the performance of quantized decoders. The perhaps simplest case
is the Gallager decoder A, \cite{Gal63} (see \cite{RiU01} for an
in-depth discussion). All messages in this case are from $\{\pm
1\}$. The initial message sent out by the variable nodes is the
received message.  At a check node, the outgoing message is the
product of the incoming messages. At variable nodes, the outgoing
message is the received message unless all incoming messages agree,
in which case we forward this incoming message.

Let $x^{(\ell)}$, $\ell \in \naturals$, be the state of the decoder,
namely the fraction of ``$-1$"-messages sent out by the variable
nodes in iteration $\ell$. We have $x^{(0)} = \epsilon$, and for
$\ell \geq 1$, the DE equations read
\begin{align*}
y^{(\ell)} & = \frac{1-\rho(1-2 x^{(\ell-1)})}{2}, \\
x^{(\ell)} & = \epsilon (1- \lambda(1-y^{(\ell)}))+ (1-\epsilon) \lambda(y^{(\ell)}).
\end{align*}
Since the state of this system is a scalar, our theory can be applied
directly. Unfortunately, as discussed in \cite{BRU04}, for most
(good) degree-distributions the threshold under the Gallager A
algorithm is determined by the behavior either at the very beginning
of the decoding process or at the very end. In neither of those
cases does spatial coupling improve the threshold.

\begin{figure}[htp]
{
\centering
\input{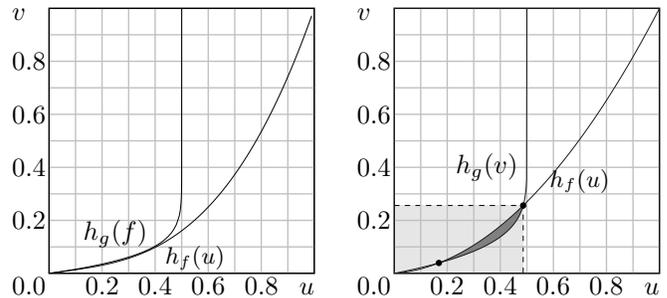}
}
\caption{\label{fig:exitgalA}
Left: EXIT charts for the $(4, 8)$-regular degree distribution
under the Gallager algorithm A with $\epsilon^{\GalAsmall}_{\text{\tiny uncoup}} = 0.0476$. The curves
do not cross. The threshold is determined by the stability condition. Right:
EXIT charts for the $(3, 6)$-regular degree distribution
under the Gallager algorithm A with $\epsilon^{\GalAsmall}_{\text{\tiny uncoup}} =  0.0395$. The
threshold is determined by the behavior at the start of the algorithm.
}
\end{figure}

In more detail, consider Figure~\ref{fig:exitgalA}.
The left picture shows the two EXIT functions for the $(4, 8)$-regular
ensemble under the Gallager algorithm A and
$\epsilon^{\GalAsmall}_{\text{\tiny uncoup}} = 0.0476$.  As one can
see from this picture, this is the threshold for the uncoupled case.
This threshold is determined by the stability condition, i.e., the
behavior of the decoder towards the end of the decoding process.
In other words, the functions $\hg(\xf)$ and the inverse of $\hf(\xg)$
have the same derivative at $0$.  If we increase the channel parameter
then the resulting EXIT curves no longer fulfill the positive gap
condition (since they cross already at $0$).  This implies that the
threshold of the spatially coupled ensemble is the same as for the
uncoupled one.

The right picture in Figure~\ref{fig:exitgalA} shows the two EXIT
functions for the $(3, 6)$-regular ensemble under the Gallager
algorithm A and $\epsilon = 0.0395$, the threshold for the uncoupled
case.  In this case the threshold is determined by the behavior at
the beginning of the decoding process.  As one can see from the
picture, there are two non-zero FPs. The ``smaller'' one is unstable
and the ``larger'' one is stable. If the initial state of the system
is below the small FP then the decoder converges to $0$, i.e., it
succeeds. But if it starts above the small FP, then the decoder
converges to the large and stable non-zero FP, i.e., it fails. As
one can see from the picture, already for the channel parameter
which corresponds to the threshold of the uncoupled these two EXIT
curves do not fulfill the positive gap condition -- the total area
enclosed by the two curves is negative. And if we increase the
channel parameter, the area would become even more negative. Hence,
also in this case spatial coupling does not help.

\begin{figure}[htp]
{
\centering
\input{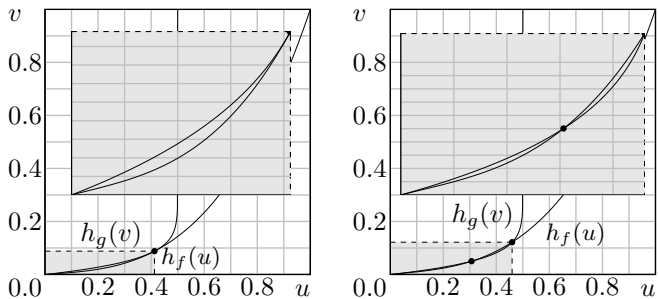}
}
\caption{\label{fig:exitgalB}
Left: EXIT charts for the
the $(4, 10)$-regular ensemble and the Gallager algorithm B with $b=3$
and $\epsilon^{\GalBsmall}_{\text{\tiny uncoup}} = 0.02454$. The curves do not cross.
Right: The same example but with
$\epsilon^{\GalBsmall}_{\tiny \text{coup}} = 0.0333$.  For
this channel parameter the positive gap condition is fulfilled and
the two areas are in balance. In both cases, the inlets show a
magnified version of the gray box.}
\end{figure}

Let us therefore consider the Gallager algorithm B, \cite{Gal63,RiU01}.
As for the Gallager algorithm A, all messages are from the set
$\{\pm 1\}$. The initial message and the message-passing rule at
the check nodes are identical. But at variable nodes we have a
parameter $b$, an integer. If at least $b$ of the incoming messages
agree, then we send this value, otherwise we send the received
value. This threshold $b$ can be a function of time. Initially the
internal messages are quite unreliable. Therefore, $b$ should be
chosen large in this stage (if we choose $b$ to be the degree of
the node minus one we recover the Gallager algorithm A). But as
time goes on, the internal messages become more and more reliable
and a simple majority of the internal nodes will be appropriate.
The DE equations for this case are
\begin{align*}
y^{(\ell)} & = \frac{1-\rho(1-2 x^{(\ell-1)})}{2}, \\
x^{(\ell)} = & (1-\epsilon) \sum_{k=b}^{\dl-1}
\binom{\dl-1}{k} (y^{(\ell)})^k (1-y^{(\ell)})^{\dl-1-k}\\
& + \epsilon \sum_{\dl-1-b}^{\dl-1}
\binom{\dl-1}{k} (y^{(\ell)})^k (1-y^{(\ell)})^{\dl-1-k}.
\end{align*}
Assume at first that we keep $b$ constant over time.  Consider
the $(4, 10)$-regular ensemble and choose $b=3$.
The left picture in Figure~\ref{fig:exitgalB} shows this example for
$\epsilon^{\GalBsmall}_{\text{\tiny uncoup}} = 0.02454$. As we can see, this is the largest
channel parameter for which the two curves do not cross, i.e., this is
the threshold for the uncoupled case.
The right picture in Figure~\ref{fig:exitgalB} shows the same example
but for $\epsilon^{\GalBsmall}_{\tiny \text{coup}} = 0.0333$.  For
this channel parameter the strictly positive gap condition is fulfilled and
the two areas are exactly in balance, i.e., this is the threshold
for the coupled ensemble.
We see that the increase in the threshold is substantial for this
case.

We can do even better if we allow $b$ to vary as a function of the state of the system.  The optimum
choice of $b$ as a function of the state $x$ was already determined
by Gallager and we have
\[
b(\epsilon, x) = \Big\lceil \Bigl(\frac{\log \frac{1-\epsilon}{\epsilon}}{\log \frac{1-x}{x}} + (\dr-1) \Bigr)/2 \Big\rceil.
\]
Assume that at any point we pick the optimum $b$ value. For the
EXIT charts this corresponds to looking at the minimum of the EXIT
chart at the variable node over all admissible values of $b$.  If
we apply this to the $(4, 10)$-regular ensemble then we get a
threshold of $\epsilon^{\GalBsmall, \text{\tiny opt}}_{\tiny
\text{coup}}(4, 8) = 0.04085$, another marked improvement. As a
second example, consider the $(6, 12)$-regular ensemble.  For this
ensemble no fixed-$b$ decoding strategy improves the threshold under
spatial coupling compared to the uncoupled case.  But if we admit
an optimization over $b$ then we get a substantially improved
threshold, namely the threshold is now $\epsilon^{\GalBsmall,\text{\tiny
opt}}_{\tiny \text{\tiny coup}}(6, 12) = 0.0555$.  For comparison,
$\epsilon^{\GalBsmall}_{\text{\tiny uncoup}}(6, 12) = 0.0341$.

{\em Discussion:} The optimum strategy assumes that at the
decoder we know at each iteration (at at each position if we consider
spatially coupled ensembles) the current state of the system. Whether
or not this is realistic depends somewhat on the circumstances. For
very large codes the evolution of the state is well predicted by
DE and can hence be determined once and for all. For smaller systems
the evolution shows more variation. One option is to measure e.g.
the number unsatisfied check nodes given the current decisions and
to estimate from this the state.

\subsection{CDMA  Demodulation}

Spatially coupling has been applied to CDMA demodulation in
 \cite{ScT11} and  \cite{TTK11}.
We will follow \cite{ScT11} in our exposition.

The basic (real, uncoded) CDMA  transmission model is
\[
y=\sum_{k=1}^Kd_k \bold{a}_k  + \sigma \bold{n}
\]
where there are $K = \alpha N$ users, each transmitting a single bit $d_k = \pm 1$  
using random spreading sequence $\bold{a}_k$ of unit energy and length $N$,
and $\bold{n}$ is a vector of length $N$ of independent $N(0,1)$ random variables
(for further details see \cite{ScT11}).

In \cite{Tan2002} statistical mechanical methods were used to analyze randomly spread synchronous CDMA detectors over the additive white Gaussian noise channel.
The non-rigorous replica method was used to predict the asymptotic (in system size) performance of various detectors.
In this setting the solution states that
the symbol-wise marginal-posterior-mode detector in the large $K$ and $N$ limit (with $\alpha=K/N$ held fixed) has posterior probabilities with signal to interference ratio $(1/z)$ satisfying the equation
\begin{equation}\label{eqn:cdmaFP}
z= 
\sigma^2 + \alpha \expectation \Biggl(1-\tanh\Bigl(\frac{1}{z}+\sqrt{\frac{1}{z}} \xi\Bigr) \Biggr)^2 
\end{equation}
where the expectation is over $\xi \sim N(0,1).$
Here $z$ represents the variance of the posterior equivalent Gaussian channel $d_k + \sqrt{z} n.$

For $\alpha < \alpha_{\text{crit}} \simeq 1.49$ (numerically determined) this
equation has single solution
(including the case $z=0$ for $\sigma^2=0.$)
For $\alpha \ge \alpha_{\text{crit}}$ it is observed
that the equation has one, two, or three solutions depending 
on $\sigma^2.$

In  \cite{ScT11}, a message passing scheme was developed such that the associated density evolution gives rise to
\eqref{eqn:cdmaFP} as a fixed point equation.
The scheme requires a modification of the transmission setup which we will now describe.
First consider repeating each bit $M$ times so $d_{k} \boldmath{a}_k$ is simply rewritten as
$\frac{1}{M} \sum_{m=1}^M d_{k,m} \boldmath{a}_k$ where $d_{k,m} = d_k.$
Now, take $l=1,2,...,L$ instances (e.g. successive  transmissions) of this system, so we write
$d_{k,m,l},$ and permute indices  so that the $l$th signal for user $k$ is
$\frac{1}{\sqrt{M}}\sum_{m=1}^M d_{k,\pi_k(m,l)} \boldmath{a}_k$ 
where $\pi_k$ is a (randomizing) permutation on $[M]\times [L].$
Note the change in scaling with respect to $M$ due to
non-coherent addition of the bit values.
(This may require $L \gg  M$ and/or some constraint on $\pi_k.$)
The received signal for instance $l$ is now given by
\[
y_l=\sum_{k=1}^K \frac{1}{\sqrt{M}} \sum_{m=1}^M d_{k\pi_k(m,l)} \boldmath{a}_k  + \sigma \boldmath{n}
\]
In  \cite{ScT11} belief propagation is applied to this setup and the analysis
leads to the density evolution fixed point equation \eqref{eqn:cdmaFP}. 

The DE system can be expressed in our framework as follows.  Define 
$\Psi:[0,\infty] \rightarrow [0,1].$
\begin{align*}
\Psi(z) & = \expectation{(1-\tanh{(z+\sqrt{z}\xi))}^2} 
\end{align*}
where $\xi \sim N(0,1).$
Now, further define
\begin{align*}
\hf(u)&=\alpha \Psi(u)+\sigma^2\\
\hg(v)&=1/v
\end{align*}
where, we note, $\hf(u) \in [\sigma^2,\sigma^2 + \alpha].$
The fixed point equation \eqref{eqn:cdmaFP} can now be written
\[
z = \hf(\hg(z))\,.
\]
The function $\hf$
corresponds to updating the LLRs  of the bits taking into account the repetition of the bits and the function
$\hg$ corresponds to a soft cancellation step.  In each case the resulting message LLR values are (symmetric) 
Gaussian distributed and the density evolution update corresponds to the input-output map of the effective
variances of the equivalent AWGN channel.  The iterations can be initialized with $z = \infty$ although a single iteration will reduce it to $\sigma^2 + \alpha.$

The DE corresponding to the message passing decoder will converge to the solution of
\eqref{eqn:cdmaFP} having the largest magnitude.
Hence for $\alpha \ge \alpha_{\text{crit}}$ the BP decoder will not generally achieve optimal performance.

In \cite{ScT11} the authors further modify the scheme to
introduce spatial coupling.  The basic construction uses a chain of
instances of the above system and couples them by exchanging bits between neighboring instances

The spatially coupled version of \eqref{eqn:cdmaFP}  (corresponding to local uniform coupling of width $W$) 
appearing in \cite{ScT11} reads
\begin{align*}
z_i^t & = \sigma^2 + \frac{\alpha}{2W+1}\sum_{j=-W}^W \Psi\Bigl( \frac{1}{2W+1} \sum_{l=-W}^W \frac{1}{z_{i-1}^{t+j+l}}\Bigr)\, \\
&=  \frac{1}{2W+1}\sum_{j=-W}^W \Biggl(\sigma^2 +\alpha \Psi\Bigl( \frac{1}{2W+1} \sum_{l=-W}^W \frac{1}{z_{i-1}^{t+j+l}}\Bigr)\Biggr)\,.
\end{align*}
Termination is accomplished by setting bits outside some finite region of the chain to be known which in effect sets $z$ to $0.$

We are now in the regime where our results may be applied.
We will discuss only the continuum case and
we assume the non-trivial conditions, i.e. we assume $\alpha$ large enough and $\sigma^2$ small enough so that
there a three fixed point solutions to DE equations.
Let $z_2 = v_2 = 1/u_2$ be the smallest solution and  let $z_1 = v_1 = 1/u_1$ be the largest solution (the solution found by DE for the component system).
Let us first consider the case $\sigma^2>0$ where $z_1$
and $z_2$ are necessarily finite  and the
component DE is essentially
confined to the region $[\sigma^2,\sigma^2+\alpha] \times [(\sigma^2+\alpha)^{-1},\sigma^{-2}].$

To make closer contact with our framework
it is helpful to make a change of variables.
Let $\bar{u} > \frac{1}{\sigma^2}$ 
be a large value and define $u'=\bar{u}-u$
and $v' = v.$  Consider
\begin{align*}
\thf(u')&=\alpha \Psi(\bar{u}-u')+\sigma^2\\
\thg(v')&=\bar{u}-1/v'
\end{align*}
In this equivalent formulation of the system $\thf$ and $\thg$ are increasing
and the extreme fixed points are $(u'_i,v'_i)=(\bar{u} - u_i,v_i)$ for $i=1,2.$
Slightly abusing notation,
let us identify the potential $\altPhi(\hf,\hg;u,v)$  with
$\altPhi(\thf,\thg;u'_1,v'_1).$  Then we have 
\begin{align*}
&\altPhi(\hf,\hg;u_1,v_1) -\altPhi(\hf,\hg;u_2,v_2)
\\&=\altPhi(\thf,\thg;u'_1,v'_1)  - \altPhi(\thf,\thg;u'_2,v'_2)
\\&=\int_{u'_2}^{u'_1}\thg^{-1}(u') - \thf(u')  \, du'
\\&=\int_{u'_2}^{u'_1} \frac{1}{\bar{u}-u'} -\thf(u')  \, du'
\\&=\int_{u_1}^{u_2} \Bigl( \frac{1}{u} - \hf(u)   \Bigr) \, du
\\&=\int_{1/z_1}^{1/z_2} \Bigl(\frac{1}{u} - \hf(u)  \Bigr) \, du
\end{align*}
Hence the potential $\altPhi$
is uniquely minimized at $(u_2,v_2)$ when
\[
\int_{1/z_1}^{1/z_2} \Bigl( \frac{1}{u} -\hf(u)   \Bigr) \, du > 0\,.
\]
If, under this condition, we initialize the 
unterminated continuum system with $f^0$ such that $f^0(\minfty)\le z_2$
and $f^0(\pinfty)\ge z_2$ then we can now
conclude from Theorem \ref{thm:globalconv} that $f^t(x) \rightarrow z_2$ for all $x.$
For the terminated case the limiting solution can be only smaller.

The case $\sigma^2=0$ ($z_2 = 0$) is special because $\hg(v)$ is unbounded in this case.  We can easily handle this case by degrading the
system slightly by replacing $\thf$ with $\thf  \wedge 1/\bar{u}.$
That is, we limit $\hf$ to be at least $1/\bar{u},$ effectively saturating
$\hg$ at $\bar{u}.$  Assuming $\bar{u}$ large enough
the maximal fixed point of the modified system is
at $(\bar{u},\alpha \Psi(\bar{u}))$ (where $\Psi(\bar{u}) \ll 1/\bar{u}.$)
For $\bar{u}$ large enough we have
\[
\int_{u_1}^{\bar{u}} \Bigl(  \frac{1}{u}- \hf(u)\Bigr) \, du > 0\,.
\]
If we initialize the 
unterminated continuum system with $f^0$ such that $f^0(\minfty) = 0$
then, applying Theorem \ref{thm:globalconv}, the limit will be the constant function $f(x)=1/\bar{u}.$
Since $\bar{u}$ is arbitrarily large we see that the unmodified system
converges to $f(x)=0.$
This was the main claim in \cite{ScT11}.
The case $\sigma^2 >0$ was treated more recently in \cite{6620553}.

%


\subsection{Compressed Sensing}

In a typical compressed sensing scenario one observes a ``sparse" vector $x$
through a underdetermined linear system as
\[
y = Ax +n\,.
\]
where $n$ is an additive noise vector.
The matrix  $A$ is  $m \times n$ typically with 
$m \ll n$ where $\delta= m/n$ is termed the undersampling ratio.
The vector $x$ is constrained to be sparse, or, alternatively, to have
entries distributed according to a distribution $p_X$ with small R\'{e}nyi information dimension
\cite{VWcs}. 
In the setup we consider here the entries of $A$ are independently sampled zero mean Gaussians random variables.
Letting $V$ denote the $m \times n$ all-1 matrix, the
 variances of the entries of $A$ are component-wise given by $\frac{1}{m} V$ so that columns of 
$A$ have (approximately and in expectation) unit $L_2$ norm.
The problem is to estimate $x$ from knowledge of $y$ and $A.$
Here we also assume knowledge of $p_X.$
The problem can be scaled up by letting $n$ and $m$ tend to infinity while keeping
$\delta$ fixed.  
Asymptotic performance is characterized in terms of the large system limit.

One can associate a bipartite graph to $A$ in which one set of nodes corresponds to 
the columns (and the entries of $x$) and the other set of nodes corresponds to the rows (and the entries of $y$).
The graphical representation suggests the use of  message passing algorithms for
this problem and they have indeed been proposed and studied, see \cite{DMM} and references therein.
In \cite{DMM} a reduced complexity variation, AMP (Approximate Message Passing), is developed in which there are
only $n$ or $m$ distinct messages, depending on the direction.
An additional term, the so-called Onsager reaction term, is brought into the algorithm to compensate
of the feedback inherent in AMP (due to the violation of the extrinsic information principle and the denseness of the graph).
In \cite{DMM} an analysis of AMP is given that leads in the large system limit
to an iterative function system called
state evolution, which is analogous to density evolution.
The large system limit analysis is quite different from the usual density evolution analysis in that,
rather than relying on sparseness and tree-like limits, the state evolution analysis relies on the central limit theorem and the fact that contributions from single edges are asymptotically negligible.
In the large system limit, messages (or their errors) in the AMP algorithm are normally distributed 
(this is the important consequence of the including the Onsager reaction term)
and state evolution captures the variance (SNR) associated to the messages.
For our current setup: a $m\times n$ sensing matrix with independent $\frac{1}{\sqrt{m}}N(0,1)$ Gaussian entries 
and known $p_X,$
the state evolution equations take the form
\cite{DJM11}
\begin{align*}
z_{t+1} = \sigma^2+\frac{1}{\delta}\text{mmse}(z_t^{-1})
\end{align*}
where $z$ is the estimation error variance.  In this expression
\[
\text{mmse}(s) =
\expectation (X-\expectation(X\mid Y))^2
\]
is the minimum mean square error of an estimator of $X$ given $Y$
where $X$ is distributed as $p_X$ and $Y = \sqrt{s}X+Z$ where $Z$ is $N(0,1)$
and independent of $X.$
The main aspects of $\text{mmse}$ that are relevant here are
\begin{align*}
\bar{D}_{p_X} &\defeq \limsup_{s\rightarrow \infty} s\,\text{mmse}(s),\,\,\,
\underline{D}_{p_X} \defeq \liminf_{s\rightarrow \infty} s\,\text{mmse}(s)   \\
\intertext{and the closely related quantity}
\bar{d}_{p_X}
&\defeq \limsup_{\ell \rightarrow \infty} \frac{H\lfloor \ell X \rfloor}{\log \ell}
\end{align*}\marginpar{check}
where  $H$ denotes the Shannon entropy and $\lfloor \cdot \rfloor$ is the integer-valued floor function.
The quantity $\bar{D}_{p_X}$ is termed the mmse dimension \cite{VWcs} 
and $\bar{d}_{p_X}$ is the upper information dimension \cite{VWcs} of
$p_X.$
Under some regularity conditions one has $\bar{d}_{p_X} = \bar{D}_{p_X}.$
Further, under some mild regularity condition on $p_X$ we have
\begin{align}\label{eqn:CSbnd}
\limsup_{s\rightarrow \infty} \frac{1}{\log(s)}\int_0^s \text{mmse}(u) du = \bar{d}_{p_X}
\end{align}
(see \cite{DJM11}[Prop. 7.15]).

Spatial coupling can be introduced by imposing additional structure on $A.$
Let us first consider a collection of parallel systems.
Thus, let $\tilde{A}$ be a doubly infinite array of $m\times n$ matrices in 
consisting of i.i.d. Gaussian samples with entry-wise variance matrix $\frac{1}{m}V.$
The variance matrix associated to matrix $\tilde{A}$ is 
$\tilde{V}$ with $\tilde{V}_{i,i}= \frac{1}{m}V$ and $\tilde{V}_{i,j}=0$ for $i\neq j.$
Spatial coupling is achieved by setting $V_{i,j} =  w_{i-j} \frac{1}{m} V.$
Termination can be effected by providing additional measurements for variables associated
to the termination.
Spatially coupled constructions of this type and resulting performance improvements were first presented
in \cite{KMSSZ11}. 
The analytical results on information theoretic optimal performance that we reproduce here were presented in
\cite{DJM11}.

The spatially coupled system can be understood within our framework as having the following exit functions.
\begin{align*}
\hf(u) &= \sigma^2+\frac{1}{\delta} \text{mmse}(u)\\
\hg (v) &= 1/v 
\end{align*}
We see that the form is very similar to the CDMA detection case
and we assume a definition of $\altPhi$ in an analgous fashion.
The behavior of $\text{mmse}$  is potentially more complicated then that of $\Psi$ but the basic analysis is similar.
Assuming $\expectation_{p_x}(X^2) < \infty$ we have $\text{mmse}$ is bounded above and
so $\hf(u)$ is bounded.
There is a crossing point $(u_1,v_1)$ where $u_1$ is minimal and $v_1$ is maximal.
It is easy to see that we have the bound $u_1 \ge \frac{\delta}{ \expectation(X^2)}$
since $\hf$ is decreasing in $u$ and $\hf(0) = \frac{1}{\delta} \expectation(X^2).$

We can now easily recover the main results in \cite{DJM11}.
Consider first the noiseless case $\sigma^2 = 0.$  The FP of interest in the
component system above occurs at $(\infty,0).$
If $\bar{d}_{p_X} < \delta$ then we have by \eqref{eqn:CSbnd}
\[
\int_{u_1}^{\infty} \bigl(\frac{1}{u}-\frac{1}{\delta} \text{mmse}(u)\bigr) du
=
\infty\,.
\]
The spatially coupled system with $f(x)$ initialized to $0$ for $x\le 0$
then converges  to $f(x)=0.$
(The mmse error, $f,$ converges to $0.$) 
Some simple adjustment of our arguments, as in the CDMA case, are needed to handle this unbounded case.

Consider now $\sigma^2 > 0.$ 
As in the CDMA case we can write
\begin{align*}
&\altPhi(\hf,\hg;u_1,v_1) - \altPhi(\hf,\hg;u,\hf(u))
\\= &
\int_{u_1}^{u} \bigl(\frac{1}{u'} - (\sigma^2 + \frac{1}{\delta} \text{mmse}(u')\bigr) du'
\\ = &
\log(u/u_1) - \sigma^2 (u-u_1) -\frac{1}{\delta}\int_{u_1}^{u}  \text{mmse}(u) du \,.
\end{align*}

It is clear that for any $z$ and all $\sigma^2>0$ $\altPhi(\hf,\hg;u,\hf(u))$
is bounded below for $u \le z.$  Assuming $\bar{d}_{p_X}<\delta$ and $\sigma^2$ small
enough then $\altPhi(\hf,\hg;u,\hf(u))$ will be minimized for some $u>z.$
It follows that for the spatially coupled system arbitrarily small error can be achieved.

Let 
$(u^*(\sigma^2),v^*(\sigma^2))$ denotes the crossing point
with maximal $u$ and minimal $v.$
Assume the stronger condition that $\bar{D}_{p_X}<\delta$  then for all $\sigma^2$ small enough we have
$(u^*(\sigma^2),v^*(\sigma^2))$ minimizes $\altPhi(\hf,\hg).$
Furthermore it follows that 
$1-\delta^{-1}\bar{D}_{p_X}\lesssim \sigma^2 u^*(\sigma^2)   \lesssim 1-\delta^{-1}\underline{D}_{p_X}\,.$ 
In this case it follows that the unterminated spatially coupled system (suitably initialized)
will converge to this
minimal crossing point.

\section{Higher-Dimensional Systems and the Gaussian Approximation}\label{sec:gaussapprox}
We have discussed in the previous section several scenarios where
the state of the system is one dimensional and the developed theory
can be applied directly and gives precise predictions on the threshold
of coupled systems. But we can considerably expand the field of
applications if we are content with {\em approximations}.  For
uncoupled systems a good example is the use of EXIT functions.  EXIT
functions are equivalent to DE for the case of the BEC, where the
state is indeed one dimensional. For transmission over general BMS
channels they are no longer exact but they are very useful engineering
tools which give accurate predictions and valuable insight into
the behavior of the system.

The idea of EXIT functions is to replace the unknown message densities
appearing in DE by Gaussian densities. If one assumes that the
densities are symmetric (all densities appearing in DE are symmetric)
then each Gaussian density has only a single degree of freedom and
we are back to a one-dimensional system. Clearly, the same approach
can be applied to coupled systems.  Let us now discuss several
concrete examples. We start with transmission over general BMS
channels.

\subsection{Coding and Transmission over General Channels}
As we have just discussed, for transmission over general BMS channels
it is natural to use EXIT charts as a one-dimensional approximation
of the DE process \cite{teB99a,teB99b,teB00,teB01}.  This strategy
has been used successfully in a wide array of settings to approximately
predict the performance of the BP decoder. As we have seen, whereas
for the BP decoder the criterion of success is that the two EXIT
curves do not overlap, for the performance of spatially coupled
systems the criterion is the positive gap condition and the area condition.

We demonstrate the basic technique by considering the simple setting
of point-to-point transmission using irregular LDPC ensembles. It
is understood that the same ideas can be applied to any of the many
other scenarios where EXIT charts have been used to predict the
performance of the BP decoder of uncoupled systems.

In the sequel, let $\psi(m)$ denote the function which gives the
entropy of a symmetric Gaussian of mean $m$ (and therefore standard
variation $\sigma=\sqrt{2/m}$). Although there is no elementary
expression for this function, there are a variety of efficient
numerical methods to determine its value, see \cite{RiU08}.

Define the two functions
\begin{align*}
\hg(\xf) & = 1-\sum_{i} \rho_i \psi\bigl((i-1)\psi^{-1}(1-\xf)\big), \\
\hf(\xg)  & = \sum_{i} \lambda_i \psi\bigl((i-1)\psi^{-1}(\xg)+\psi^{-1}(c) \big).
\end{align*}
Note that $\hg(\xf)$ describes the entropy at the output of a check
node assuming that the input entropy is equal to $\xf$ and $\hf(\xg)$
describes the entropy at the output of a variable node assuming
that the input entropy is equal to $\xg$ and that the entropy of the
channel is $c$. Both of these functions are computed under the
assumption that all incoming densities are symmetric Gaussians (with
the corresponding entropy). In addition, for the computation of the
function $\hg(\xf)$ we have used the so-called ``dual'' approximation,
see \cite[p. 236]{RiU08}.

\begin{figure}[htp]
{
\centering
\input{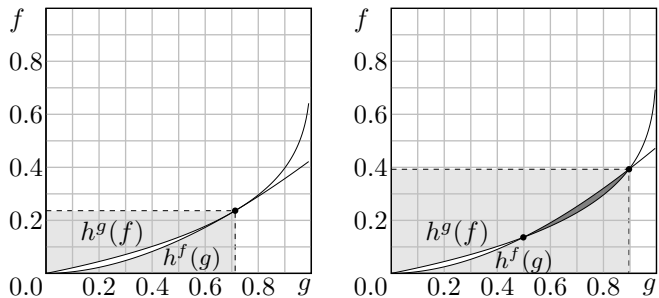}
}
\caption{\label{fig:positivegapbawgnc36} Left: Determination of the
BP threshold according to the EXIT chart paradigm for the $(3,
6)$-regular ensemble and transmission over the BAWGNC. The two
curves are shown for $\ent^{\BPsmall, \EXITsmall}=0.42915$. As one
can see from this picture, the two curves touch but do not cross.
Right: Determination of the BP threshold for the coupled ensemble
according to the EXIT chart paradigm and the positive gap condition.
The two curves are shown for  $\ent^{\BPsmall, \EXITsmall}_{\text{\tiny
coupled}}=0.4758$.  For this parameter the ``white'' and the ``dark
gray'' area are in balance.  } \end{figure}

Fig.~\ref{fig:positivegapbawgnc36} plots the EXIT charts for the
$(3, 6)$-regular ensemble and transmission over the BAWGNC. The
plot on the left shows the determination of the BP threshold for
the uncoupled system according to the EXIT chart paradigm. The
threshold is determined by the largest channel parameter so that
the two curves do not cross. This parameter is equal to $\ent^{\BPsmall,
\EXITsmall}=0.42915$.  Note that according to DE the BP threshold
is equal to $\ent^{\BPsmall} = 0.4293$, see \cite[Table 4.115
]{RiU08}, a good match.

The plot on the right show the determination of the BP threshold
for the coupled ensemble according to the positive gap condition.
Since for this case we only have a single nontrivial FP, this
threshold is given by the maximum channel entropy so that the gap
for the largest FP is equal to $0$.  This means, that for this
channel parameter the ``white'' and the ``dark gray'' area are
equally large.  This parameter is equal to $\ent^{\BPsmall,
\EXITsmall}_{\text{\tiny coupled}}=0.4758$.  Note that according
to DE, the BP threshold of the coupled system is equal to
$\ent^{\BPsmall}_{\text{\tiny coupled}} = 0.4794$, see \cite[Table
II]{KRU12}, again a good match.

\subsection{Min-Sum Decoder}
As a second application let us consider the min-sum decoder.  The
message-passing rule at the variable nodes is identical to the one
used for the BP decoder.  But at a check nodes the rule differs --
for the min-sum decoder the sign of the output is the product of
the signs of the incoming messages (just like for the BP decoder)
but the absolute value of the outgoing message is the minimum of
the absolute values of the incoming messages. 

For, e.g., the $(3, 6)$-regular ensemble DE predicts a min-sum
decoding threshold on the BAWGNC of $\ent^{\MinSumsmall}_{\text{\tiny
uncoup}}=0.381787$, \cite{Chu00}. For the coupled case this threshold
jumps to $\ent^{\MinSumsmall}_{\text{\tiny
coupled}}=0.429$.\footnote{Strictly speaking it is not known that
min-sum {\em has} a threshold, i.e., that there exists a channel
parameter so that for all better channels the decoder converges
with high probability in the large system limit and that for all
worse channels it does not. Nevertheless, one can numerically compute
``thresholds'' and check empirically that indeed they behave in the
expected way. }

In order to derive a one-dimensional representation of DE , we
restrict the class of densities to symmetric Gaussians. Of course,
this introduces some error. Contrary to BP decoding, the messages
appearing in the min-sum decoding are not in general symmetric (and
neither are they Gaussian).

The DE rule at variable nodes is identical to the one used when we
modeled the BP decoder. The DE rule for the check nodes is more
difficult to model but it is easy to compute numerically.  

Rather than plotting EXIT charts using entropy, we use the error
as our basic parameter.  There are two reasons for this choice.
First, our one-dimensional theory does not depend on the choice of
parameters and so it is instructive see an example which uses a
parameter other than entropy. Second, the min-sum decoder is
inherently invariant to a scaling, whereas entropy is quite sensitive
to such a scaling. Error probability on the other hand is also
invariant to scaling.

Figure~\ref{fig:minsum36} shows the predictions we get by applying our
one-dimensional model.
\begin{figure}[htp]
{
\centering
\input{ps/minsum36}
}
\caption{\label{fig:minsum36}
Left: Determination of the MinSum threshold according to the EXIT
chart paradigm for the $(3, 6)$-regular ensemble and transmission
over the BAWGNC. The two curves are shown for $\ent^{\BPsmall,
\EXITsmall}=0.401$. As one can see from this picture, the two curves
touch but do not cross.  Right: Determination of the MinSum threshold
for the coupled ensemble according to the EXIT chart paradigm and
the positive gap condition.  The two curves are shown for
$\ent^{\MinSumsmall, \EXITsmall}_{\text{\tiny coupled}}=0.436$.  For
this parameter the ``white'' and the ``dark gray'' area are in
balance.  } \end{figure} 
The predicted thresholds are $\ent^{\MinSumsmall, \EXITsmall}_{\text{\tiny
uncoup}}= 0.401$, $\ent^{\MinSumsmall, \EXITsmall}_{\text{\tiny
coupled}}=0.436$.  These predictions are less accurate than the
equivalent predictions for the BP decoder.  Most likely this is due
to the lack of symmetry of the min-sum decoder.  But the predictions
still show the right qualitative behavior.

\section{Analysis and Proofs}\label{sec:proof}

\subsection{Outline}

In this section we present the analysis that
leads to the proof of existence of wave-like solutions to the
spatially coupled system.
In section \ref{sec:notation} we introduce some notation 
and some elementary results.
In section \ref{sec:SFPW} we give a more genearlized
characterization of solutions to  \eqref{eqn:gfrecursion}.
In general a pair of functions can be consistent with the fixed point
equations without being a solution of the recursion and we will refer to 
such pairs as consistent solutions.
The distinction between consistent solutions and proper solutions
can arise only in the case of discontinuous exit functions.
It is relatively easy given an interpolating pair $(f,g)$ to give exit functions for
which the pair are a consistent spatial fixed point.  We use this connection to obtain
necessary conditions on spatial fixed points associated to exit functions. 
In section \ref{sec:SFPI}, we present a fundamental technical result
that integates spatial fixed points to recover local evaluation of
$\altPhi$ in the spatial fixed point.
This result easily shows the necessity of the positive gap condition.
In section \ref{sec:BTR} we provide some upper and lower
bounds the translation speed of wave-like solutions.
This along with results in sections \ref{sec:MGC} and \ref{sec:limitthms},
which, respectively, look at properties of the component potential function under iteration and
compactness properties of solution spaces to \eqref{eqn:gfrecursion}
form the foundation for the existence proofs.
In section \ref{sect:inverse} we present a formulation of \eqref{eqn:gfrecursion}
in terms of inverse functions.  This formulation underlies the existence proof 
for the piecewise constant case, which is presented in Section \ref{sec:PCcase}.
The proof uses a method of continuation in which we obtain the desired spatial fixed point
as the solution of a differential equation.
This basic results is extended to show the
existence of consistent spatial waves under the strictly positive gap condition
in Section \ref{sec:ECSW}.
Finally, in Section \ref{ESWS} we show that proper
solutions of the recursion can be obtain.

\subsection{Notation\label{sec:notation}}
In the analysis we allow discontinuous update (EXIT) functions.
This is not merely for generality
but also for modeling of termination and to allow discontinuous perturbations.
We will require some notation for dealing with this.

Given a monotonically non-decreasing function $\ff$ we write
\[
v \veq \ff(u)
\]
to mean $v \in [\ff(u-),\ff(u+)].$
Given $\fg\in\sptfns,$ continuous $\ff \in \sptfns,$ and $h\in\exitfns,$ we write
\[
\fg \veq h\circ\ff
\]
to mean 
$\fg(x)\veq  h(\ff(x)),$
for all $x\in \reals.$ We write
\[
\fg = h\circ\ff
\]
to mean $\fg(x) = h(\ff(x))$ for all $x.$
In some contexts we may have equality holding
up to a set of $x$ of measure $0.$
To distinguish this we write
\[
\fg \equiv h\circ\ff
\]
to mean $\fg(x) = h(\ff(x))$ for all $x$
up to a set of measure $0.$
Note that modifying $\fg$ on a set of measure $0$ has no impact on $\gS$
so there is little significant difference between $\equiv$ and $=$ in this case.
We use $\equiv$ generally to indicate equality up to sets of measure $0.$

Given a real number $\ashift$ we use the notation
$\gSa$ to denote the reverse shift of $\gS$ by $\ashift,$ i.e.,
\[
\gSa(x) = \gS(x+\ashift) \,
\]
and $\gSa_x(x)$ do denote $\frac{d}{dx}\gSa(x).$
Ultimately we are interested in interpolating functions such that
\(
\fg = \hg\circ\fS\,,
\)
and
\(
\ff = \hf\circ\gSa\,,
\)
since this represents a wave-like solution to system \ref{eqn:gfrecursion}.
The mathematical arguments, however, sometimes only guarantee functions
{\em consistent} with the equations, i.e., such that
\(
\fg \veq \hg\circ\fS\,,
\)
and
\(
\ff \veq \hf\circ\gSa\,.
\)

An important role is played by the area bound between the EXIT functions.
We introduce some notation to characterize that area. Let 
\begin{align*}
G^+(\hf,\hg) \defeq &\{(u,v):v > \hf(u+) \text{ and } u>\hg(v+)\}\\
\intertext{and}
G^-(\hf,\hg) \defeq &\{(u,v):v < \hf(u-) \text{ and } u<\hg(v-)\}.
\end{align*}
Then, by \eqref{eqn:altPhiuhfu} for example,
\begin{equation}\label{eqn:AGG}
A(\hf,\hg) = \mu(G^+) - \mu(G^-)
\end{equation}
where $\mu(G)$ denotes the 2-D Lebesgue measure of $G.$

\begin{lemma}\label{lem:stposbound}
Let $(\hf,\hg) \in \exitfns^2$   satisfy the strictly positive gap condition.
Then $\mu(G^+) >0$ and $\mu(G^-) >0.$
\end{lemma}
\begin{IEEEproof}
If the strictly positive gap condition is satisfied then there
exists $(u^*,v^*) \in \intcross(\hf,\hg)$ with
$\altPhi(\hf,\hg;u^*,v^*) > \max \{0,A(\hf,\hg)\}.$
Let $R = [0,u^*]\times[0,v^*].$ Now, it follows from \eqref{eqn:altPhiuhfu} that
\[
\altPhi(\hf,\hg;u^*,v^*)
=
 \mu(G^+ \cap R ) - \mu(G^- \cap R )
\]
hence $\mu(G^+ \cap R ) >  \max \{0,A(\hf,\hg)\}$
and it follows from \eqref{eqn:AGG} that $\mu(G^-) > 0.$
\end{IEEEproof}

\subsection{Converse of Spatial Fixed Points and Waves \label{sec:SFPW}}

Although is typically difficult to analytically determine an interpolating spatial fixed point solution $(\ff,\fg)$ 
for a given pair $(\hg,\hf),$ 
the reverse determination is relatively straightforward.
In particular, given a putative $(0,1)$-interpolating spatial fixed point $(\ff,\fg)$
the corresponding  $(\hf,\hg)$ is essentially determined by the requirement that
$\fg(x) \veq \hg(\fS(x))$ and $f(x) \veq \hf(\gS(x)).$
The trace of the parametric curve $(\gS(x),f(x))$ is essentially a subset of the graph
of $\hf$ and if $f$ and $g$ are $(0,1)$-interpolating then $\hf$
is essentially deteremined.
Some degeneracy is possible if, for example, $\gS$ is constant over some interval on which $\ff$ varies. 
In such a case $\hf$ is necessarily discontinuous. 
Even in this degenerate case, however, the equivalence class of $\hf$ is uniquely determined.
Thus, given $(0,1)$-interpolating $\ff$ and $\fg$  where $\fg$ is continuous, we define
$h_{[\ff,\fg]}$ to be any element of the uniquely determined equivalence class such that
\[
\ff \veq h_{[\ff,\fg]}\circ\fg\,.
\]
(A simple argument shows that the equivalence class is indeed uniquely determined.)
In general, if $f$ and $g$ are not $(0,1)$-interpolating, then we still consider 
$h_{[\ff,\fg]}$ to be defined on $[\fg(\minfty),\fg(\pinfty)]$ and the inverse to be defined on
$[\ff(\minfty),\ff(\pinfty)].$ 

If $(\ff,\fg)$ is a $(0,1)$-interpolating spatial fixed point solution to \eqref{eqn:gfrecursion}
then we have $h_{[\ff,\gS]} \equiv \hf$ and $h_{[\fg,\fS]} \equiv \hg.$
In the reverse direction, $h_{[\ff,\gS]} \equiv \hf$ and $h_{[\fg,\fS]} \equiv \hg$ implies,
and, (assuming $\ff$ and $\fg$ are $(0,1)$-interpolating) is in fact equivalent to,
\begin{align} \label{eqn:consistentFP}
\fg \veq \hg\circ \fS,\quad
\ff \veq \hf\circ \gS 
\end{align}
but does not in general imply the stronger condition
\begin{align}\label{eqn:equivFP}
\fg \equiv \hg\circ \fS,\quad
\ff \equiv \hf\circ \gS \,.
\end{align}
If $\hf$ and $\hg$ are continuous then this equivalence, and in fact equality, is implied.

\subsection{Interpolating spatial fixed point integration.\label{sec:SFPI}}

Consider a $(0,1)$-interpolating spatial fixed point $(\ff,\fg).$
Then, at $v=\fS(x_1)$ the integral $\int_0^v \hg$
can be expressed as
\[
\int_0^v \hg(z) \text{d}z=
\int_{-\infty}^{x_1} \fg(x) \bigl(\frac{d}{dx} \fS(x )\bigr) \text{d}x
=
\int_{-\infty}^{x_1} \fg(x) \text{d}\fS(x)\,.
\]
Similarly, since $g(x_1+) = \hg(\fS(x_1)+)$
\[
\int_0^{g(x_1+)} \hginv(u) \text{d}u=
\int_{-\infty}^{x_1+} \fS(x) \text{d}g(x)\,
\]
and since $f(x_2+) = \hf(\gS(x_2)+)$
\[
\int_0^{f(x_2+)} \hfinv(z)\text{d}z = \int_{-\infty}^{x_2+} \gS(x) \text{d}\ff(x)\,.
\]
The product rule of calculus reads $\fg(x) \text{d}\ff(x) + \ff(x) \text{d}\fg(x) =  \text{d}(\fg(x)\ff(x))$ and, were it not for the
spatial smoothing, this would solve directly the sum of the above two integrals in terms of the product
$\fg(x)\ff(x).$  By properly handling the spatial smoothing we can accomplish something similar, and the result is
presented in Lemma \ref{lem:twofint}.
We find a succinct formula for the evaluation of
$\altPhi(\hf,\hg;\fg(x_1),\ff(x_2))$ that is {\em local}
in its dependence on $\ff$ and $\fg.$  
This formula captures valuable information concerning
the $(0,1)$-interpolating spatial fixed point solution and its relation to $\altPhi.$ 

Given $(\ff, \fg) \in \sptfns^2$ and  an even averaging kernel $\smthker$ we define
\begin{align}\label{eqn:definexi}
\PhiSI (\smthker;f,g;x_1,x_2) & \defeq
(\fS(x_1) - f(x_2+))
(\gS(x_2) - g(x_1+))\nonumber
\\ &\quad
+
\altPhiSI(\smthker;f,g;x_1,x_2)
\end{align}
where
\begin{align*}
\altPhiSI&(\smthker; f,g;x_1,x_2) \defeq 
\\&
\iint dg(y) df(x) (\ind_{T_1}\Omega (x-y)
+ \ind_{T_2}\Omega (y-x))
\end{align*}
where 
\[T_1 = \{ (x,y): x\le x_2, y>x_1 \}\]
 and
\[T_2 = \{ (x,y): x > x_2, y\le x_1\}.\]  Note that in $(x,y)\in T_1$ implies
$x-y < x_2-x_1$ and  $(x,y)\in T_2$ implies
$y-x < x_1-x_2.$  Since $\Omega(z) = 0$ for $z<-W$
we see that the expression is local up to $W.$
The integrand has positive support only in the region $|x-y|<W$
and $T_1$ and $T_2$ are each quadrants with vertex at 
$(x_2,x_1).$
Note that $\altPhiSI$ is non-negative.

Note that if $\fg$ is discontinuous at $x_1$
then $\altPhiSI(\smthker; f,g;x_1,x_2)$ is discontinuous at $x_1,$ 
and similarly for $\ff$ at $x_2.$
However, $\PhiSI(\smthker;f,g;x_1,x_2)$ is continuous in $x_1$ and $x_2$
as can be verified directly.

\begin{lemma}[Spatial fixed point integration]\label{lem:twofint}
Let $f,g \in \sptfns$ satisfies
\begin{align*} 
\fg \veq \hg\circ \fS,\quad
\ff \veq \hf\circ \gS 
\end{align*}
where $\smthker$ is an even averaging kernel, then
\begin{align*}
&\quad \int_{\fg(\minfty)}^{\fg(x_1+)} \hginv (u)\text{d}u
+ \int_{\ff(\minfty)}^{\ff(x_2+)} \hfinv (v)\text{d}v \\
&\quad-\ff(x_2+)\fg(x_1+) + \ff(\minfty)\fg(\minfty)\\
&= \altPhiSI(\smthker;f,g;x_1,x_2) 
\end{align*}
for all $x_1$ and $x_2.$
\end{lemma}
The proof of this key lemma can be found in appendix \ref{app:A}.

There are many important consequences of this result.
One fundamental consequence of Lemma \ref{lem:twofint} is that if $\ff,\fg \in \sptfns$ 
satisfies \eqref{eqn:consistentFP} then

\begin{align}
\begin{split}
\altPhi(\hf,\hg;\fg(x_1+),\ff(x_2+))
&-\altPhi(\hf,\hg;\fg(\minfty),\ff(\minfty))
\\=& \altPhiSI(\smthker;f,g;x_1,x_2) \label{eqn:altPhiXi}
\end{split}
\end{align}
and
\begin{align*}
\Phi(\hf,\hg;\gS(x_2),\fS(x_1))
&-\Phi(\hf,\hg;\gS(\minfty),\fS(\minfty))
\\=& \PhiSI(\smthker;f,g;x_1,x_2) \label{eqn:PhiXi}
\end{align*}
for all $x_1$ and $x_2.$
Note that $\altPhiSI$ and $\PhiSI$ are local in the sense that if $\smthker$ is
finitely supported then the functionals above depend on the interpolating spatial fixed 
point solution only in a finite neighborhood of the interval between $x_1$ and $x_2.$
Thus, an intepolating spatial fixed point carries information about $\altPhi$ locally.

\begin{lemma}[Necessity of Positive Gap for FP]\label{lem:FPAzero}
Let $(f,g)$ be $(0,1)$-interpolating functions
satsifying 
\begin{align} 
\fg \veq \hg\circ \fS,\quad
\ff \veq \hf\circ \gS 
\end{align}
where
 $\smthker$  is an averaging kernel.
Then $(\hf,\hg)$ satisfies the positive gap condition and 
$A(\hf,\hg) = 0.$
\end{lemma}
\begin{IEEEproof}
First note that $\altPhi(\hf,\hf;0,0)=0.$
Since  $\altPhiSI(\smthker;f,g;x_1,x_2) \ge 0$ for all $x_1,x_2$ we see 
from Lemma \ref{lem:twofint} (and Lemma \ref{lem:monotonic} for the discontinuous case)
that
$\altPhi(h_{[\ff,\gS]},h_{[\fg,\fS]};u,v) \ge 0$ for all $(u,v) \in [0,1]^2.$
Letting $x_1,x_2 \rightarrow \infty$ we see $\altPhiSI(\smthker;f,g;x_1,x_2)\rightarrow 0$
and we obtain $A(\hf,\hg) = 0$ from \eqref{eqn:altPhiXi}. 
\end{IEEEproof}
\begin{lemma}[Partial Positive Gap for Waves]\label{lem:WaveGap}
Let $(f,g)$ be $(0,1)$-interpolating functions 
satsifying 
\begin{align} 
\fg \veq \hg\circ \fS,\quad
\ff \veq \hf\circ \gSa 
\end{align}
where
 $\smthker$  is an averaging kernel and $\ashift$ is real.
Then 
\[
\altPhi(\hf,\hg;\cdot,\cdot) \ge \min \{0, A(\hf,\hg) \}\,
\]
and 
\[
|\altPhi(\hf,\hg;u,v) 
-
\altPhi(h_{[\ff,\gS]},h_{[\fg,\fS]};u,v)|
\le 
|A(\hf,\hg)|\,.
\]
\end{lemma}
\begin{IEEEproof}
We have $\hf \equiv h_{[\ff,\gSa]}$ and $\hg \equiv h_{[\fg,\fS]}$ and
\begin{align*}
\altPhi(&h_{[\ff,\gSa]},h_{[\fg,\fS]};u,v) 
-
\altPhi(h_{[\ff,\gS]},h_{[\fg,\fS]};u,v)
\\ &=\int_0^v (h^{-1}_{[\ff,\gSa]}(v') - h^{-1}_{[\ff,\gS]}(v'))dv'\,,
\end{align*}
which is easily seen to be monotonic in $v$ (with direction depending on $\ashift$)
and independent of $u.$
At $(u,v)=(0,0)$ it evaluates to $0$ and at $(1,1)$ it evaluates to
$A(h_{[\ff,\gSa]},h_{[\fg,\fS]})$ since $A(h_{[\ff,\gS]},h_{[\fg,\fS]})=0$ by Lemma \ref{lem:FPAzero}.
\end{IEEEproof}

\subsubsection{Fixed Point Potential Bounds}

Lemma \ref{lem:twofint} provides information on the local structure of fixed point solutions
and relates it to the value of the potential function for the component systems.
In particular  we can extract information on the
spatial transition between the two underlying component fixed points.
More specifically, flatness of the spatial fixed point implies a relatively low
potential value.

Let us introduce the notation
\[
\Delta_L f(x) = f((x+L)-) - f((x-L)+)
\]
and the recall the definition
\(
\Omega(-L) = \int_{-\infty}^{-L} \smthker(x)\text{d}x\,= \int_L^\infty \smthker(x)\text{d}x 
\)

\begin{lemma}\label{lem:transitionPhiBounds}
Let $\ff,\fg$ be functions in $\sptfns.$
For any $L>0$ the following inequalities hold
\begin{align*}
\altPhiSI(\smthker,\ff,\fg;x,x)
& \le
\Delta_L f(x)\Delta_L g(x)+  \Omega(-L)
\end{align*}
\begin{align*}
|\fS(x)-f(x)|
 \le
\Delta_L f(x) +  \Omega(-L) 
\end{align*}
and for $(0,1)$-interpolating $\ff,\fg \in \sptfns$ we have
%
\begin{align*}
\altPhi(h_{[\ff,\gS]},h_{[\fg,\fS]};\fg(x),\ff(x))
& \le
\Delta_L f(x)\Delta_L g(x)+  \Omega(-L)
\\
\altPhi(h_{[\ff,\gS]},h_{[\fg,\fS]};\gS(x),\ff(x)) & \le
2\Delta_L g(x) + 2\Omega(-L)
\\
\altPhi(h_{[\ff,\gS]},h_{[\fg,\fS]};\fg(x),\fS(x)) &  \le
2\Delta_L f(x) + 2\Omega(-L)
\\
\altPhi(h_{[\ff,\gS]},h_{[\fg,\fS]};\gS(x),\fS(x)) &  \le
2\Delta_L f(x) + 2\Delta_L g(x)  + 3\Omega(-L)
\end{align*}
\end{lemma}
The Lemma is proved in appendix \ref{sec:FPbounds}.

\subsubsection{Transition length.}

In this section our aim is to show that fixed point solutions arising from systems satisfying
the strictly positive gap condition have bounded transition regions.
We show that the transition of solutions from one value to another
is confined to a region whose width can be bound from above using properties of $\altPhi$

\begin{lemma}\label{lem:transitionBounds}
Let $\ff,\fg$ be  $(0,1)$-interpolating functions satisfying \eqref{eqn:consistentFP}. 
Let $0 < a < b < 1$ and let $x_a,x_b$ satisfy $a=\gS(x_a)$
and $b=\gS(x_b).$
Define
\begin{align*}
\delta &= \inf \{ \altPhi(\hf,\hg;\gS(x),\ff(x)) : x\in [x_a,x_b]\}
\\
& = \inf \{ \altPhi(\hf,\hg;u,\hf(u)) : u\in [a,b]\}
\end{align*}
then
\[
\Bigl(\frac{1}{2}\delta-\Omega(-L)\Bigr)\lfloor \frac{x_b-x_a}{2L} \rfloor \le
1
\]
and
\begin{align*}
\Bigl( \frac{1}{2}{\delta- \Omega(-L)}\Bigr) 
\lfloor
\frac{x_b-x_a-2L}{2L}
\rfloor 
\le b-a\,.
\end{align*}
\end{lemma}
\begin{IEEEproof}
For any $x \in [x_a,x_b]$ we have
\(
\Delta_L g(x) \ge \frac{1}{2}{\delta- \Omega(-L)}\,
\)
by Lemma \ref{lem:transitionPhiBounds}.
In the interval $[x_a,x_b]$ we can find
\(
\lfloor
\frac{x_b-x_a}{2L}
\rfloor
\)
non-overlapping intervals of length $2L.$
From this we obtain
\[
\lfloor
\frac{x_b-x_a}{2L}
\rfloor \Bigl( \frac{1}{2}{\delta- \Omega(-L)}\Bigr) \le g(x_b-)-g(x_a+) \le  1\,.
\]
A similar argument considering $x_a+L$ and $x_b-L$ gives
\begin{align*}
\lfloor
&\frac{x_b-x_a-2L}{2L}
\rfloor \Bigl( \frac{1}{2}{\delta- \Omega(-L)}\Bigr) 
\\ \le &
g((x_b-L)-)-g((x_a+L)+)
\\ \le &
\gS(x_b)-\gS(x_a)
\\ \le &  b-a\,.
\end{align*}
\end{IEEEproof}

\subsubsection{Equality of End Point Potential}

The existence of an interpolating spatial fixed point implies a delicate balance in
the potential function of the underlying component system. The limit values of the
spatial fixed point must be crossing points with equal potential.
In this section we extend the result to the travelling wave case.

\begin{lemma}\label{lem:FPequal}
Assume $(\hf,\hg) \in \exitfns^2$ and $\smthker$ an averaging kernel.
Let $(\ff,\fg) \in \sptfns^2$ 
satisfy
\begin{align*}
\ff &\veq \hf \circ \fg^{\smthker,\ashift}  \text{ and } \,\,
\fg \veq \hg \circ \ff^{\smthker} \,
\end{align*}
for finite $\ashift.$
 Then
\begin{itemize}
\item[A.] $(\ff(\minfty),\fg(\minfty)) \in \cross(\hf,\hg).$
\item[B.] $(\ff(\pinfty),\fg(\pinfty)) \in \cross(\hf,\hg).$
\item[C.] If $\ashift = 0$ then 
\[
\altPhi(\hf,\hg;\fg(\minfty),\fg(\minfty)) =
\altPhi(\hf,\hg;\fg(\pinfty),\ff(\pinfty))\,.
\]
\end{itemize}
\end{lemma}
\begin{IEEEproof}
By definition we have $\ff(x) \veq \hf(\fg^{\smthker,\ashift}(x))$ for each $x \in \reals.$
Taking limits we have $\ff(\minfty) \veq \hf(\fg^{\smthker,\ashift}(\minfty)).$
Since $\fg^{\smthker,\ashift}(\minfty)=\fg^{\smthker}(\minfty)=\fg^{}(\minfty)$ we have
$\ff(\minfty) \veq \hf (\fg(\minfty)).$ Similarly, 
$\fg(\minfty) \veq \hg (\ff(\minfty))$
and part A follows.  Part B can be shown similarly.

If $(\ff(\minfty),\fg(\minfty)) = (\ff(\pinfty),\fg(\pinfty))$ then part C is immediate,
so assume  $(\ff(\minfty),\fg(\minfty)) < (\ff(\pinfty),\fg(\pinfty)).$
In this case part C follows from Lemma \ref{lem:FPAzero} by affine rescaling.
\end{IEEEproof}

\subsubsection{Discrete Spatial Integration}

Perhaps somewhat surprisingly, a version of Lemma \ref{lem:twofint} that
applies to spatially discrete systems also holds.
If $\ff,\fg$ are spatially discrete functions and $\tff,\tfg$ are their
piecewise constant extensions, then
Lemma \ref{lem:twofint} can be applied to
these extensions.  If we then restrict $x_1$ and $x_2$ to points in
$\Delta \integers,$ then $\altPhiSI$ can be written as discrete sums.

Let $\discsmthker$ be related to $\smthker$ as in \eqref{eqn:kerdiscretetosmth}
and let $x_1,x_2 \in \Delta \integers,$ denoted
$x_{i_1},x_{i_2}.$
Then 
\begin{align*}
&\altPhiSI(\smthker;\tff,\tfg;x_{i_1},x_{i_2}) \\&=
 \sum_{i=-\infty}^{i_2} \sum_{j=i_1+1}^\infty \partial\dv{f}_i  \partial\dv{g}_{j}  {\cal W}_{i-j}
+ 
\sum_{i=i_2+1}^\infty \sum_{j=-\infty}^{i_1}  \partial\dv{f}_i  \partial\dv{g}_{j}  {\cal W}_{j-i}
\end{align*}
where $\partial f_i \defeq f_i-f_{i-1}$ and 
${\cal W}_k \defeq \frac{1}{2}w_k+\sum_{i=-\infty}^{k-1} w_i .$

Lemma \ref{lem:twofint} continues to hold and a proof using entirely discrete
summation can be found in appendix \ref{app:Aa}.

{\em Discussion:} The proof of Lemma \ref{lem:twofint} as well as the spatially discrete version found in appendix \ref{app:Aa} are entirely algebraic in character.  Consequently, they apply to spatially coupled systems generally and not only those with a one dimensional state.  In a follow-up paper we apply the result to the arbitrary binary memoryless symmetric channel case to obtain a new proof that spatially coupled regular ensembles achieve capacity universally on such channels.

%


\subsection{Bounds on Translation Rates\label{sec:BTR}}

\begin{lemma}\label{lem:shiftlowerbound}
Let $f,g$ 
be $(0,1)$-interpolating and let $\smthker$ be an averaging kernel.
Then 
\[
|A(h_{[\ff,\gSa]},h_{[\fg,\fS]})| \le |\ashift|\|\smthker\|_\infty\,.
\]
\end{lemma}
\begin{IEEEproof}
We have $A(h_{[\ff,\gS]},h_{[\fg,\fS]})=0$ and hence
\begin{align*}
A(h_{[\ff,\gSa]},h_{[\fg,\fS]})
&=
A(h_{[\ff,\gSa]},h_{[\fg,\fS]})
-
A(h_{[\ff,\gS]},h_{[\fg,\fS]})
 \\
&= \int_0^1 h_{[\ff,\gS]}(u) - h_{[\ff,\gSa]}(u) \,\text{d}u \\
&= \int_{\minfty}^{\pinfty} (\ff(x) - \ff(x-\ashift)) \gS_x(x) dx \,.
\end{align*}
Since $|\gS_x(x)| \le \| \smthker \|_{\infty}$ we obtain
$|A(h_{[\ff,\gSa]},h_{[\fg,\fS]})|\le |\ashift|\|\smthker\|_\infty.$
\end{IEEEproof}

In general this estimate can be weak.  In Section \ref{sec:pathology} we gave an example
of a system with $A(\hf,\hg) =0$ and irregular $\smthker$ that can exhibit both left
and right moving waves by changing the value $\hf$ and $\hg$ at a point of discontinuity.
Further, given $(0,1)$-interpolating $\ff,\fg$ and positive $\smthker$ the system
 $(h_{[\ff,\fg^{\smthker,a+\ashift}]},h_{[\fg,\ff^{\smthker,-a}]})$ (with real parameter $a$)
has a 
traveling solution with shift $\ashift$ and yet
$A(h_{[\ff,\fg^{\smthker,a+\ashift}]},h_{[\fg,\ff^{\smthker,-a}]})$
can be made arbitrarily close to $0$ by choosing $a$ with large enough magnitude.

Now we consider upper bounds on $|\ashift|.$
If $\smthker$ has compact support then
the width of the support is an upper bound.  Consider a $\smthker$ that is strictly positive
on $\reals.$  Let $\hf(x)=\hg(x)=\unitstep(x-(1-\epsilon))$ for small positive $\epsilon.$
A traveling wave solution for this system is $\ff^t(x)=\unitstep(x-t\ashift)$
and $\fg^t(x)=\unitstep(x-t\ashift-\ashift/2)$
where $\ashift$ is given by
$\intsmthker(-\ashift/2) = (1-\epsilon).$
This example motivates the following bound.

\begin{lemma}\label{lem:shiftupperbound}
Let $\ff,\fg\in\sptfns$ be $(0,1)$-interpolating and assume
\[
f \veq\hg \circ \gSa\text{  and  } g\veq\hf \circ\fS\,.
\]
Given $(u,v) \in G^-(\hf,\hg)$ we have the bound
\[
\ashift \le \Omega^{-1}\Bigl(\frac{v}{\hf(u-)}+\Bigr)
+
\Omega^{-1} \Bigl(\frac{u}{\hg(v-)}+\Bigr)\,.
\]
and given  $(u,v) \in G^+(\hf,\hg)$  we have the bound
\[
-\ashift \le \Omega^{-1}\Bigl(\frac{1-v}{1-\hf(u+)}+\Bigr)
+
\Omega^{-1} \Bigl(\frac{1-u}{1-\hg(v+)}+\Bigr)\,.
\]
\end{lemma}
\begin{IEEEproof}
We will show the first bound, the second is similar.
For any $x,z\in\reals$ we have 
\(
f(x) \ge f(z+) \unitstep_0(x-z)
\)
from which we obtain
\begin{align*}
\fS(x) \ge f(z+)\Omega(x-z)
\end{align*}

Therefore, for any $x_1,x_2$ we have the inequality
\[
x_1-x_2 \le \Omega^{-1} \Bigl(\frac{\fS(x_1)}{\ff(x_2+)}+\Bigr)
\]
Choose $x_1$ so that $\fS(x_1)=v.$
Then we have $\fg(x_1+) \ge \fg(x_1) \ge \hg(v-).$
Choose $x_2$ so that $\gSa(x_2)=\gS(x_2+\ashift)=u.$
Then we have $\ff(x_2+) \ge \ff(x_2) \ge \hf(u-).$

Applying the above inequality we obtain
\[
x_1-x_2 \le \Omega^{-1} \Bigl(\frac{\fS(x_1)}{\ff(x_2+)}+\Bigr)
 \le \Omega^{-1} \Bigl(\frac{v}{\hf(u-)}+\Bigr)
\]
and
\[
x_2+\ashift-x_1 \le \Omega^{-1}\Bigl(\frac{\gS(x_2+\ashift)}{\fg(x_1+)}+\Bigr)
\le \Omega^{-1}\Bigl(\frac{u}{\hg(v-)}+\Bigr)
\]
Summing, we obtain
\[
\ashift \le \Omega^{-1}\Bigl(\frac{v}{\hf(u-)}+\Bigr)
+
\Omega^{-1} \Bigl(\frac{u}{\hg(v-)}+\Bigr)\,.
\]
\end{IEEEproof}

\begin{corollary}\label{cor:regshiftbound}
Let $\ff,\fg \in \sptfns$ be $(0,1)$-interpolating.
If $( h_{[\ff,\gSa]},h_{[\fg,\fS]})$  satisfies the strictly positive gap condition
and $\smthker$ is regular
then $\ashift < 2W.$
\end{corollary}
\begin{IEEEproof}
This combines Lemma \ref{lem:shiftupperbound} with Lemma \ref{lem:stposbound}.
\end{IEEEproof}

\subsection{Monotonicity of $\altPhi$ and the Gap Conditions\label{sec:MGC}}

In this section we collect some basic results on $\altPhi$ and the component DE that are useful for constructing spatial wave solutions.

\begin{lemma}\label{lem:descend}
Let $\hf,\hg \in \exitfns.$
If  $(u,v) \in G^-$
then there exists a minimal element $(u^*,v^*) \in \cross(\hf,\hg)$ with
$(u^*,v^*)>(u,v)$ component-wise and $\altPhi(u^*,v^*) < \altPhi(u,v).$

Similarly, if $(u,v) \in G^+$
then there exists a maximal element $(u^*,v^*) \in \cross(\hf,\hg),$
with $(u^*,v^*)<(u,v)$ (component-wise) and $\altPhi(u^*,v^*) < \altPhi(u,v).$
\end{lemma}
\begin{IEEEproof}
We show only the first case since the other case is analogous.
Assuming  $(u,v)\in G^-$ 
we  have $\hginv(u+) < v < \hf(u-)$  and we
see that there is no crossing point $(u',v')$  with $u'=u.$ Similarly,
there is no crossing point with $v'=v.$ 
Since $\cross(\hf,\hg)$ is closed, the set $(u,1]\times(v,1] \cap \cross(\hf,\hg)$ is closed.
By Lemma \ref{lem:crossorder} $\cross(\hf,\hg)$ is ordered, so there exists a minimal element $(u^*,v^*)$ in
$(u,1]\times(v,1] \cap \cross(\hf,\hg).$
Set $(u^0,v^0)=(u,v)$ and consider the sequence of points
$(u^0,v^0),(u^0,v^1),(u^1,v^1),(u^1,v^2),(u^2,v^2),\ldots$ as determined by \eqref{eqn:DE}.
It follows easily from  \eqref{eqn:DE} that this sequence is non-decreasing.
If $u^t < u^*$ then $v^{t+1} \le v^*$
and
if $v^t < v^*$ then $u^{t} \le u^*.$
Thus we have either $(u^t,v^t) < (u^*,v^*)$ for all $t$ or there is some minimal $t$ where at least one of the coordinates
is equal.
If $(u^t,v^t) < (u^*,v^*)$ for all $t$ then the sequence must converge to $(u^*,v^*)$
since the limit is in $\cross(\hf,\hg)$ by continuity and $(u^*,v^*)$ is minimal.
It then follows by continuity of $\altPhi(\hf,\hg;)$ and Lemma \ref{lem:monotonic} that 
\[
\altPhi(\hf,\hg;u^*,v^*) \le \altPhi(\hf,\hg;u^0,v^1) < \altPhi(\hf,\hg;u^0,v^0)\,.
\]
Assume now that $u^t = u^*$ for some $t.$  Then $t>0$ and  Lemma \ref{lem:monotonic}  gives
\[
\altPhi(\hf,\hg;u^*,v^*) = \altPhi(\hf,\hg;u^t,v^{t+1}) < \altPhi(\hf,\hg;u^0,v^0)\,.
\]
Finally, assume that $v^t = v^*$ for some $t.$  Then $t>0$ and  Lemma \ref{lem:monotonic}  gives
\[
\altPhi(\hf,\hg;u^*,v^*) = \altPhi(\hf,\hg;u^t,v^{t}) < \altPhi(\hf,\hg;u^0,v^0)\,.
\]
This completes the proof.
\end{IEEEproof}

\begin{lemma}\label{lem:crosspointmono}
Let $(\hf,\hg)\in \exitfns^2$ and 
let $(u,v) \in [0,1]^2.$
We then have the following trichotomy:
\begin{itemize}
\item
If $\hg(\hf(u))=u$ then $(u,\hf(u))\in\cross(\hf,\hg).$
\item
If $\hg(\hf(u))>u$ then 
\(
\altPhi(\hf,\hg;u^*,v^*)
\le
\altPhi(\hf,\hg;u,v)\,
\)
where $(u^*,v^*)\in\cross(\hf,\hg)$ is coordinate-wise minimal with
 $(u^*,v^*)\ge(u,\hf(u)).$ 
\item
If $\hg(\hf(u))<u$ then
\(
\altPhi(\hf,\hg;u^*,v^*)
\le
\altPhi(\hf,\hg;u,v)\,
\)
where $(u^*,v^*)\in\cross(\hf,\hg)$
is coordinate-wise maximal with $(u^*,v^*)\le(u,\hf(u)).$ 
\end{itemize}
\end{lemma}
\begin{IEEEproof}
If $u \veq \hg(\hf(u))$ then $(u,\hf(u))\in\cross(\hf,\hg)$ by definition.
Thus, the first case holds and the other two hold under this condition.
We assume henceforth that $u\not\veq\hg(\hf(u)).$
Assuming $\hg(\hf(u))>u,$ we now have 
\begin{align}\label{eqn:uitineq}
\hg(\hf(u)-)>u
\end{align}
and $\hginv(u+) < \hf(u).$

Let $(u^*,v^*)\in\cross(\hf,\hg)$ be the minimal element such that  $(u^*,v^*)\ge(u,\hf(u)).$
It follows that  $u^* >u$ and $v^* \ge \hf(u+).$
For all $\epsilon >0$ sufficiently small we claim $(u+\epsilon,\hf(u)-\epsilon)\in G^-.$
Indeed $\hf(u-)-\epsilon <  \hf((u+\epsilon)-)$ and for $\epsilon$ small enough
$u+\epsilon < \hg((\hf(u)-\epsilon)-)$
by \eqref{eqn:uitineq}.
Assuming $\epsilon$ sufficiently small $(u^*,v^*)$ is the minimal element in $\cross(\hf,\hg)$
with $(u^*,v^*)> (u+\epsilon,\hf(u)-\epsilon)$ and
by Lemma \ref{lem:descend} we have
$\altPhi(u^*,v^*) < \altPhi(u+\epsilon,\hf(u)-\epsilon).$
Letting $\epsilon$ tend to $0$ we obtain
$\altPhi(u^*,v^*) \le \altPhi(u,\hf(u)) \le \altPhi(u,v).$

The argument for the case $\hg(\hf(u))<u$ is similar and we omit it.
\end{IEEEproof}

\begin{lemma}\label{lem:zocontinuity}
If $(\hf,\hg) \in \exitfns^2$ satisfies the strictly positive gap condition
then $\hf$ and $\hg$ are continuous at $0$ and at $1.$
\end{lemma}
\begin{IEEEproof}
Assume $(\hf,\hg) \in \exitfns^2$ satisfies the strictly positive gap condition
and assume $\hf(0+)>0.$
Without loss of generality we can assume that $\hf(0)=\hf(0+).$
We have $\altPhi(\hf,\hg;0,v)=0$ for all
$v \in [0,\hf(0)]$ so the strictly positive gap condition (no interior crossing point
where $\altPhi$ is $0$) implies that $\hginv(0+)=0,$ which gives $\hg(\hf(0))>0.$
For all $u\in[0,\hg(\hf(0)))$ we now have $\hg(\hf(u))>u.$
For such $u>0$ we have $\altPhi(\hf,\hg;u,\hf(u))<0$
by \eqref{eqn:altPhiuhfu}.
By Lemma \ref{lem:crosspointmono} the minimal crossing point $(u^*,v^*) \ge (u,\hf(u))$
($(u^*,v^*) $ is the same for all choices of $u$)
satisfies $\altPhi(\hf,\hg;u^*,v^*)<0.$  By the strictly positive gap condition
$(u^*,v^*) \neq (1,1)$ and we obtain a contradiction.
Therefore, we must have $\hf(0+) = 0.$

All other conditions, $\hg(0+) = 0, \hf(1-) = 1,$ and $\hg(1-) = 1$ can be shown similarly.
\end{IEEEproof}

One useful consequence of Lemma \ref{lem:zocontinuity}
 is that is $\hf,\hg$ satisfies the strictly positive gap condition and
$(u,v) \in \intcross(\hf,\hg)$ then we have
$(0,0) < (u,v) < (1,1)$ component-wise.

\begin{lemma}\label{lem:Sstructure}
Let $(\hf,\hg) \in \exitfns^2$ satisfy the strictly positive gap condition.
If $A(\hf,\hg) \ge 0$ then $\altPhi(\hf,\hg;u,v) >0$ 
for $(u,v)\in [0,1]^2\backslash\{(0,0),(1,1)\}.$
If $A(\hf,\hg) > 0$ then there exists a minimal point $(u^*,v^*) \in \intcross(\hf,\hg)$
and the set
\[
S(\hf,\hg) = \{(u,v): \altPhi(\hf,\hg;u,v) < A(\hf,\hg)\}
\]
is simply connected 
and  $\closure{S(\hf,\hg)} \subset[0,u^*)\times[0,v^*).$
Moreover,
\[
\{(u,v): \altPhi(\hf,\hg;u,v) \le A(\hf,\hg)\}
= \closure{S(\hf,\hg)}\cup \{(1,1)\}.
\]
\end{lemma}
\begin{IEEEproof} 
Assume $(\hf,\hg) \in \exitfns^2$ satisfies the strictly positive gap condition and that
$A(\hf,\hg) = 0.$
It follows from Corollary \ref{lem:miniscross} that $\altPhi(\hf,\hg;u,v)$
achieves its minimum on $\cross(\hf,\hg),$
hence,  the strictly positive gap condition implies $\altPhi(\hf,\hg;\cdot,\cdot) \ge 0.$
Corollary \ref{lem:miniscross} further implies that if there exists $(u,v)$ with
$\altPhi(\hf,\hg;u,v) = 0$ then $(u,v) \in \cross(\hf,\hg).$
Thus, we have $\altPhi(\hf,\hg;u,v) > 0$ for $(u,v) \not\in \{(0,0),(1,1)\}.$

Assume now that $A(\hf,\hg) > 0.$ 
Let $(u^*,v^*)$ be the infimum of $\intcross(\hf,\hg).$
Since $\cross(\hf,\hg)$ is closed we have  $(u^*,v^*)\in \cross(\hf,\hg)$
and by continuity and
the strictly positive gap condition we have 
$(u^*,v^*) \in \intcross(\hf,\hg)$
and $\altPhi(\hf,\hg;u^*,v^*) > A.$
By Lemma \ref{lem:zocontinuity} the strictly positive gap condition implies $(u^*,v^*) > (0,0).$ 
%

Let $(u,v) \in S(\hf,\hg)$ and assume that $u \neq 0$ and $v\neq 0.$
By Lemma  \ref{lem:monotonic} and Lemma \ref{lem:crosspointmono}
we see that we must have $\hg(\hf(u))<u.$ 
Setting $(u^0,v^1) = (u,\hf(u))$
then the sequence of points
$(u^0,v^1),(u^1,v^1),(u^1,v^2),(u^2,v^2),\ldots$ as determined by \eqref{eqn:DE}
are coordinate-wise non-increasing and must converge to $(0,0).$
Hence, all these points are in $S(\hf,\hg).$
Furthermore, by Lemma \ref{lem:monotonic} (coordinate-wise convexity) the line segments joining successive points are all in $S(\hf,\hg).$
The line segment joining $(u,v)$ to $(u,\hf(u))$ is also in $S(\hf,\hg).$
By continuity of $\altPhi,$ $S(\hf,\hg)$ contains a neighborhood of $(0,0).$
Thus, $S(\hf,\hg)$ is pathwise connected.  For any fixed $u$ the
set $S(\hf,\hg)$ is an open interval in $v$ and $S(\hf,\hg)$ is itself open. 
It is now easy to see that $S(\hf,\hg)$ is simply connected.

Lemma \ref{lem:monotonic} implies that $\altPhi(\hf,\hg;u^*,v) > A(\hf,\hg)$ for all $v\in [0,1]$
and  $\altPhi(\hf,\hg;u,v^*) > A(\hf,\hg)$ for all $u\in [0,1]$
so by continuity of $\altPhi$ we have  $\closure{S(\hf,\hg)} \subset [0,u^*) \times [0,v^*).$


Assume there exists $(u,v)\not\in \closure{S(\hf,\hg)} \cup \{ (1,1)\}$ with $\altPhi(\hf,\hg;u,v) = A(\hf,\hg).$
Then $(u,v)$ is a local minimum of $\altPhi(\hf,\hg;u,v)$
which, by Corollary \ref{lem:miniscross}, implies $(u,v) \in \cross(\hf,\hg),$
contradicting the strictly positive gap condition.
\end{IEEEproof}

In the case where $A(\hf,\hg)<0$ Lemma \ref{lem:Sstructure} gives $\altPhi(\hf,\hg;\cdot,\cdot) > A$ on $[0,1]^2 \backslash \{(0,0),(1,1)\}$ and the set
\[
S(\hf,\hg) = \{(u,v): \altPhi(\hf,\hg;u,v) < 0\}
\]
will be a simply connected open set containing $(1,1).$

\subsection{Limit Theorems \label{sec:limitthms}}

In this section we prove certain closure properties of wavelike solutions under various limit
processes.  The results are used later to extend existence results established for special cases
to more general cases.  

Let us recall the notation $g^{{\smthker},\ashift}(x) = g^{\smthker}(x+\ashift).$
We have the bound
\begin{align}
\begin{split}\label{eqn:diffbound}
|g^{{\smthker},\ashift}(x)
-
g^{\smthker',\ashift'}(x)|
\le
|\ashift-\ashift'| \|\smthker\|_\infty
+
\|\smthker-\smthker'\|_1 \,
\end{split}
\end{align}
from
\begin{align*}
&g^{{\smthker},\ashift}(x)
-
g^{\smthker',\ashift'}(x)
=
\\ &
\int_{-\infty}^\infty (g(y-\ashift)-g(y-\ashift')) \smthker(x-y) \,dx
\\&+
\int_{-\infty}^\infty g(y-\ashift') (\smthker(x-y)-\smthker'(x-y)) \,dx\,.
\end{align*}

%

\begin{theorem}\label{thm:mainlimit}
Let $f_i,g_i,\ashift_i,\smthker_i,\; i=1,2,3,...$ be sequences where $(f_i,g_i) \in \sptfns^2$
are $(0,1)$-interpolating, 
$\ashift_i \in \reals,$ and $\smthker_i$ are averaging kernels. 
Assume
\[
f_i  \rightarrow f,\;
g_i  \rightarrow g,\;
\ashift_i  \rightarrow \ashift,\;\text{ and }
\smthker_i  \rightarrow \smthker \text{ (in $L_1$) },
\]
where $|\ashift| < \infty$
and $\smthker$ is an averaging kernel.
(Note that we do not assume $f$ and $g$ are interpolating or that $\smthker$ is regular.)
Further assume
\[
h_{[\ff_i,\gSiai_i]}\rightarrow \hf
,\quad
h_{[\fg_i,\fSi_i]}\rightarrow \hg
\]
for some  $\hf,\hg \in \exitfns$ respectively.
Then we have the following
\begin{itemize}
\item[A.] 
\begin{align*}
\ff \veq \hf\circ\gSa
\text{  and   }
\fg \veq \hg\circ\fS
\end{align*}
\item[B.]
\[
(f(\minfty),g(\minfty)),(f(\pinfty),g(\pinfty)) \in \cross (\hf,\hg)
\]
\item[C.]
If $\ashift=0$ then
\begin{align*}
0 = &\altPhi(\hf,\hg;g(\minfty),f(\minfty)) \\ 
 = &\altPhi(\hf,\hg;g(\pinfty),f(\pinfty)) \,,
\end{align*}
and, for all $x_1,x_2$
\[
\altPhi(\hf,\hg;\fg(x_2+),\ff(x_1+)) = 
\altPhiSI (\smthker;f,g;x_1,x_2).
\]
\item[D.]
For $(u,v)\in\{(\ff(\minfty),\fg(\minfty)),(\ff(\pinfty),\fg(\pinfty))\}$
\[
\!\!\!\!\!\!\min\{0,A(\hf,\hg)\}\le\altPhi(\hf,\hg;u,v) \le \max\{0,A(\hf,\hg)\}.
\]
\end{itemize}
\end{theorem}
\begin{IEEEproof}
Since $\fg_i \rightarrow \fg,$ $\smthker_i \rightarrow \smthker$
and $\ashift_i \rightarrow \ashift$ we have from \eqref{eqn:diffbound}
that $\fg^{{{\smthker}_i},\ashift_i}_i  \rightarrow \fg^{{\smthker},\ashift}$ point-wise.

If $x$ is a point of continuity of $f$ then $f_i(x) \rightarrow f(x)$ and we have
$(g^{{\smthker}_i}(x+\ashift_i),f_i(x))\rightarrow (g^{\smthker}(x+\ashift),f(x))$ which implies
$f(x) \veq \hf( \gSa(x)).$ 
Since $\gS$ is continuous we can extend this to all $x$ by taking limits.
This shows part A.

Part B follows from part A by Lemma \ref{lem:FPequal}.

Now we consider part C where we assume $\ashift=0.$
By Lemma \ref{lem:FPequal} it is sufficient for the first part to show that
$\altPhi(\hf,\hg;g(\pinfty),f(\pinfty))=0.$
Since $\ashift_i \rightarrow 0$ if follows from Lemma \ref{lem:WaveGap}
and Lemma \ref{lem:shiftlowerbound}
that $(\hf,\hg)$ satisfies the positive gap condition, hence
$\altPhi(\hf,\hg;g(\pinfty),f(\pinfty)) \ge 0.$
We now prove the opposite inequality.
For any $\epsilon>0$ 
we can find $L$ large enough so that $\int_L^\infty  \smthker_i(x) dx < \epsilon$ for all $i$ since $\smthker_i\rightarrow\smthker.$
Now choose $z$ large enough so that 
$f^\smthker(z-L),f(z-L) > f(\pinfty)-\epsilon$
and
$g^{\smthker}(z-L),g(z-L) > g(\pinfty)-\epsilon.$  
It follows that 
$\Delta_L g(z) < \epsilon$ and $\Delta_L f(z) < \epsilon.$
For all $i$ large enough we have $\Delta_L g_i^{\smthker_i}(z) < 2\epsilon$ and 
$\Delta_L f_i^{\smthker_i}(z) < 2\epsilon.$
By Lemma \ref{lem:transitionPhiBounds} this implies
\[
 \altPhi(h_{[\ff_i, \fg_i^{\smthker_i}]},h_{[\fg_i, \fg_i^{\smthker_i}]}; \fg_i^{\smthker_i}(z),\ff_i^{\smthker_i}(z)) < 11\epsilon\,
\]
and applying Lemma \ref{lem:WaveGap} and Lemma \ref{lem:shiftlowerbound}
we have
\[
 \altPhi(h^i_\ff,h^i_\fg; \fg_i^{\smthker_i}(z),\ff_i^{\smthker_i}(z)) < 11\epsilon + |\ashift_i|\|\smthker\|_\infty\,
\]
It follows from \eqref{eqn:diffbound} that $\ff_i^{\smthker_i} \rightarrow \fS$ 
and  $\fg_i^{\smthker_i} \rightarrow \gS$ 
(as well as $\fg_i^{\smthker_i,\ashift_i} \rightarrow \gS$)
point-wise
so we have for all $x_1,x_2,$
\[
\begin{split}
& \altPhi(\hf^i,\hg^i; g^{\smthker_i}_i(x_2),f^{\smthker_i}_i(x_1))
\\ &
\rightarrow
\altPhi(\hf,\hg;g^{\smthker}(x_2),f^{\smthker}(x_1))\,.
\end{split}
\]
We now obtain
\[
 \altPhi(\hf,\hg; g^{\smthker}(z),f^{\smthker}(z)) \le 11\epsilon\,.
\]
By Lipschitz continuity of $\altPhi$ we have
\[
 \altPhi(\hf,\hg; g^{\smthker}(\pinfty),f^{\smthker}(\pinfty)) < 13\epsilon
\]
and since $\epsilon$ is arbitrary we obtain
\[
 \altPhi(\hf,\hg; g^{\smthker}(\pinfty),f^{\smthker}(\pinfty)) = 0\,.
\]
It now follows from Lemma \ref{lem:twofint} (see \eqref{eqn:altPhiXi}) and part A that
\[
\altPhi(\hf,\hg;g(x_2+),f(x_1+)) = \altPhiSI(\smthker;\ff,\fg;x_1,x_2)
\]
for all $x_1,x_2.$

Finally, we show part D. 
If $\ashift=0$ then part $C$ gives part $D.$
By choosing a subsequence if necessary, we can assume that
\(
h_{[\ff_i,\gSi_i]}
\)
converges to some $\thf \in \exitfns.$

We assume $\ashift > 0,$ the case $\ashift<0$ is analogous.
Since $\thfinv \le \hfinv$ almost everywhere we have
\(
\altPhi(\thf,\hg;u,v) - \altPhi(\hf,\hg;u,v) =
\int_0^{v} (\thfinv(x) -\hfinv(x))dx \le 0\,.
\)
For $(u,v)\in\{(\ff(\minfty),\fg(\minfty)),(\ff(\pinfty),\fg(\pinfty))\}$
we have $\altPhi(\thf,\hg;u,v)=0$ 
by part C,
and therefore  $\altPhi(\hf,\hg;u,v) \ge 0.$

Now 
\(
A(\hf,\hg)-A(\thf,\hg) =
\int_0^{1} (\hfinv(x) -\thfinv(x))dx
\ge 
 \int_0^{v} (\hfinv(x) -\thfinv(x))dx
\)
and since $A(\thf,\hg)=0$  by Lemma \ref{lem:FPAzero}, we have
$\altPhi(\hf,\hg;u,v) \le A(\hf,\hg)$ for
$(u,v)\in\{(\ff(\minfty),\fg(\minfty)),(\ff(\pinfty),\fg(\pinfty))\}.$
This completes the proof.
\end{IEEEproof}

The following result is largely a corollary of the above but it is
more convenient for us to apply.

\begin{lemma}\label{lem:limitexist}
Let $(\hf,\hg) \in \exitfns^2$ satisfy the strictly positive gap condition.
If there exists a sequence of $(0,1)$-interpolating $f_i,g_i \in \sptfns$ and $\ashift_i$ such that 
$(\hf^i,\hg^i) \rightarrow (\hf,\hg)$
and $\smthker_i \rightarrow \smthker$ in $L_1,$
where $\hf^i \equiv h_{[\ff_i,\fg^{\smthker_i,\ashift_i}_i]}$ and
$\hg^i \equiv h_{[\fg_i,\ff^{\smthker_i}_i]},$
then
there exists $(0,1)$-interpolating $\ff,\fg \in \sptfns$ and finite $\ashift,$ all limits of some translated subsequence,
such that $\hf \equiv h_{[\ff,\gSa]}$ and
$\hg \equiv h_{[\fg,\fS]}.$
\end{lemma}
\begin{IEEEproof}
Since $\smthker_i \rightarrow \smthker$ in $L_1$ and
 $(\hf^i,\hg^i) \rightarrow (\hf,\hg)$  we conclude from 
Lemma \ref{lem:stposbound} and Lemma \ref{lem:shiftupperbound}
that $|\ashift_i|$ is bounded.

By translating $\ff$ and $\fg$ as necessary, we can assume that $\ff^{\smthker_i}(0) = 1/2$ for each $i.$
Taking subsequences as necessary, we can now assume that $f_i \rarrowi f,$
$g_i \rarrowi g,$ and $\ashift_i \rarrowi \ashift,$ for some finite $\ashift.$

We claim that $\ff$ and $\fg$ are $(0,1)$-interpolating.
For all $(u,v) \in \intcross(\hf,\hg)$ we have
$\altPhi(\hf,\hg;u,v) >\max \{0,A(\hf,\hg)\}$
by assumption.
By Theorem \ref{thm:mainlimit} parts B and D we now have
$(\ff(\minfty),\fg(\minfty)) \in \cross(\hf,\hg) \backslash \intcross(\hf,\hg) = \{ (0,0),(1,1) \}.$
Since $\fS(0) = \frac{1}{2}$ we must have $(\ff(\minfty),\fg(\minfty)) = (0,0)$
and $(\ff(\pinfty),\fg(\pinfty)) = (1,1),$
proving the claim.
\end{IEEEproof}

\subsection{Inverse Formulation.\label{sect:inverse}}

It is instructive in to the analysis to view the system in terms of inverse functions.
Let $\fg(x) = \hg((\ff\otimes \smthker) (x))$ with $f \in \sptfns.$
Then, for almost all $u\in [0,1]$ we have
\(
\hg^{-1}(u) = \int_0^1 \intsmthker (\fg^{-1}(u)-\ff^{-1}(v)) \text{d}v\,.
\)
To show this we first integrate by parts to write
\(
(\ff\otimes \smthker) (x) = \int_{-\infty}^\infty \intsmthker(x-y) d\ff(y)
\)
and then make the substitutions $v=f(y)$ and $u=g(x).$
It follows that, up to equivalence, the recursion \eqref{eqn:gfrecursion} may be expressed as
\begin{equation}\label{eqn:gfrecursionInv}
\begin{split}
\hginv(u)  & =  \int_0^1 \intsmthker ((\fg^t)^{-1}(u)-(\ff^t)^{-1}(v)) \text{d}v, \\
\hfinv(v) & = \int_0^1 \intsmthker ((\ff^{t+1})^{-1}(v)-(\fg^t)^{-1}(u)) \text{d}u\,.
\end{split}
\end{equation}
Since $\smthker$ is even we have $\Omega(x) = 1-\Omega(-x),$ so we immediately observe
that if $(\ff,\fg)\in\sptfns^2$ is a $(0,1)$-interpolating  fixed point of the above system then
\begin{align*}
1 & =
\int_0^1 \hginv(u) \text{d}u + \int_0^1 \hfinv(v)\text{d}v \\
& =
\int_0^1 \hg(u) \text{d}u + \int_0^1 \hf(v)\text{d}v \,.
\end{align*}
This is the area condition that we already established in Lemma \ref{lem:FPAzero}
but the derivation here is particularly elegant.

Assume that $\ff$ and $\fg,$ both in $\sptfns,$ form a $(0,1)$-interpolating spatial fixed point.
Consider perturbing the inverse functions by $\ffinv\rightarrow \ffinv+\delta \ffinv$ and 
$\fginv\rightarrow \fginv+\delta \fginv$ respectively.
We could then perturb $\hfinv$ and $\hginv,$ by $\delta\hfinv$ and $\delta\hginv$ respectively so that
the perturbed system would remain a fixed point.  To first order we will have from \eqref{eqn:gfrecursionInv},
\begin{equation}\label{eqn:delgfrecursionInv}
\begin{split}
\delta\hginv(u)  & =  \int_0^1 \smthker (\fginv(u)-\ffinv(v)) (\delta\fginv(u)-\delta\ffinv(v)) \text{d}v,  \\
\delta\hfinv(v) & = \int_0^1 \smthker (\ffinv(v)-\fginv(u)) (\delta\ffinv(v)-\delta\fginv(u))\text{d}u\,.
\end{split}
\end{equation}
This formulation is at the heart of the analysis in the next section.  In a more recent work
\cite{KNTU14} this formalism is used in an analysis that shows uniquess of the 
spatial fixed point solutions developed here.

\subsection{Existence: The Piecewise Constant Case\label{sec:PCcase}}
In this section we focus on the case where $\hf$ and $\hg$ are piecewise constant.
In this case the spatially coupled system is finite dimensional, which simplifies the analysis
significantly.
We further assume that $\smthker$ is strictly positive on $\reals$ and Lipschitz continuous.
Strict positivity ensures in a simple way that no degeneracy occurs when determining EXIT functions
from spatial functions since $\gS_x(x) > 0$ for any non-constant $g \in \sptfns.$

We will write piecewise constant functions $\hf,\hg \in \exitfns$ as
\begin{align*}
\hf(u) &= \sum_{j=1}^{\Kf}  \delhf_j \,\unitstep(u - \uf_j) \\
\hg(u) &= \sum_{i=1}^{\Kg}  \delhg_i \,\unitstep(u - \ug_i)
\end{align*}
where $\unitstep$ is the unit step (Heaviside)
 function\footnote{The regularity assumptions on $\smthker$ ensure that the precise value of $\hf$ and $\hg$
at points of discontinuity has no impact on the analysis.}
and where we assume $\delhf_j,\delhg_i > 0,$ and $\sum_{j=1}^{\Kf}  \delhf_j = 1$
and $\sum_{i=1}^{\Kg}  \delhg_i = 1.$

Generally we will have
$0 < \uf_1 \le \uf_2 \le \cdots \le \uf_{\Kf} < 1$
and
$0 < \ug_1 \le \ug_2 \le \cdots \le \ug_{\Kg} < 1$
but the ordering is actually not critical to the definition.
We view the vectors $\delhf$ and $\delhg$ as fixed and 
to explicitly indicate the dependence on $\uf = (\uf_1,\ldots,\uf_{\Kf})$ and $\ug$ we will
write $\hf(u;\uf)$ and $\hg(u;\ug).$

Piecewise constant $\hf$ and $\hg$ also have piecewise constant inverses.
Given $\hf$ as above we have
\[
\hfinv (v) = \sum_{j=1}^{\Kf} (\uf_j-\uf_{j-1})\unitstep(v-\sum_{k=1}^{j} \delhf_j)
\]
where  we set $\uf_0=0.$

If $\fg$ is a continuous, strictly increasing, $(0,1)$-interpolating function
and $\hf$ is piecewise constant as above then $\ff \in\sptfns$ defined by
$\ff(x) = \hf(\fg(x))$ is also piecewise constant and
can be written as
\[
\ff(x) = \sum_{i=1}^{\Kf}  \delhf_i \,\unitstep(x - \zf_i)
\]
with
$-\infty < \zf_1 \le \zf_2 \le \cdots \le \zf_{\Kf} < \infty$
given by  $\uf_i = \fg^{-1}(\zf_i).$ 
The inverse of $\ff$ is then given by 
\[
\ffinv(v) =  \sum_{j=1}^{\Kf} (\zf_j-\zf_{j-1})\unitstep(v-\sum_{k=1}^{j} \delhf_j) \,.
\]
where we set $\zf_0 =0$

The purpose of this section is to prove a special case of Theorem
\ref{thm:mainexist} under piecewise constant assumptions on the EXIT functions and
regularity conditions on $\smthker.$ 
In this special case we obtain in addition uniqueness and continuous dependence of the solution.
For convenience we state the main result here.
\begin{theorem}\label{thm:PCexist}
Assume $\smthker$ is a strictly positive and Lipschitz continuous
averaging kernel.
Let $(\hf,\hg)$ be a pair of piecewise constant functions in $\exitfns$
satisfying the strictly positive gap condition.
Then there exists unique (up to translations) $(0,1)$-interpolating functions
$\tmplF,\tmplG \in\sptfns$ and $\ashift \in \reals$ satisfying $\sgn (\ashift) = \sgn (A(\hf,\hg)),$ such that
setting
$\ff^t(x)  = \tmplF(x-\ashift t)$ and
$\fg^t(x)  = \tmplG(x-\ashift t)$
solves \eqref{eqn:gfrecursion}.
Further,
$\tmplF^{-1}(v)-\tmplG^{-1}(u)$ depends continuously on the vectors $\uf,\ug.$
\end{theorem}

The remainder of this section is dedicated to the proof of this result.
Our proof constructs the solutions $\tmplF$ and $\tmplG$ by a method of continuation.
In the case where $\hf$ and $\hg$ are unit step functions it is easy to find
the solution: $\tmplF$ and $\tmplG$ are also unit step functions and we need only correctly relatively position the steps.
Starting from this case we continuously deform the solution to
arrive at a solution for a given $\hf$ and $\hg.$
We do this in two stages where in the first stage $\ashift = 0$ and in the second
is $\ashift$ varied while $\hg$ is held fixed.
The deformation is obtained as a solution to a differential equation.
To set up the equation we require a detailed description of the dependence
of $\uf$ and $\ug$ on $\zf,\zg$ and $\ashift.$


Let us first consider the case $\ashift=0.$
Let $\ff(x;\zf)$ and $\fg(x;\zg)$ be piecewise constant functions
parameterized by their jump point locations
$\zf$ and $\zg$ as
\begin{align}
\begin{split}\label{eqn:fgtdef}
g(x;\zg) & = \sum_{i=1}^{\Kg} \delhg_i \,\unitstep(x - \zg_{i}) \\
f(x;\zf) & = \sum_{j=1}^{\Kf} \delhf_j \,\unitstep(x - \zf_{j})
\end{split}
\end{align}
and let us then define
\begin{align}
\begin{split}\label{eqn:discreteInv}
\ug_i &\defeq \fS(\zg_i;\zf) = \sum_{j=1}^{\Kf} \delhf_j \Omega (\zg_i-\zf_j) \\
\uf_j &\defeq  \gS(\zf_j;\zg) =\sum_{i=1}^{\Kg} \delhg_i \Omega (\zf_j-\zg_i)\,.
\end{split}
\end{align}
It follows that $f(x;\zf),g(x;\zg)$ is a spatial fixed point for the system
$\hf(\cdot;\uf),\hg(\cdot;\ug).$ Hence, by Lemma \ref{lem:FPAzero}
we have $A(\hf(\cdot;\uf),\hg(\cdot;\ug)) = 0.$

\newcommand{\tvar}{\tau}
Now, suppose we introduce smooth dependence on a real parameter $\tvar,$
i.e., we are given smooth vector valued  functions $\zf(\tvar)$ and $\zg(\tvar)$
and then determine vector valued  functions $\uf(\tvar)$ and $\ug(\tvar)$ from \eqref{eqn:discreteInv}.
By differentiating \eqref{eqn:discreteInv} we obtain
\begin{align}\label{eqn:zeroAdiffeq}
\frac{d}{d\tvar}
\begin{bmatrix}
\ug(\tvar) \\
\uf(\tvar)
\end{bmatrix}
&=
H(\zf(\tvar),\zg(\tvar))
\;
\frac{d}{d\tvar}
\begin{bmatrix}
\zg(\tvar) \\
\zf(\tvar)
\end{bmatrix}
\end{align}
where $H(\zf(\tvar),\zg(\tvar))$ is a $(\Kg+\Kf)\times(\Kg+\Kf)$ matrix
\begin{align}
\label{eqn:matrixdif}
H(\zf,\zg)
& =
\begin{bmatrix}
 \Df & -\Bf \\
-\Bg &  \Dg
\end{bmatrix},
\end{align}
which we rewrite as $H=D(I-M),$
and where
\[
D =
\begin{bmatrix}
 \Df & 0     \\
  0  & \Dg
\end{bmatrix}
\text{  and  }
M =
\begin{bmatrix}
 0 & (\Df)^{-1}\Bf \\
(\Dg)^{-1}\Bg & 0
\end{bmatrix}
\]
and where 
\begin{itemize}
\item $\Df$ is the $\Kg \times \Kg$ diagonal matrix with
\[
\Df_{i,i} = \fS_x(\zg_i;\zf) = \sum_{j=1}^{\Kf} \smthker(\zf_j-\zg_i) \delhf_j,
\]
\item $\Dg$ is the $\Kf \times \Kf$ diagonal matrix with
\[
\Dg_{j,j} = \gS_x(\zf_j;\zg) = \sum_{i=1}^{\Kg} \smthker(\zg_i-\zf_j) \delhg_i,\]
\item $\Bf$ is the $\Kg \times \Kf$ matrix with 
\[
\Bf_{i,j} = 
-\frac{\partial \fS(\zg_i;\zf)}{\partial \zf_j} 
= \smthker(\zf_j-\zg_i) \delhf_j\,,
\]
\item $\Bg$ is the $\Kf \times \Kg$ matrix with 
\[
\Bg_{j,i} = -\frac{\partial \gS(\zf_j;\zg)}{\partial \zg_i} = \smthker(\zg_i-\zf_j) \delhg_i\,.
\]
\end{itemize}

Since $\Dg_{j,j} = \sum_{i=1}^{\Kg} \Bg_{j,i}$ and
$\Df_{i,i} = \sum_{j=1}^{\Kf} \Bf_{i,j}$
we observe that $M$ is a stochastic matrix
(non-negative with rows that sum to $1.$)

Our strategy to construct spatial fixed points for a given pair $\hf,\hg$ is to solve
\eqref{eqn:zeroAdiffeq} for $\zg(\tvar),\zf(\tvar)$ for a specified pair $\uf(\tvar),\ug(\tvar).$
The main difficulty we face is that $H(\zf,\zg)$ is not invertible.
In particular, $(I-M)\vec{1}_{\Kg+\Kf} = 0,$ where  $\vec{1}_k$ denotes
the all-$1$ vector of length $k.$
This is a consequence of the fact that
translating $\zf$ and $\zg$ together does not alter $\ug$ and $\uf$ as defined
by \eqref{eqn:discreteInv}.
The corresponding left null eigenvector of $H(\zf,\zg)$ 
arises from the fixed point condition $A(\hf(\cdot;\uf),\hg(\cdot;\ug)) = 0$
which reduces to
\begin{align}\label{eqn:discA}
1 = \sum_{j=1}^{\Kf} \uf_j \delhf_j
+ \sum_{i=1}^{\Kg} \uf_i \delhg_i,
\end{align}
hence
\[
\sum_{j=1}^{\Kf} \delhf_j \frac{d\uf_j}{d\tvar}
+ \sum_{i=1}^{\Kg} \delhg_i \frac{d\ug_i}{d\tvar}= 0
\]
as can be verified directly.

Let us consider the matrix
\[
H(\zf,\zg) + \vec{1}_{\Kg+\Kf}\vec{\delta}^T
\]
where $\vec{\delta}$ is the column
vector obtained by stacking $\delhg$ on $\delhf.$ 
We claim that this matrix is invertible, i.e., its determinant is non-zero. 
To see this note that $M^2$ is a block diagonal matrix where the diagonal blocks
are positive stochastic matrices. It follows from the Perron-Frobenius theorem
that $M^2$ has eigenvectors
\( 
\vec{1}_{\Kf+\Kg} =\begin{bmatrix}\vec{1}_{\Kg} \\ \vec{1}_{\Kf} \end{bmatrix}
\)
and
\( 
\begin{bmatrix}\vec{1}_{\Kg} \\ -\vec{1}_{\Kf} \end{bmatrix}
\)
both with eigenvalue $1$ and that all other eigenvalues have magnitude strictly less than $1.$
Correspondingly, $M$ has the above eigenvectors with eigenvalues $1$ and $-1$
respectively and all other eigenvalues have magnitude less than $1.$
It follows that $\vec{1}_{\Kf+\Kg}$ is the unique right null vector of $H(\zf,\zg)$ (up to scaling)
and that $\vec{\delta}$ is the corresponding left null vector.
The left subspace orthogonal to $\vec{1}_{\Kg+\Kf}$ is invariant under $H(\zf,\zg).$
It now follows that $H(\zf,\zg) + \vec{1}_{\Kg+\Kf}\vec{\delta}^T$ has no left null vector and
it is therefore invertible.


Now, consider the differential equation
\begin{align}
\frac{d}{d\tvar}
\begin{bmatrix}
\zg(\tvar) \\
\zf(\tvar)
\end{bmatrix}
&=
(H(\zf(\tvar),\zg(\tvar))+ \vec{1}_{\Kg+\Kf}\vec{\delta}^T)^{-1}
\;
\frac{d}{d\tvar}
\begin{bmatrix}
\ug(\tvar) \\
\uf(\tvar)
\end{bmatrix}
\label{eqn:diffEQR}
\end{align}
If 
\(
\frac{d}{d\tvar}
\vec{\delta}^T 
\begin{bmatrix}
\ug(\tvar) \\
\uf(\tvar)
\end{bmatrix}=0
\)
then we obtain
\(
\frac{d}{d\tvar}
\vec{\delta}^T 
\begin{bmatrix}
\zg(\tvar) \\
\zf(\tvar)
\end{bmatrix}=0
\)
and we see that \eqref{eqn:zeroAdiffeq} is satisfied.

\begin{lemma} \label{lem:PCexitcont}
Let $\smthker$ be a strictly positive Lipschitz continuous smoothing kernel.
Let $\uf(\tvar)$ and $\ug(\tvar)$ be $C^1$ ordered vector valued functions on $[0,1]$
such that $(\hf(\cdot;\uf(\tvar)),\hg(\cdot;\ug(\tvar)))$ satisfies the strictly positive
gap condition and $A(\hf(\cdot;\uf(\tvar)),\hg(\cdot;\ug(\tvar)))=0$ for all $\tvar\in [0,1].$

Assume further that $\zf(0)$ and $\zg(0)$
are given so that 
\begin{align}
\begin{split}\label{eqn:deqsoln}
\fg(x;\zg(\tvar)) &= \hg(\fS(x;\zf(\tvar));\ug(\tvar))\\ \ff(x;\zg(\tvar)) &= \hf(\gS(x;\zg(\tvar));\uf(\tvar))
\end{split}
\end{align}
holds for all $x\in\reals$ at $t=0$ where 
$\ff(\cdot;\zf)$ and $\fg(\cdot;\zg)$ are defined as in 
\eqref{eqn:fgtdef}.
Then there exist unique bounded $C^1$ ordered vector valued functions
$\zf(\tvar)$ and $\zg(\tvar)$ on $[0,1],$
with $\zf(0)$ and $\zg(0)$ as specified,
such that  \eqref{eqn:deqsoln} holds
for all $x\in \reals$ and $\tvar\in [0,1].$
\end{lemma}
\begin{IEEEproof}
The idea of the proof is to solve \eqref{eqn:diffEQR} and conclude that
\eqref{eqn:deqsoln}  is satisfied.  By assumption \eqref{eqn:deqsoln} is satisfied
at $\tvar=0$ and if \eqref{eqn:zeroAdiffeq} is satisfied on $[0,1]$ then we can conclude
that \eqref{eqn:deqsoln} holds on $[0,1].$
Since $A(\hf(\cdot;\uf(\tvar)),\hg(\cdot;\ug(\tvar)))=0$ for $\tvar \in [0,1]$
we have 
\(
(\sum_i\delhg_i \ug_i(\tvar) +\sum_j\delhf_j \uf_j(\tvar)) = 1
\)
by \eqref{eqn:discA} so 
\(
\frac{d}{d\tvar}(\sum_i\delhg_i \ug_i(\tvar) +\sum_j\delhf_j \uf_j(\tvar))=0\,.
\)
Thus, on $[0,1]$ \eqref{eqn:diffEQR}  implies \eqref{eqn:zeroAdiffeq} and
we see that solving \eqref{eqn:diffEQR} is sufficient.
For sake of argument we can extend $\uf(\tvar),\ug(\tvar)$
for all $\tau \in \reals$ so that 
\(
\frac{d}{d\tvar}(\sum_i\delhg_i \ug_i(\tvar) +\sum_j\delhf_j \uf_j(\tvar))=0
\)
holds and $\frac{d}{d\tvar}\ug(\tvar)$ and  $\frac{d}{d\tvar}\uf(\tvar)$
are bounded continuous functions. 


Let us define the region $R_D \defeq \{ (\zf,\zg) : \|\zf\|_\infty < D; \|\zg\|_\infty < D\}.$
For any fixed $D$ we have that the entries of 
\(
(H(\zf,\zg) + \vec{1}\vec{\delta}^T)^{-1}
\)
are Lipschitz on $R_D.$
By standard results on differential equations (e.g. \cite{DIFFEQREF}), the equation \eqref{eqn:diffEQR} 
has a unique continuous solution $(\zg(\tvar),\zf(\tvar))$ in some neighborhood of $\tvar=0$ and the solution extends uniquely
as long as it does not approach the boundary of the region $R_D.$
Since $D$ is arbitrary  the solution extends uniquely as long as 
$\zg(\tvar),\zf(\tvar)$ remain finite.

By assumption \eqref{eqn:deqsoln} holds for $\tvar=0.$
 By translating (adding a constant to both $\zf$ and $\zg$) we can assume
\begin{align}\label{eqn:sumzzero}
\sum_{i=1}^{\Kg} \delhg_i \zg_i(\tvar)  +\sum_{j=1}^{\Kf} \delhf_j \zf_j(\tvar)  = 0
\end{align}
at $\tvar=0$ and it then follows that
\eqref{eqn:sumzzero} holds along the solution.
We claim that there exists a constant $Z$ such that \eqref{eqn:deqsoln} implies 
$\max \{\| \zf(\tvar)\|_\infty,\|\zg_j(\tvar)\|_\infty\} \le Z.$
This claim then implies the existence of a unique solution to \eqref{eqn:diffEQR}
for $\tvar \in [0,1],$ completing the proof.

We now prove the claim.
By continuity we see that $\altPhi$
satisfies the strictly positive gap condition {\em unifomly} for  $\tvar \in [0,1].$
 More specifically, for all $(u,v) \in  [0,1]^2\backslash \{(0,0),(1,1)\},$
we have
\[
\altPhi_{\min}(u,v) \defeq \min_{\tvar\in [0,1]} \altPhi\bigl(\hf(\cdot;\uf(\tvar)),\hg(\cdot;\ug(\tvar));u,v\bigr)> 0\,.
\]
Note, moreover, that $\altPhi_{\min}(u,v)$ is Lipschitz continuous.
Hence, there exists $\eta > 0$  such that
$\altPhi_{\min}(u,v)  \ge \eta$ for $(u,v) \in  [\delhg_1,1- \delhg_{\Kg}]\times [0,1]$ and for
$(u,v) \in  [0,1]\times [\delhf_1,1- \delhf_{\Kf}]$ and such that
$\uf_1(\tvar),\ug_1(\tvar),1-\uf_{\Kf}(\tvar),1-\ug_{\Kg}(\tvar)  \ge \eta$
for all $\tvar \in [0,1].$  For convenience we also assume $\eta \le \frac{1}{2}.$

Assume that \eqref{eqn:deqsoln} holds for some $\tvar\in [0,1].$
Then we have $\eta \le \ug_1(\tvar) = \fS(\zg_1(\tvar)) \le \Omega(\zf_1(\tvar)-\zg_1(\tvar))$
and 
$\eta \le \gS(\zf_1(\tvar)) \le \Omega(\zg_1(\tvar)-\zf_1(\tvar)).$
Hence 
\[
|\zg_1(\tvar)-\zf_1(\tvar)| \le -\Omega^{-1}(\eta-).
\]
Let $x = (\zf_i(\tvar) + \zf_{i+1}(\tvar))/2$ and let $L =  (\zf_{i+1}(\tvar) - \zf_{i}(\tvar))/2,$ then $\Delta_L f(x) = 0.$ 
Since $f(x) \in [\delhf_1,1- \delhf_{\Kf}]$ 
we have by Lemma \ref{lem:transitionPhiBounds} (the first inequality) 
$\eta \le \Omega(-L).$  Hence we obtain 
 \[
|\zf_{i+1}(\tvar) - \zf_{i}(\tvar)|  \le -2\Omega^{-1}(\eta-)\,.
\]
A similar argument applies to $\zg$ and so, 
by \eqref{eqn:sumzzero} we see that we can take 
 $Z = -2(\Kf+\Kg)\Omega^{-1}(\eta-).$
\end{IEEEproof}

Now we extend the above analysis to the case where $\ashift \neq 0.$
We modify \eqref{eqn:discreteInv} as follows
\begin{align}
\begin{split}\label{eqn:discreteInvShift}
\ug_i &\defeq \fS(\zg_i;\zf) = \sum_{j=1}^{\Kf} \delhf_j \Omega (\zg_i-\zf_j) \\
\uf_j &\defeq  \gS(\zf_j+\ashift;\zg) =\sum_{j=1}^{\Kf} \delhf_i \Omega (\zf_j+\ashift-\zg_i)\,.
\end{split}
\end{align}
Introducing smooth dependence on $\tvar$ we now have
\begin{align}
\begin{split}
\fg(x;\zg(\tvar)) &= \hg(\fS(x;\zf(\tvar));\ug(\tvar))\\ \ff(x;\zg(\tvar)) &= \hf(\gS(x+\ashift(\tvar);\zg(\tvar));\uf(\tvar))\,
\end{split}
\end{align}
and by differentiating we obtain
\begin{align}
\begin{split}
\frac{d}{d\tvar}
\begin{bmatrix}
\ug(\tvar) \\
\uf(\tvar)
\end{bmatrix}
=&
H(\zf(\tvar),\zg(\tvar))
\;
\frac{d}{d\tvar}
\begin{bmatrix}
\zg(\tvar) \\
\zf(\tvar)
\end{bmatrix}
\\ & +
\begin{bmatrix}
0 \\
\Dga \vec{1}_{\Kf}
\end{bmatrix}
\frac{d}{d\tvar}\ashift(\tvar)
\label{eqn:diffEQ}
\end{split}
\end{align}
where $H(\zf(\tvar),\zg(\tvar))$ is a $(\Kg+\Kf)\times(\Kg+\Kf)$ matrix
\begin{align}
\label{eqn:matrixdif}
H(\zf,\zg)
& =
\begin{bmatrix}
 \Df & -\Bf \\
-\Bga &  \Dga
\end{bmatrix}
\\
& =
D
\begin{bmatrix}
I-M
\end{bmatrix}
\end{align}
where
\[
D =
\begin{bmatrix}
 \Df & 0     \\
  0  & \Dga
\end{bmatrix}
\text{  and  }
M =
\begin{bmatrix}
 0 & (\Df)^{-1}\Bf \\
(\Dga)^{-1}\Bga & 0
\end{bmatrix}\,.
\]

The form of $\Df$ and $\Bf$  are as before and
\begin{itemize}
\item $\Dga$ is the $\Kf \times \Kf$ diagonal matrix with
\[
\Dga_{j,j} = \gS_x(\zf_j+\ashift) = \sum_{i=1}^{\Kg} \smthker(\zg_i-(\zf_j+\ashift)) \delhg_i,
\]
\item $\Bga$ is the $\Kf \times \Kg$ matrix with 
\[
\Bga_{j,i} = 
-\frac{\partial \gSa(\zf_j;\zg)}{\partial \zg_i} = 
\smthker(\zg_i-(\zf_j+\ashift)) \delhg_i\,.
\]
\end{itemize}
Since $\Dga_{j,j} = \sum_{ij=1}^{\Kg} \Bga_{j,i}$ and
$\Df_{i,i} = \sum_{j=1}^{\Kf} \Bf_{i,j}$
we observe that $M$ is a stochastic matrix: $\sum_{j=1}^{\Kf+\Kg} M_{i,j} = 1.$


Let $P$ be the projection matrix which is the $(\Kf+\Kg) \times (\Kf+\Kg)$
identity matrix except that $P_{\Kf+\Kg , \Kf+\Kg}=0.$
It follows that $I-PMP$ is invertible and $PMP$ has spectral radius less than one.
Indeed, let
$\tilde{B_1}$ denote the matrix obtained from $(\Df)^{-1}\Bf$ be removing the rightmost column and
let $\tilde{B_2}$ denote the matrix obtained from $(\Dga)^{-1}\Bga$ be removing the bottom row.
Let $\tilde{M}$ denote the upper left  $\Kf+\Kg -1 \times \Kf+\Kg-1$ submatrix of $M.$
Then
\[
\tilde{M}^{2} =
\begin{bmatrix}
(\tilde{B_1}\tilde{B_2})^{2} & 0 \\
0  & (\tilde{B_2}\tilde{B_1})^{2}
\end{bmatrix}\,.
\]
Let $\xi < 1$ denote the maximum row sum from $\tilde{B_1}.$
By the Perron-Frobenious theorem $\tilde{B_2}\tilde{B_1}$ has a maximal positive eigenvalue $\lambda$
with positive left eigenvector $x.$ Then $x^T \tilde{B_2}\tilde{B_1} \vec{1} = \lambda x^T  \vec{1},$
but $\tilde{B_2}\tilde{B_1} \vec{1} \le \xi \vec{1}$ (component-wise) so $\lambda \le \xi.$
We easily conclude that $\| \tilde{M}^2 \|_2 \le \xi.$
Hence $(I-PMP)^{-1}$ exists and is strictly positive.

Given $\zf(0),\zg(0),$ let $\zg(\tvar),\zf(\tvar)$ be the solution to
\begin{align}
\frac{d}{d\tvar}
\begin{bmatrix}
\zg(\tvar) \\
\zf(\tvar)
\end{bmatrix}
&=
-(I-PM(\zf(\tvar),\zg(\tvar))P)^{-1}
\;
P
\begin{bmatrix}
0 \\
\vec{1}_{\Kf}
\end{bmatrix}\,.
\label{eqn:diffEQRshift}
\end{align}
Note that the last coordinate on the right hand side is $0$ so
\[
P\frac{d}{d\tvar}
\begin{bmatrix}
\zg(\tvar) \\
\zf(\tvar)
\end{bmatrix} =
\frac{d}{d\tvar}
\begin{bmatrix}
\zg(\tvar) \\
\zf(\tvar)
\end{bmatrix}\,.
\]
Recall that we assume that $\smthker$ is Lipschitz continuous.
By standard results on differential equations  \cite{DIFFEQREF}
a unique solution exists in some neighborhood of $\tvar = 0$ and can be uniquely extended
as long as $\zg(\tvar),\zf(\tvar)$ remain finite.
Note that $\zf(\tvar)$ and $\zg(\tvar)$ are component-wise decreasing in $\tvar,$
except for the component $\zf_{\Kf}(\tvar)$ which is constant.
Thus, the solution can be extend for increasing $\tvar$ as long as 
$\zf_1(\tvar) > -\infty$ and $\zg_1(\tvar) >-\infty.$
Let $T$ denote the maximal value such that the solution exists for $\tau \in [0,T).$

If we substitute the solution into \eqref{eqn:diffEQ}
and set $\frac{d}{d\tau} \ashift(\tau) = 1$ then we obtain
\begin{align*}
P\frac{d}{d\tvar}&
\begin{bmatrix}
\ug(\tvar) \\
\uf(\tvar)
\end{bmatrix} 
=
D\Biggl[
P(I-M)
\frac{d}{d\tvar}
\begin{bmatrix}
\zg(\tvar) \\
\zf(\tvar)
\end{bmatrix}
+P
\begin{bmatrix}
0 \\
\vec{1}_{\Kf}
\end{bmatrix}
\Biggr] \\
&=
D\Biggl[
P(I-PMP)
\frac{d}{d\tvar}
\begin{bmatrix}
\zg(\tvar) \\
\zf(\tvar)
\end{bmatrix}
+P
\begin{bmatrix}
0 \\
\vec{1}_{\Kf}
\end{bmatrix}
\Biggr] \\
& = 0\,,
\end{align*}
so that only the last coordinate of $\uf(\tvar)$ is non-constant.
This coordinate is non-decreasing since
\begin{align*}
& \frac{d}{d\tvar} \gS(\zf_{\Kf}+\ashift(\tvar);\zg(\tvar)) \\
& =
\gS_x(\zf_{\Kf}+\ashift(\tvar);\zg(\tvar))\cdot \\
&\qquad\Bigl(1 - \sum_{j=1}^{\Kg}
\smthker(\zf_{\Kf}+\ashift(\tvar)-\zg_j(\tvar))\, \delhg_j
\frac{d}{d\tvar} \zg_j(\tvar)
\Bigr)
\end{align*}
and $\frac{d}{d\tvar} \zg_j(\tvar) \le 0.$

We require the following auxilliary Lemma.
\begin{lemma}\label{lem:satSPGC}
Assume $\hf,\hg$ satisfies the strictly positive gap condition with $A>0.$
Given $r\in [0,1]$ define $\hf(v;r)$ by
$\hf^{-1}(v;r)  = \hf^{-1}(v) \wedge r,$
i.e. saturate $\hf^{-1}$ at $r.$
Let $r_0$ satisfy $A(\hf(\cdot;r_0),\hg)=0$
($r_0$ is uniquely determined).
Then $\hf(\cdot;r),\hg$ satisfies the strictly positive gap condition for
all $r \in [r_0,1].$
\end{lemma}
\begin{IEEEproof}
Clearly $\int_0^v \hfinv(v';r) dv'$ is non-decreasing in $r$ for all $v \in [0,1].$
Since $A(\hf,\hg) >0$ we have $\int_0^1 \hfinv(v) dv > 1 -\int_0^1 \hginv(u) du$ and
since $\int_0^1 \hfinv(v;0) dv = 0$
there exists a unique positive $r_0 < 1$ such that
$\int_0^1 \hfinv(v;r_0) dv =  1-\int_0^1 \hginv(u) du,$
i.e. such that $A(\hf(\cdot;r_0),\hg)=0.$

To prove the lemma we need to show that
$\altPhi(\hf(;r),\hg) > A(\hf(;r),\hg )$ on $\intcross(\hf(;r),\hg ).$
We have
\begin{align*}
&A(\hf(;r),\hg)-\altPhi(\hf(;r),\hg;u,\hf(u;r))
\\&= \int_u^1 (\hg^{-1}(u')-\hf(u';r)) du'
\\&\le \int_u^1 (\hg^{-1}(u')-\hf(u')) du'
\\&=A(\hf,\hg)-\altPhi(\hf,\hg;u,\hf(u))
\end{align*}
Let $(u,v) \in \intcrossing (\hf(;r),\hg )$ and note that this implies $u \le r$
since $\hg$ is continuous at $1$ (Lemma \ref{lem:zocontinuity})
and we may assume that $v = \hf(u;r).$
If $u<r$ then $(u,v) \in \intcrossing (\hf,\hg )$ and we have
$A(\hf,\hg)-\altPhi(\hf(;r),\hg;u,\hf)<0$ giving
$A(\hf(;r),\hg)-\altPhi(\hf(;r),\hg;u,\hf(u;r))<0.$
If $u=r$ then $r<1$ and we have 
$\int_u^1 (\hg^{-1}(u')-\hf(u';r)) du' =\int_u^1 (\hg^{-1}(u')-1) du' < 0$
again by continuity of $\hg$ at $1.$
\end{IEEEproof}

%

\begin{lemma}\label{lem:PCshiftexist}
Let $\zf(0),\zg(0)$ and $\ashift(0) \ge 0$ be given, thereby defining
$(0,1)$-interpolating functions
$f(\cdot;\zf(0))$ and $g(\cdot;\zg(0)).$
Let $\zf(\tvar),\zg(\tvar)$
be the solution to \eqref{eqn:diffEQRshift} defined on $[0,T),$
set $\ashift(\tvar) = \ashift(0)+\tvar,$
and define
\begin{align*}
\ug_i(\tvar) &= \fS(\zg_i(\tvar);\zf(\tvar))\\
\uf_j(\tvar) & =\gS(\zf_j(\tvar)+\ashift(\tvar);\zg(\tvar))\\
r(\tvar) & = \uf_j(\tvar)-\uf_j(0)\,.
\end{align*}
Assume $\hf(\cdot;\uf+r e_{\Kf}),\hg$ satisfies the strictly positive gap condition for some $r\in (0,1-\uf_{\Kf})$ where 
$\vec{e}_{Kf} = (0,\ldots,0,1)^T$ is of length $\Kf.$
Then there exists $\tvar < T$ such that
$r(\tvar) = r.$
\end{lemma}
\begin{IEEEproof}
First we show that $T=\infty$ implies the lemma.
Assume  $\uf_{\Kf}(\tvar) \le \uf_{\Kf}(0)+r$
and consider $u>\uf_{\Kf}(0)+r$  and $v > \fS(\zg_{\Zg}(0);\zf(0)) = \fS(\zg_{\Zg}(\tvar);\zf(\tvar)).$
We have $(u,v) \in G^-(\hf(\cdot;\tau),\hg)$ 
and we obtain
a finite upper bound on $\ashift(\tvar)$ from Lemma \ref{lem:shiftupperbound}
that is independent of $\uf_{\Kf}(\tvar).$
This then gives a finite upper bound on $\tvar.$

Assume that $T < \infty$ and that $\uf_{\Kf}(\tvar) \le \uf_{\Kf}(0)+r$ for $\tvar \le T.$
Now, since  $f(\cdot;  \zf(\tau))$ and $g(\cdot; \zg(\tau))$ are non-increasing in $\tvar$ we  see that the limits as $\tau \rightarrow T$,
which we denote $f(\cdot;  T)$ and $g(\cdot; T),$
are well defined.
Since  $\uf_{\Kf}(\tvar)$ is non decreasing and bounded above $\uf_{\Kf}(T)$
is also well defined as a limit and $\hf(\cdot;\uf(\tvar))$ has limit $\hf(\cdot;\uf(T)).$
From Theorem \ref{thm:mainlimit}
it follows that $(f(\minfty;T),g(\minfty;T)) \in \cross(\hf(\cdot;\uf(T)),\hg)$
and 
\[
0 \le \altPhi(\hf(\cdot;T),\hg;g(\minfty;T),f(\minfty;T))
\le A(\hf(\cdot;r(T)),\hg)\,.
\]
Since $\zf_{\Kf}$ is constant
we have $f(\minfty)<1$ and $g(\pinfty;T)=f(\pinfty;T))=1.$ 
By Lemma \ref{lem:satSPGC} the strictly positive
gap condition holds for $\hf(\cdot;T),\hg$ and
hence we must have $(f(\minfty),g(\minfty)) = (0,0)$ 
which contradicts the definition of $T.$
We conclude that there must exist $\tvar < T$ such that
$\uf(\tvar) = \uf(0)+r.$
\end{IEEEproof}

We are now ready to prove the main result.
\begin{IEEEproof}[Proof of Theorem \ref{thm:PCexist}]
We first consider the case $A(\hf,\hg)= 0.$
Let the piecewise constant target EXIT functions be $\hf=\hf(\cdot;\uf)$ and $\hg=\hg(\cdot;\ug).$
Let $B_{\hf} = \int_0^1 \hfinv(x) dx = \sum_j \delhf_{j} \uf_{j},$
and $B_{\hg} = \int_0^1 \hginv(x) dx =  \sum_i \delhg_{i} \ug_{i}= 1-B_{\hf}.$ 
The last equality encapsulates $A(\hf,\hg)=0.$
For $\tvar \in [0,1]$ define the vector valued functions
\begin{align}
\begin{split}
\uf (\tvar) & = (1-\tvar) B_{\hf}\vec{1}_{\Kf} + \tvar \uf \\
\ug (\tvar) & = (1-\tvar) B_{\hg}\vec{1}_{\Kg} + \tvar \ug\,,\label{eqn:pcscale}
\end{split}
\end{align}
Note that we have $\uf(1)=\uf$ and $\ug(1)=\ug.$
Note that $\hf(;\uf(\tvar))$ and $\hg(;\ug(\tvar))$ are in $\exitfns$ for all $\tvar,$
 that $\int_0^1 \hfinv(v;\uf(\tvar)) dv = B_{\hf},$
and that $\int_0^1 \hginv(u;\ug(\tvar)) du = B_{\hg}$
so $A (\hf(;\uf(\tvar)),\hg(;\ug(\tvar))) = 0$ for all $\tvar \in [0,1].$

Let $h \in \exitfns$ be arbitrary and let $(u,v)$ be in the graph of $h.$  Then
\begin{align*}
&\altPhi(h,\hg(\cdot;\ug(\tvar));u,v) - \altPhi(h,\hg(\cdot;\ug(1));u,v)\\
= &
\int_0^u (\hginv(z;\ug(\tvar))-\hginv(z;\ug(1)))\text{d}z \\
= &
(1-\tvar)\Bigl(u B_{\hg} - \int_0^u \hginv(z;\ug(1))\text{d}z\Bigr) \\
\ge & 0
\end{align*}
where the last inequality holds since we have equality at $u=0$ and $u=1$
and $(u B_{\hg} - \int_0^u \hginv(z;\ug(1)))\text{d}z$ is concave in $u.$
By Corollary \ref{lem:miniscross} we have $\altPhi(h,\hg(\cdot;\ug(1));u,v)>0$
for $(u,v)\not\in \{(0,0),(1,1)\}$ if  $h,\hg(\cdot;\ug(1))$ satisfies the 
strictly positive gap condition with $A=0.$
The above then implies that 
$(h,\hg(\cdot;\ug(\tvar)))$ also satisfies the strictly positive gap condition
with $A(h,\hg(\cdot;\ug(\tvar))) = 0$ for all $\tvar \in [0,1].$

The above argument shows that $(\hf(\cdot;\uf(1)),\hg(\cdot;\ug(\tvar)))$
satisfies the strictly positive gap condition for all $\tvar\in [0,1].$
Applying the argument analogously to $\hf$ we can deduce that
$(\hf(\cdot;\uf(s)),\hg(\cdot;\ug(\tvar)))$
satisfies the strictly positive gap condition for all $s,\tvar\in [0,1],$
and, in particular, 
$(\hf(\cdot;\uf(\tvar)),\hg(\cdot;\ug(\tvar)))$
satisfies the strictly positive gap condition for all $\tvar\in [0,1].$

All that remains to apply Lemma \ref{lem:PCexitcont} over $\tvar \in [0,1]$ and
conclude the proof for the case $A(\hf,\hg)=0$
is to find $\zf(0)$ and $\zg(0).$
Set $\zf(0) = 0$  so that
$f(x;\zf(0)) = \,\unitstep(x).$
Let $y$ be the unique point such that $\fS(y;\zf(0)) = B_{\hg}$ and
set each component of $\zg(0)$ to $y$ so that $\fg(x;\zg(0)) = \unitstep(x-y).$
It follows that $\fg(x;\zg(0)) = \hg(\fS(x;\zf(0));\ug(0))$ and that
$\ff(x;\zf(0)) = \hf(\gS(x;\zg(0));\uf(0)).$

Applying Lemma \ref{lem:PCexitcont} for $\tvar \in [0,1]$ we obtain
$\ff(;\zf(\tvar))$ and $\fg(;\zg(\tvar))$ such that
$\fg(x;\zg(\tvar)) = \hg(\fS(x;\zf(\tvar));\ug(\tvar))$ and
$\ff(x;\zf(\tvar)) = \hf(\gS(x;\zg(\tvar));\uf(\tvar))$
completing the proof for the $A(\hf,\hg)=0$ case.

We now consider the case $A(\hf,\hg) \neq 0.$
Without loss of generality we assume $A(\hf,\hg)  > 0.$
The case $A(\hf,\hg)  < 0$ is equivalent to the case $A(\hf,\hg)  > 0$
under the affine symmetry that allows us to exchange $(0,0)$
and $(1,1).$

Let us introduce a modification of $\uf,$ denoted $\uf(r).$ as follows.
For $r \in (0,1)$ define $\uf_i(r) = r \wedge  \uf_i.$
Then $\int_0^v \hfinv(x;\uf(r)) dx$ is non-decreasing in $r$ for all $v \in [0,1].$
Since $\int_0^1 \hfinv(x) dx > 1 -\int_0^1 \hginv(x) dx$ and
$\int_0^1 \hfinv(x;\uf(t;0)) dx = 0$
there exists a unique positive $r_0 < \uf_{\Kf}$ such that
$\int_0^1 \hginv(x;\uf(t;r_0)) dx =  1-\int_0^1 \hginv(x) dx.$

We claim that for all $r \in [r_0,\uf_{\Kf}]$ the pair
$(\hf(;r),\hg)$ satisfies the strictly positive gap condition.
To establish the claim we need to show that
$\altPhi(\hf(;r),\hg) > A(\hf(;r),\hg )$ on $\intcross(\hf(;r),\hg ).$
Let $ (u,v) \in \intcrossing (\hf(;r),\hg ).$
Note that this implies $u \le r$ since $\hg$ is continuous at $1$ 
by Lemma \ref{lem:zocontinuity}.
If $u < r$ then $v \le \hf^{-1}(r-)$ and
we have $(u,v) \in \intcrossing (\hf,\hg).$
This yields
\(
\altPhi(\hf(;r),\hg;u,v) =
\altPhi(\hf,\hg;u,v) > A(\hf,\hg) \ge A(\hf(;r),\hg)
\,.
\)
If $u = r$ then we have from \eqref{eqn:altPhivhgv}
\(
\altPhi(\hf(;r),\hg;1,1)-\altPhi(\hf(;r),\hg;u,v)
= \int_u^1 (\hg^{-1}(u')-1) du' < 0,
\)
where the last inequality uses continuity of
$\hg$ at $1$ (Lemma \ref{lem:zocontinuity}).

Applying our above result for the $A = 0$ case
we can find $f(\cdot;\zf),g(\cdot;\zg)$  that
form a $(0,1)$-interpolating spatial fixed point pair for $(\hf(\cdot;\uf(r_0)), \hg(\cdot;\ug)).$

We now apply Lemma \ref{lem:PCshiftexist} in a series of stages.
Let $j'$ be the least $j$ such that $r_0 < \uf_j.$
Then we have a stage for $j',j'+1,\ldots,\Kf.$
Let us first consider the stage $j'.$
We effectively initialized \eqref{eqn:diffEQRshift} with $\zf,\zg$ but, to be precise,
we must reduce the coordinate system.
Note that $\uf(r_0)$ can be interpreted as a vector of length $j'$
by collapsing $r_0 = \uf_{j'}(r_0) = \uf_{j'+1}(r_0)  = \ldots = \uf_{\Kf}(r_0)$
to a single coordinate.
Consequently, we have $\zf_{j'}= \zf_{j'+1}  = \ldots = \zf_{\Kf}.$
We introduce $\tzf(0)$ a vector with $j'$ coordinates with
$\tzf_{j'}(0) = \zf_{j'}.$
Correspondingly, we redefine $\delhf$ as $\tilde{\delta}^f$ with
$\tilde{\delta}^f_j =\delhf_j$ for $j<j'$ and
 $\tilde{\delta}^f_{j'} = \sum_{j=j'}^{\Kf} \delhf_j.$
We leave $\zg$ unchanged, i.e., $\zg(0)=\zg.$
We now apply Lemma \ref{lem:PCshiftexist} to the reduced system with initial condition
$\zg(0),\tzf(0)$ to obtain
$\zg(\tvar)$ and $\tzf(\tvar)$ such that $\ug$ is invariant with $\tvar$ and only the $j'$th
coordinate of $\tuf$ changes (increases) with $\tvar.$
For $j<j'$ we have $\tuf_j(\tvar) = \uf_j$ and
Lemma \ref{lem:PCshiftexist} guarantees the existence of $\tvar'$ such that
 $\tuf_{j'}(\tvar') = \uf_{j'}$ and 
the functions $\tzf(\tvar'),\zg(\tvar')$ form a travelling wave solution for
$\hf(\cdot,\tuf(\tvar')),\hg(\cdot;\ug))$ with 
$\ashift(\tvar') = \ashift(0)+\tvar'.$ 
Note that $\hf(\cdot,\tuf(\tvar')) =\hf(\cdot,\uf(r))$ with $r=\uf_{j'}.$

We now take this traveling wave solution to construct the initial condition for the $j'+1$th stage.
In this stage we increase the dimension of $\tzf$ by $1.$ 
Thus, for this stage we set $\tzf_j(0)$ equal to $\tzf_j(\tvar')$ from the previous stage
for $j<j'+1$ and we set $\tzf_{j'+1}(0) = \tzf_{j'}(0).$
Correspondingly we now have 
 $\tilde{\delta}^f_{j'+1} = \sum_{j=j'+1}^{\Kf} \delhf_j$
and $\tilde{\delta}^f_j =\delhf_j$ for $j<j'+1.$
Recall that $\zg$ has remained unchanged, and we reinitialize $\zg(0).$
We also reinitialize $\ashift(0)$ setting it equal to $\ashift(\tvar')$ from the previous stage.
We now again apply Lemma \ref{lem:PCshiftexist}
to the reduced system with initial condition
$\zg(0),\tzf(0)$ to obtain
$\zg(\tvar)$ and $\tzf(\tvar)$ such that $\ug$ is invariant with $\tvar$ and only the $j'+1$th
coordinate of $\tuf$ changes (increases) with $\tvar.$
For $j<j'+1$ we have $\tuf_J(\tvar) = \uf_j$ and
Lemma \ref{lem:PCshiftexist} guarantees the existence of $\tvar''$ such that
 $\tuf_{j'+1}(\tvar'') = \uf_{j'+1}$ and 
the functions $\tzf(\tvar''),\zg(\tvar'-)$ determine a traveling wave solution for
$\hf(\cdot,\tuf(\tvar'')),\hg(\cdot;\ug))$ with 
$\ashift(\tvar'') = \ashift(0)+\tvar''.$ 
Note that now $\hf(\cdot,\tuf(\tvar'')) =\hf(\cdot,\uf(r))$ with $r=\uf_{j'+1}.$

We can now take the resulting solution at $\tvar''$ as the initial condition 
for $j'+2$th stage.  Using the same argument as above we obtain a travelling wave
solution for $\hf(\cdot,\uf(r)),\hg(\cdot;\ug))$ with $r=\uf_{j'+2}.$
Continuing the stages in this fashion we ultimately arrive at a travelling wave solution
for the original system ($r=\uf_{\Kf}.$)

By standard results on differential equations we see that the solutions $\zf,\zg$ we have
obtained depend continuously on $\uf,\ug.$  Moreover, the differential equations that
we used to construct $\zf,\zg$ are reversible, and the reverse equations also have unique solutions.
This implies that the found vectors $\zf,\zg$ are unique.
\end{IEEEproof}

\subsubsection{Convergence}

In the piecewise constant case with strictly positive averaging kernel 
we can also show convergence to the solution constructed above for all initial conditions.
Define $f_\lambda$ by 
\[
f^{-1}_\lambda = \lambda f^{t+1,-1} + (1-\lambda) f^{t,-1}
\]
and set
$g_\lambda = \hf \circ f^\smthker_\lambda.$
Then by applying \eqref{eqn:delgfrecursionInv} we  obtain
\begin{align*}
&g^{t+1,-1}(u) -  g^{t,-1}(u) 
\\
=&
\int_0^1 \int_0^1 M(\lambda,u,v)(f^{t+1,-1}(v) - f^{t,-1}(v)) \,dv\,d\lambda
\end{align*}
where
\begin{align*}
M(\lambda,u,v)=\frac{\int_0^1 \smthker(g^{-1}_\lambda(u)-f^{-1}_\lambda(v))}{\int_0^1 \smthker(g^{-1}_\lambda(u)-f^{-1}_\lambda(v)) \,dv}
\end{align*}
Since $\smthker$ is strictly positive we have
 for each $\lambda$ and $u$ that $M(\lambda,u,v)>0$ and $\int_0^1 M(\lambda,u,v)\,dv = 1.$
So we obtain
\begin{align*}
\sup_u (g^{t+1,-1}(u) -  g^{t,-1}(u)) &\le \sup_v (f^{t+1,-1}(v) -  f^{t,-1}(v))\, ,
\\
\inf_u (g^{t+1,-1}(u) -  g^{t,-1}(u)) &\ge \inf_v (f^{t+1,-1}(v) -  f^{t,-1}(v))\,.
\end{align*}
In the piecewise constant case the inverse functions are bounded and with strictly positive $\smthker$ we see that
the inequalities are strict unless  $f^{t+1,-1}(v) -  f^{t,-1}(v)$ is a constant.
It is easy to conclude in this case that $f^{t+1,-1}(v) -  f^{t,-1}(v)$ converges in $t$ to a constant in $v.$
From this it follows that the $f^t$ converges to the unique solution given above (with suitable translation).


\subsection{Existence of Consistent Spatial Waves\label{sec:ECSW}}

In Section \ref{sec:PCcase} we proved Theorem
\ref{thm:PCexist},
a special case of Theorem \ref{thm:mainexist}
in which $\hg$ and $\hf$ are piecewise constant
functions and $\smthker$ is Lipschitz continuous and strictly positive.
In this section we show how to remove the special conditions 
to arrive at the general results.
We make repeated use of the limit theorems of Section \ref{sec:limitthms} and develop
some approximations for functions in $\exitfns.$
It is quite simple to approximate $h \in \exitfns$ using piecewise constant functions.
The challenge is to approximate a pair $(\hg,\hf)$ so that the strictly positive gap
condition is preserved.

\subsubsection{Approximation by Tilting}

In a manner analogous to \eqref{eqn:pcscale} we define a perturbation of
$\hf,\hg$ as $\hf(;\tvar),\hg(;\tvar)$ for $\tvar \in [0,1]$ by
\begin{align}
\begin{split}
\hfinv (v;\tvar) & = (1-\tvar) B_{\hf}  + \tvar \hfinv(v) \\
\hginv (u;\tvar) & = (1-\tvar) B_{\hg}  + \tvar \hginv(u)\,,\label{eqn:genscale}
\end{split}
\end{align}
where we recall $B_h = \int_0^1 h^{-1}(x) \,dx.$
This can also be expressed as
\begin{align}
\begin{split}
\hf (u;\tvar) & = \hf\Bigl(\frac{u-B_{\hf}}{\tvar} + B_{\hf} \Bigr) \\
\hg (v;\tvar) & =  \hg\Bigl(\frac{u-B_{\hg}}{\tvar} +  B_{\hg}\Bigr) \,,\label{eqn:genscalefor}
\end{split}
\end{align}
with appropriate extension of $\hf$ and $\hg$ outside of $[0,1],$
$\hf(x)=\hg(x)=0$ for $x<0$ and
$\hf(x)=\hg(x)=1$ for $x>1.$

Letting $h$ denote either $\hf$ or $\hg,$ we clearly have
\begin{align}
\int_0^1 h^{-1} (x;\tvar) dx = B_h\label{eqn:slanteq}
\end{align}
for all $\tvar.$
Note also that 
\(
h^{-1}(v;\tvar) -h^{-1} (v) =
(1-\tvar)(B_h-h^{-1} (v))
\)
is non-increasing  in $v.$
It follows that
\(
\int_0^v h^{-1}(x;\tvar) dx \ge 
\int_0^v h^{-1} (x) dx 
\)
for all $v\in[0,1]$
and we obtain
\begin{align}
\altPhi(\hf(;\tvar),\hg(;\tvar);)\ge \altPhi(\hf,\hg;)\label{eqn:altslbound}
\end{align}
for all $\tvar\in [0,1].$

\begin{lemma}\label{lem:smoothcompress}
Let $(\hg,\hf) \in \exitfns^2$ 
satisfy the strictly positive gap condition.
Then, there exists $\epsilon > 0$ such that
$(\slanta{\hg},\slanta{\hf})$ 
satisfies the strictly positive gap condition
for any $t \in (1-\epsilon,1].$
\end{lemma}
\begin{IEEEproof}[Proof of Lemma \ref{lem:smoothcompress}]
For the case $A(\hf,\hg)=0$ equation \eqref{eqn:slanteq}, inequality \eqref{eqn:altslbound} and
Lemma \ref{lem:monotonic} gives
the result immediately.
By symmetry we now need only consider the case $A(\hf,\hg) > 0.$

By Lemma \ref{lem:Sstructure} and \eqref{eqn:altslbound}
it is sufficient to show that 
$\intcross(\hf(;\tvar),\hg(;\tvar)) \cap  \closure{S(\hf,\hg)}=\emptyset$
for $\tvar\in[1-\epsilon,1].$

Also by Lemma \ref{lem:Sstructure}, there exists a
minimal and positive element $(u^*,v^*) \in \intcross(\hf,\hg).$
There exists a neighborhood ${\cal N}$ of $(0,0),$ which we take to be a subset of
$ [0,u^*)\times [0,v^*),$
in which
$\hginv(u;\tvar) \ge \hginv(u)$ and 
$\hfinv(v;\tvar) \ge \hfinv(v).$ 
It follows that  
$\intcross(\hf(;\tvar),\hg(;\tvar)) \cap {\cal N} =\emptyset$
for all $\tvar.$

Let $\delta>0$ be small enough so that $\neigh{(0,0)}{\delta} \subset {\cal N}$
and $\neigh{(u^*,v^*)}{\delta} \cap \closure{S(\hf,\hg)}=\emptyset.$
For $\epsilon$ small enough and $t\in[1-\epsilon,1]$ we have
$\cross(\hf(;\tvar),\hg(;\tvar))\subset \neigh{\cross(\hf,\hg)}{\delta}$
by Lemma \ref{lem:crosspointlimit}
and it now follows that
$\intcross(\hf(;\tvar),\hg(;\tvar)) \cap  \closure{S(\hf,\hg)}=\emptyset.$
\end{IEEEproof}

\subsubsection{Piecewise Constant Approximation}
Given $h \in \exitfns$ let us define a sequence of piecewise constant approximations
$Q_n(h),$ $n=1,2,...$  by
\begin{align*}
Q_n(h) (x) &= \sum_{j=1}^n \frac{1}{n} \,\unitstep(x- u_{n,j}) 
\end{align*}
where we define the non-decreasing (in $j$) sequence $u_{n,j}$ by
\begin{align*}
u_{n,j} 
& = n\int_{(j-1)/n}^{j/n}h^{-1}(v)dv
\end{align*}
and we have
\begin{align*}
\int_{(j-1)/n}^{j/n} (Q_n(h))^{-1} (x) dx &
=\frac{u_{n,j}}{n}  = \int_{(j-1)/n}^{j/n} h^{-1}(x) dx 
\end{align*}
which gives in particular
\(
\int_0^1 (Q_n(h))^{-1} (x) dx = \int_0^1 h^{-1} (x) dx.
\)
Note that $(Q_n(h))^{-1} (x) -  h^{-1}(x)$ non-increasing on $((j-1)/n,j/n)$ so
it follows that $\int_0^z (Q_n(h))^{-1} (x) dx \ge \int_0^z h^{-1} (x) dx$ for all
$z \in [0,1].$

\begin{lemma}\label{lem:PCapprox}
Let $(\hg,\hf)$ be pair of functions in $\exitfns$ satisfying the strictly positive gap condition 
such that for some $\eta>0$ we have
$\hg (x) =\hf(x)= 0$ for
$x \in [0,\eta)$ and $\hg (x) =\hf(x)= 1$ for $x \in (1-\eta,1].$
Then, for all $n$ sufficiently large
$(Q_n(\hg),Q_n(\hf))$ satisfies the strictly positive gap condition.
\end{lemma}
\begin{IEEEproof}
We have
\(
A(Q_n(\hf),Q_n(\hg)) =
A(\hf,\hg)
\)
and
\(
\altPhi(Q_n(\hf),Q_n(\hg);\cdot,\cdot) \ge
\altPhi(\hf,\hg;\cdot,\cdot)
\)
so it suffices to show that 
\(
\intcrossing (Q_n(\hf),Q_n(\hg)) \cap \closure{S(\hf,\hg)}=\emptyset.
\)

Since $\hg$ and $\hf$ are $0$ on $[0,\eta)$ and $1$  on $(1-\eta,1]$ it follows that 
$\intcrossing (\hf,\hg) \subset [\eta,1-\eta]^2$ and that
$\intcrossing (\hf,\hg)$ is closed and by Lemma \ref{lem:Sstructure} it is disjoint from
$\closure{S(\hf,\hg)}.$
Thus, for $\delta$ sufficiently small we have $\neigh{\intcrossing(\hf,\hg)}{\delta}\cap  \closure{S(\hf,\hg)}=\emptyset.$

By Lemma \ref{lem:crosspointlimit}  we now have
\(
\intcrossing (Q_n(\hf),Q_n(\hg))  \cap 
\closure{S(\hf,\hg)}=\emptyset
\)
for all $n$ sufficiently large.
\end{IEEEproof}

We are now ready to prove the main result of this section.

\begin{lemma}\label{lem:weakexistence}
Let $(\hf,\hg)$ satisfy the strictly positive gap condition.
Then there exists $(0,1)$-interpolating $(\ff,\fg) \in \sptfns^2$ and $\ashift$ such that
\[
f\veq \hf \circ\gSa \text{  and  }
g\veq \hg \circ\fS\,.
\]
\end{lemma}
\begin{IEEEproof}
The simplest case is already established in Theorem \ref{thm:PCexist}
and we first generalize to arbitrary $\smthker.$
Assume that $(\hg,\hf)$ are both piecewise constant.
Define $\smthker_k = \smthker \otimes G_k$ where 
$G_i(x) = \frac{k}{\sqrt{2\pi}} e^{- (kx)^2/2}.$
It follows that $\smthker_k \rightarrow \smthker$ in $L_1$
and $\| \smthker_k \|_\infty \le \| \smthker\|_\infty.$
For each $\smthker_k$ we apply Theorem \ref{thm:PCexist}
to obtain piecewise constant $f_k,g_k \in \sptfns$ 
(with corresponding $z^{f_k},z^{g_k}$) and constants $\ashift_k$
such that $ h_{[\fg_k,\ff_k^{{\smthker}_k}]} = \hg$ and $h_{[\ff_k,g_k^{{\smthker}_k,\ashift_k}]} = \hf.$
We can now apply Lemma \ref{lem:limitexist} to conclude
that the theorem holds for piecewise constant $\hf,\hg$ and general $\smthker.$

Let now assume first that for some $\eta>0$ we have $\hf(x)=\hg(x)=0$ for $x\in [0,\eta)$
and $\hf(x)=\hg(x)=0$ for $x\in (1-\eta,1].$
Consider $Q_n(\hf)$ and $Q_n(\hg).$ 
We apply Lemma \ref{lem:PCapprox} and the preceding case already established
to conclude that for all $n$ sufficiently large
there exists (piecewise constant) $(0,1)$-interpolating $f_n,g_n \in \sptfns$ and finite constants $\ashift_n$  such that
$h_{[\fg_n,\fS_n]} = Q_n(\hg)$ and $h_{[\ff_k,\fg_k^{{\smthker}_k,\ashift_n}]}= Q_n(\hf).$
Since $Q_n(\hg)$ and $Q_n(\hf)$ converge to $\hg$ and $\hf$ respectively,
we can apply Lemma \ref{lem:limitexist} to conclude that the theorem holds for this case.

For arbitrary $(\hg,\hf)$ we consider $(\slanta{\hg},\slanta{\hf})$
as in \eqref{eqn:genscale}.

By Lemma \ref{lem:smoothcompress} we can find a sequence $\tvar_i \rightarrow 1$
such $(\hg(;\tvar_i),\hf(;\tvar_i))$ satisfies the strictly positive gap condition for each $i.$
By the preceding case, there exists $f_{\tvar_i},g_{\tvar_i} \in \sptfns$ and finite constants $\ashift_i$ such that
$h_{[\fg_{\tvar_i},\fS_{\tvar_i}]} = \hg(;\tvar_i)$ and $h_{[\ff_{\tvar_i},\fg_{\tvar_i}^{{\smthker},\ashift_i}]} = \hf(;\tvar_i).$
Since $(\hf(;t_i),\hg;(t_i))\rightarrow (\hf,\hg)$ we can apply
Lemma \ref{lem:limitexist} to obtain the desired $(\ff,\fg) \in \sptfns^2$
and $\ashift.$
%
%
\end{IEEEproof}
\subsection{Existence of Spatial Wave Solutions\label{ESWS}}

In the preceeding section we estabished the existence of consistent spatial waves
under general conditions.  In this section we refine the results  to obtain full
spatial wave solutions.  Thus, in this section we complete the proof of 
\ref{thm:mainexist}.

\subsubsection{Analysis of Consistent Spatial Waves}

Let $\smthker$ be regular and assume  $\frac{d}{dx}\fS(x)=0$ at $x=x_1.$
Then $\int_{-W}^{W} \smthker(x_1-x) df(x) = 0$
from which we obtain $\Delta_W f(x_1)=0.$ 

\begin{lemma}\label{lem:phiflat}
Let $(\ff,\fg)$ be $(0,1)$-interpolating and let $\smthker$ be regular.
Assume $f\veq \hf \circ \gSa$ and $g\veq \hg \circ \fS$
with $\ashift \ge 0.$
If $\frac{d}{dx}\gS(x)=0$ or 
$\frac{d}{dx}\fS(x)=0$
at $x=z$
then 
\[
\altPhi(\hfs,\hg;g(z),f(z)) \in [0,A(\hfs,\hg)].
\] 
\end{lemma}
\begin{IEEEproof}
Assume $\frac{d}{dx}\gS(x)=0$ or $\frac{d}{dx}\fS(x)=0$ at $x=z.$
Then, since $\smthker$ is regular, $g(x)=\gS(z)$ if
$\|x-z\| <W.$
Using $L=W$ in Lemma \ref{lem:transitionPhiBounds} we obtain 
$\altPhi(\hfz,\hg;\fg(z),\ff(z)) = 0.$

Now 
\begin{align*}
&   \altPhi(\hfs,\hg;u,v) -\altPhi(\hfz,\hg;u,v) \\
=&
\int_0^v ( \hfsinv(x) -\hfzinv(x))\,dx
\end{align*}
which, since $\ashift\ge 0,$ is non-decreasing in $v$ and independent of $u.$
Hence
\begin{align*}
0
&\le
 \altPhi(\hfs,\hg;g(z),f(z)) - \altPhi(\hfz,\hg;g(z),f(z))  \\
&\le
 \altPhi(\hfs,\hg;1,1) - \altPhi(\hfz,\hg;1,1)  \\
&=
A(\hfs,\hg) - A(\hfz,\hg)
\end{align*}
from which we obtain 
\[
0
\le
 \altPhi(\hfs,\hg;\fg(z),\ff(z))
\le
A(\hfs,\hg)
\]
since $A(\hfz,\hg)  = 0$ by Lemma \ref{lem:FPAzero}.
\end{IEEEproof}



For continuous $\ff\in \sptfns$ let $\flats{\ff}$ denote the set of positive-length maximal intervals on which $\ff$ is effectively constant:
\[
\flats{\ff} = \{ [ \ff^{-1}(v-),\ff^{-1}(v+)]    : \ff^{-1}(v-)<\ff^{-1}(v+)\}\,.
\]
Let $\intflats{\ff}$ denote the subset of such intervals,
for which $v  \in (0,1).$

For $h \in \exitfns$ we use $\jump{h}$ to denote the set of discontinuity points of
$h,$ i.e.,
\[
\jump{h} = \{ u\in[0,1]:h(u-)<h(u+) \}.
\]

\begin{lemma}\label{lem:notequal}
Let $\ff,\fg \in\sptfns$ with $\fg$ continuous.
Assume that $\ff \veq \hf\circ\fg$ for $\hf \in \exitfns$
that is continuous at $0$ and $1.$ 
Then
\[
\ff(x) \neq \hf(\fg(x))\Rightarrow \fg(x) \in \jump{\hf}
\]
and if $\ff \not\equiv \hf\circ\fg$ 
then there exists $I'\in\intflats{\fg}$
such that $\fg(I') \in \jump{\hf}$
and such that $\ff(x)\neq \hf(\fg(I'))$
on a subset of $I'$ of positive Lebesgue measure.
\end{lemma}
\begin{IEEEproof}
By definition,  $\ff \veq \hf\circ\fg$ means $\ff(x) \veq  \hf(\fg(x))$
for all $x$  so we can have $\ff(x) \neq  \hf(\fg(x))$
only if $\fg(x) \in \jump{\hf} \subset (0,1).$
It follows that
\[
\{x:\ff(x) \neq  \hf(\fg(x))\}
\subset \cup_{u \in \jump{\hf}} [\fginv(u-),\fginv(u+)]\,.
\]
If $\mu\{x:\ff(x) \neq  \hf(\fg(x))\} >0$ (where $\mu$ is Lebesgue measure)
then we have
$\mu (\{x:\ff(x) \neq  \hf(\fg(x))\}\cap [\fginv(u-),\fginv(u+)]) >0$
for some $u\in \jump{\hf}.$
Then we take $I' = [\fginv(u-),\fginv(u+)] \in \intflats{\fg}.$
\end{IEEEproof}

\begin{lemma}\label{lem:rightincrease}
Let $(\hf,\hg)$ satisfy the strictly positive gap condition with $A(\hf,\hg)>0$ and let $\smthker$ be regular.
Assume $(\ff,\fg)$ are $(0,1)$-interpolating functions such that
$\ff \veq \hf \circ \gSa$ and 
$\fg \veq \hg \circ \fS.$ 
If   $I \in \intflats{\fS}$ then $\lowerx{I}-W$
is disjoint from $\flats{\gSa}$ and
If   $I' \in \intflats{\gS}$ then $\lowerx{I'}-W$
is disjoint from $\flats{\fS}$.
\end{lemma}
\begin{IEEEproof}
We will show the first case, the second is analogous.
Assume $I \in \intflats{\fS}$ and let $V= \fS(I) \in (0,1).$ 
Assume there exists $I' \in \gSa$ that
with $\lowerx{I} \in I'$
and let $U=\gSa(I').$

Then $g=U$ on $\neigh{I'+\ashift}{W}$ and
$f=V$ on $\neigh{I}{W}.$
The closure of $\neigh{I}{W}$   intersects $I'$ so
we obtain $V \veq \hf(U).$
The closure of $\neigh{I'+\ashift}{W}$ intersects $I$
since $|\ashift| < 2W$ by Corollary \ref{cor:regshiftbound}.
Hence $U\veq \hg(V)$ and we now have
$(U,V) \in \cross (\hf,\hg).$

By Lemma \ref{lem:phiflat} we have $\altPhi(\hf,\hg;U,V) \in [0,A].$
which contradicts the strictly positive gap condition.
\end{IEEEproof}

\begin{lemma}\label{lem:pathology}
Let $\ff,\fg \in \sptfns$ be $(0,1)$-interpolating.
Assume 
\[
f\veq \hf \circ\gSa\text{ and } g\veq\hg\circ \fS\,.
\]
Then we have
\[
\ff \equiv \hf\circ\gSa
\text{ and } \fg \equiv \hg\circ\fS
\]
in any of the following scenarios:
\begin{itemize}
\item[A.] $\hf$ and $\hg$ are continuous.
\item[B.] $\smthker$ is positive on all $\reals.$
\item[C.] $\smthker$ is regular,
$\ashift = 0,$
and $(\hf,\hg)$ satisfies the strictly positive gap condition.
\item[D.] $\smthker$ is regular, 
$(\hf,\hg)$ satisfies the strictly positive gap condition
and
$\jump{\hf} \cap \jump{\hginv}=\emptyset$
and
$\jump{\hg} \cap \jump{\hfinv}=\emptyset.$
\end{itemize}
\end{lemma}
\begin{IEEEproof}
If $\hf$ is continuous then $\jump{\hf}=\emptyset$ and,
by Lemma \ref{lem:notequal},
$\ff \veq\hf \circ \gSa$ then implies
$\ff = \hf \circ \gSa$
Thus, case A is clear.

If $\smthker>0$  then $\gS$ and $\fS$ are stictly increasing on $\reals$
and $\flats{\gS}=\flats{\fS}=\emptyset.$
Lemma \ref{lem:notequal} now implies
$\ff \equiv \hf \circ \gSa$ 
which shows case B.

Assume $\smthker$ is regular
and  $\ashift=0.$ 
If $\gSx(x_1)  = 0$ then
Lemma \ref{lem:phiflat} gives
$\altPhi(\hf,\hg;\ff(x_1),\fg(x_1))=0$
since $A(\hf\hg)=0$ by Lemma \ref{lem:FPAzero}.
By Lemma \ref{lem:monotonic} this violates the strictly positive gap condition if $\gS(x_1) \in (0,1)$
so condition C implies that $\gS$ is strictly increasing on $\{x: 0 < \gS(x) < 1 \}.$
Similarly, it implies that $\fS$ is strictly increasing on $\{x: 0 < \fS(x) < 1 \}.$
Since $\hf$ and $\hg$ are continuous at $0$ and $1$ by Lemma
\ref{lem:zocontinuity}, part C now follows from Lemma \ref{lem:notequal}.

To show part D 
assume $\smthker$ is regular and that
$(\hf,\hg)$ satisfies the strictly positive gap condition.
Assume $\ff \not\equiv \hf \circ\gSa.$
We have $\ff \veq \hf \circ\gSa$ so
we apply Lemma \ref{lem:notequal} to obtain $I'-\ashift \in\intflats{\gSa}$ (so $I' \in \intflats{\gS}$)
such that 
$U \defeq \gSa(I-\ashift) =\gS(I')  \in \jump{\hf}$
and such that $\ff \neq \hf(U)$ on a set of positive measure in
$I'-\ashift.$
We have $g=U$ on $\neigh{I'}{W}$ and
by Lemma \ref{lem:rightincrease}
$\fS$ is not constant on $\neigh{I'}{W}$ and hence we have
$U \in \jump{\hginv}.$

\end{IEEEproof}

%

\subsubsection{Proof of Theorem \ref{thm:mainexist}}

Since we assume that $\smthker$ is regular 
Lemma \ref{lem:pathology} shows that Lemma \ref{lem:weakexistence}
implies Theorem \ref{thm:mainexist} except in the case
$\jump{\hf}\cap\jump{\hginv} \neq \emptyset$ or
$\jump{\hg}\cap\jump{\hfinv} \neq \emptyset.$
It turns out that this case can be handled  by constructing $(\hf^i,\hg^i) \rightarrow (\hf,\hg)$
with certain properties.  The argument is lengthy and is relegated to appendix  \ref{app:B}.



\appendices

\section{ Spatial Fixed Point Integration}\label{app:A}

\subsection{The Continuum Case}

\begin{IEEEproof}[Proof of Lemma \ref{lem:twofint}]
We assume that the smoothing kernel $\smthker$ has finite total variation hence 
$\|\smthker\|_\infty < \infty.$ 
For any $\ff \in \sptfns$ a simple calculation shows that $\fS(x)-\fS(y) \le \|\smthker\|_\infty |x-y|.$
This implies that  and $\fS$ is Lipschitz continuous with Lipschitz constant $\|\smthker\|_\infty.$ 

Given a continuous function $\alpha(x)$ we define 
$\int_{a}^{b} \alpha(x) df(x)$ 
as a Lebesgue-Stieltjes integral.
We adopt the conventional definition of the integral so that 
$\int_{a}^{b} df(x) = f(b+)-f(a+).$  
More generally, for $\alpha$ of finite variation,  the integral
$\int_{a}^{b} \alpha(x) df(x)$
is well defined as long as $\alpha(x)$ and $f(y)$ do not have
any discontinuity points in common.
Using this notation we have for any $f \in \sptfns,$
\begin{align*}
\fS(x) - f(\minfty) 
&  =
\int_{-\infty}^\infty \smthker(x-y)\,
(f (y) - f(\minfty))dy
\\&  =
\int_{-\infty}^\infty \Omega (x-y)\,
df (y)  
\,.
\end{align*}
Applying the above,  the Fubini theorem, and the symmetry 
$\Omega(y-x)+\Omega(x-y) = 1$ we obtain
\begin{align*}
&\int_{-\infty}^{x_2} df(x)(\gS(x)-g(\minfty)) 
+\int_{-\infty}^{x_1} dg(y)(\fS(y)-f(\minfty)) 
\\ &= 
\int_{-\infty}^{x_2} df(x) 
\int_{-\infty}^\infty dg (y) \,\Omega (x-y)\,
\\&\qquad+
\int_{-\infty}^{x_1} dg(y) 
\int_{-\infty}^\infty df (x) \,\Omega (y-x)\,
\\ &= 
\int_{-\infty}^{x_2} df(x) 
\int_{x_1}^\infty dg (y) \,\Omega (x-y)\,
\\ & \quad+ 
\int_{-\infty}^{x_1} dg(y) 
\int_{x_2}^\infty df (x) \,\Omega (y-x)\,
\\ &\quad+ 
\int_{-\infty}^{x_2} df(x) 
\int_{-\infty}^{x_1} dg (y) 
\\ &=  
(f(x_2+)-f(\minfty))(g(x_1+)-g(\minfty))
\\&\quad+ 
\iint dg(y) df(x) \bigl(\ind_{T_1}\Omega (x-y)
+ \ind_{T_2}\Omega (y-x)\bigr)\,,
\end{align*}
where $T_1 = \{ (x,y): x\le x_2, y>x_1 \}$ and
$T_2 = \{ (x,y): x > x_2, y\le x_1\}.$  Note that in $T_1$ we have
$x-y < x_2-x_1$ and in $T_2$ we have
$y-x < x_1-x_2.$  Since $\Omega(z) = 0$ for $z<-W$
we see that the expression is local up to $W$:
The integrand has positive support only in a square of
sidelength  $|x_2-x_1|+W.$

Note that the above can also be written as
\begin{align*}
&\int_{-\infty}^{x_2} df(x)\gS(x)
+\int_{-\infty}^{x_1} dg(y)\fS(y)
\\ &  
- (f(x_2+)g(x_1+)-f(\minfty)g(\minfty))
\\=& 
\iint dg(y) df(x) \bigl(\ind_{T_1}\Omega (x-y)
+ \ind_{ T_2}\Omega (y-x)\bigr)\,.
\end{align*}


In general, for any constant $c$ we have
\begin{align*}
\int_{-\infty}^{x_2} df(x)(\gS(x)-c) 
&=
\int_{f(\minfty)}^{f(x_2+)}  (h^{-1}_{[\ff,\gS]}(v) - c)\,dv
\\
\int_{-\infty}^{x_1} dg(y)(\fS(y)-c) 
&=
\int_{g(\minfty)}^{g(x_1+)}  (h^{-1}_{[\fg,\fS]}(u) -c)\,du \; 
\end{align*}
and the lemma now follows.
%
\end{IEEEproof}

\subsection{Discrete Spatial Fixed Point Integration}\label{app:Aa}

In this section we show how spatial integration can be done in the 
spatially discrete setting.
Working directly in the spatially discrete setting has the advantage that it avoids
all measure theoretic issues.  Moreover, it highlights the central algebraic
character of the spatial integration result.
Let us introduce the notation
\[
\partial\dv{g}_i \defeq \dv{g}_i-\dv{g}_{i-1}\,
\text{ and }
\Avg\dv{g}_i \defeq \frac{1}{2}(\dv{g}_i+\dv{g}_{i-1})\,.
\]

The spatially smoothed $\dv{g}$ will be denoted $\dv{\gSdisc}$ and is defined by
\[
\dv{\gSdisc}_i = \sum_{j=-W}^W \discsmthker_j \dv{g}_{i-j}
\]
where we assume evenness $\discsmthker_j = \discsmthker_{-j},$ non-negativity $\discsmthker_j \ge 0,$ and $\sum_j \discsmthker_j =1.$

Define 
\[
 {\cal W}_k \defeq \sum_{j=-\infty}^{k-1} w_j +\frac{1}{2} w_k
=\sum_{j=-\infty}^{k} \Avg w_j \,.
\]
Note that  ${\cal W}_k = 1-  {\cal W}_{-k},$ which motivates the definition.

Let $f_i,g_i$ be bounded non-decreasing sequences. Then
\begin{align*}
& 
(\Avg \dv{g}^\discsmthker_{i} - \dv{g}_{\minfty})
\\
&=
\sum_{j-\minfty}^{\infty} (\Avg\discsmthker_{i-j} )\,(\dv{g}_{j} - \dv{g}_{\minfty})
\\
&=
 \sum_{j-\minfty}^{\infty} {\cal W}_{i-j}\,\partial\dv{g}_{j}
\end{align*}
Hence
\begin{align*}
\sum_{i=-\infty}^{k} \partial\dv{f}_i  
(\Avg \dv{g}^\discsmthker_{i} - \dv{g}_{\minfty})
=
 \sum_{i=-\infty}^{k} \sum_{j-\minfty}^{\infty}\partial\dv{f}_i  \partial\dv{g}_{j}  {\cal W}_{i-j}\,.
\end{align*}
and, similarly,
\begin{align*}
 \sum_{i=-\infty}^{k} \partial\dv{g}_i ( \Avg\dv{f}^\discsmthker_{i} -\dv{f}_{\minfty})
=
 \sum_{i=-\infty}^{k} \sum_{j-\minfty}^{\infty}\partial\dv{f}_i  \partial\dv{g}_{j}  {\cal W}_{j-i}\,.
\end{align*}

Summing, using ${\cal W}_{j-i}+{\cal W}_{i-j}=1,$ we obtain
\begin{align*}
& \sum_{i=-\infty}^{i_2} \partial\dv{f}_i  (\Avg\dv{g}^\discsmthker_{i}-g_{\minfty}) +
\sum_{i=-\infty}^{i_1} \partial\dv{g}_i  (\Avg\dv{f}^\discsmthker_{i} -f_{\minfty})
\\& =
(f_{i_2}-f_{\minfty})(g_{i_1}-g_{\minfty})  +  \\&
 \sum_{i=-\infty}^{i_2} \sum_{j=i_1+1}^\infty \partial\dv{f}_i  \partial\dv{g}_{j}  {\cal W}_{i-j}
+ 
\sum_{i=i_2+1}^\infty \sum_{j=-\infty}^{i_1}  \partial\dv{f}_i  \partial\dv{g}_{j}  {\cal W}_{j-i}\, .
\end{align*}

\section{Fixed Point Bounds on Potentials\label{sec:FPbounds}}

\begin{IEEEproof}[Proof of Lemma \ref{lem:transitionPhiBounds}]
As a first bound use the evenness of $\smthker$ to write
for any $f \in \sptfns,$
\begin{align*}
\fS(x) &= \int_{-\infty}^{\infty} \smthker(y)f(x+y)\text{d}y
\\
&= \int_{-\infty}^{L} \smthker(y)f(x+y)\text{d}y
+ \int_{L}^{\infty} \smthker(y)f(x+y)\text{d}y
\\
&\le (1-\Omega(-L)) f((x+L)-) + \Omega(-L)
\\
&\le f((x+L)-) + \Omega(-L) (1-f((x+L)-))
\\
&\le  f(x) + \Delta_L f(x) + \Omega(-L) (1- f(x))
\end{align*}
and
\begin{align*}
\fS(x) &= \int_{-\infty}^{\infty} \smthker(y)f(x+y)\text{d}y
\\
&\ge
\int_{-L}^{\infty} \smthker(y)f(x+y)\text{d}y
\\
&\ge (1-\Omega(-L)) f((x-L)+)
\\
&\ge f(x) - \Delta_L f(x) - \Omega(-L) f(x)\,.
\end{align*}

To obtain a bound on
$\altPhiSI(\smthker;f,g;x_2,x_1)$ where
$f,g \in \sptfns$ we proceed similarly.
First recall that
$T_1 = \{ (x,y): x\le x_2, y>x_1 \}$ and
$T_2 = \{ (x,y): x > x_2, y\le x_1\}$
and note that since $\Omega \le 1$ we have
\begin{align*}
& 
\iint dg(y) df(x) \ind_{T_1}\indicator{x-y>-L}\Omega (x-y)
\\ \le &
\indicator{x_2-x_1 > -L} ( f(x_2+)-f((x_1-L)+))\cdot \\
&\quad \cdot ( g((x_2+L)-)-g(x_1+))
\end{align*}
and
\begin{align*}
& 
\iint dg(y) df(x) \ind_{T_2}\indicator{y-x>-L}\Omega (y-x)
\\ \le &
\indicator{x_1-x_2 > -L} (f((x_1+L)-)-f(x_2+))\cdot \\
&\quad \cdot ( g(x_1+)-g((x_2-L)+))\,.
\end{align*}
If $|x_1-x_2| \le L$ then the sum of the above two expressions is upper bounded
by $\Delta_L f(x_1) \Delta_L g(x_2).$

%
%
Since $df dg$ has total measure at most $1$ we have
\begin{align*}
& \iint dg(y) df(x) \ind_{T_1} \indicator{x-y\le-L}\Omega (x-y)
\\&+ \iint dg(y) df(x) \ind_{T_1} \indicator{x-y\ge L}\Omega (y-x)
\\ \le &\Omega (-L)\,.
\end{align*}
Hence, if $|x_1-x_2|\le L$
then we have
\begin{align*}
&\altPhiSI(\smthker;f,g;x_2,x_1)
\le\\&
 \Delta_{L} f(x_2)\Delta_{L} g(x_1)+\Omega(-L)
\end{align*}
and since, by \eqref{eqn:altPhiderivatives},
\begin{align*}
|\altPhi(&\hf,\hg;\fg(x),\fS(x))
-\altPhi(\hf,\hg;\fg(x),f(x)) |
\\&\le
|\fS(x)-f(x)|
\\
&\le
\Delta_L f(x) + \Omega(-L)
\end{align*}
the other bounds follow easily from Lemma \ref{lem:twofint}.
\end{IEEEproof}

\section{Discrete-Continuum Relation\label{app:discretecont}}

In this section we prove Theorem  \ref{thm:discreteFPDelta} and Theorem \ref{thm:discreteFPsum}.
We associate to the discrete spatial index $i$ the real valued point $x_i = i\Delta.$
We assume that $\smthker$ is the piecewise constant extension of
$\discsmthker$  so \eqref{eqn:kerdiscretetosmth} holds trivially.

Assume a spatially discrete fixed point $\ff,\fg.$
Let $\tff$ and $\tfg$ be the piecewise constant extensions of $\ff$ and $\fg.$
We can now relate the discrete spatial EXIT sum to the corresponding continuum integral
to arrive at approximate fixed point conditions for spatially discrete fixed points.

The discrete sum
\begin{align*}
&\frac{1}{2}\sum_{i=-\infty}^i
(\dv{f}_i+\dv{f}_{i-1}) (\dv{g}_i^\discsmthker-\dv{g}_{i-1}^\discsmthker)
\\& =
\int_{-\infty}^{x_i}
\dv{\tff}(x) \text{d} \dv{\tgS}(x)
\\ & =
\int_{-\infty}^{x_i}
h_{[\tff,\tgS]}
(\dv{\tgS}(x)) \text{d} \dv{\tgS}(x)
\end{align*}
and, similarly,
\begin{align*}
&\frac{1}{2}\sum_{i=-\infty}^i
(\dv{g}_i+\dv{g}_{i-1}) (\dv{f}_i^\discsmthker-\dv{f}_{i-1}^\discsmthker)
\\& =
\int_{-\infty}^{x_i}
\dv{\tfg}(x) \text{d} \dv{\tfS}(x)
\\
& =
\int_{-\infty}^{x_i}
h_{[\tfg,\tfS]}
(\dv{\tfS}(x)) \text{d} \dv{\tfS}(x)\,.
\end{align*}
We want to compare
\(
\int_{x_{i-1}}^{x_i}
h_{[\tff,\tgS]}
(\dv{\tgS}(x)) \text{d} \dv{\tgS}(x)
\)
to
\(
\int_{x_{i-1}}^{x_i}
\hf
(\dv{\tgS}(x)) \text{d} \dv{\tgS}(x)\,.
\)
Now, $\dv{\tgS}(x)$ linearly interpolates between
$\dv{g}_{i-1}^\discsmthker$ and
$\dv{g}_i^\discsmthker$ on $[x_{i-1},x_{i}]$ and
$h_{[\tff,\tgS]}(\dv{\tgS}(x)) =\dv{f}_{i-1}^\discsmthker$
on the first half of the interval and
$h_{[\tff,\tgS]}(\dv{\tgS}(x)) =\dv{f}_{i}^\discsmthker$
on the second half. 
From this we have
\begin{align*}
&\Bigl|\int_{x_{i-1}}^{x_i}
\bigl(h_{[\tff,\tgS]}
(\dv{\tgS}(x)) 
-
\hf(\dv{\tgS}(x))\bigr)
\text{d} \dv{\tgS}(x)\Bigr|
\\& \le \half
(\hf(\dv{g}_i^\discsmthker)- \hf(\dv{g}_{i-1}^\discsmthker))
(\dv{g}_i^\discsmthker-\dv{g}_{i-1}^\discsmthker)
\\&\le
\half (\hf(\dv{g}_i^\discsmthker)- \hf(\dv{g}_{i-1}^\discsmthker))
\Delta\|\omega\|_\infty\,.
\end{align*}
Summing over $i$ we obtain
\begin{align*}
&\Bigl|\int_{0}^{1}
\bigl(h_{[\tff,\tgS]}
(u) 
-
\hf(u)
\text{d}u\Bigr|
\le
\half\Delta\|\omega\|_\infty\,
\end{align*}
and, similarly,
\begin{align*}
&\Bigl|\int_{0}^{1}
\bigl(h_{[\tfg,\tfS]}
(v) 
-
\hg(v)
\text{d}v\Bigr|
\le
\half\Delta\|\omega\|_\infty\,.
\end{align*}
These inequalities prove Theorem \ref{thm:discreteFPDelta}.

To obtain a more refined bound we write
\begin{align*}
&\int_{x_{i-1}}^{x_i}
h_{[\tff,\tgS]}
(\dv{\tgS}(x)) \text{d} \dv{\tgS}(x)
\\& =
\frac{1}{2}(
\hf(\dv{g}_i^\discsmthker)+
\hf(\dv{g}^\discsmthker_{i-1}))(\dv{g}_i^\discsmthker-\dv{g}_{i-1}^\discsmthker)
\\ & = 
\int_0^1 (\alpha
\hf(\dv{g}_i^\discsmthker)+
\bar{\alpha} 
\hf(\dv{g}^\discsmthker_{i-1})) \text{d}\alpha \,(\dv{g}_i^\discsmthker-\dv{g}_{i-1}^\discsmthker)
\end{align*}
and, since $\dv{\tgS}(x)$ linearly interpolates between
$\dv{g}_{i-1}^\discsmthker$ and
$\dv{g}_i^\discsmthker$ on $[x_{i-1},x_{i}]$ we have
\begin{align*}
&\int_{x_{i-1}}^{x_i}
\hf
(\dv{\tgS}(x)) \text{d} \dv{\tgS}(x)
\\& =
\int_0^1 
\hf(\alpha \dv{g}_i^\discsmthker +
\bar{\alpha} \dv{g}^\discsmthker_{i-1}) \text{d}\alpha \,(\dv{g}_i^\discsmthker-\dv{g}_{i-1}^\discsmthker)
\end{align*}
where $\bar{\alpha}$ denotes $1-\alpha.$
Assuming $\hf$ is $C^2$ we have by a simple application of the remainder theorem
\begin{align*}
& |\alpha \hf(\dv{g}_i^\discsmthker) + \bar{\alpha} \hf(\dv{g}_{i-1}^\discsmthker)
-
\hf (\alpha  \dv{g}_i^\discsmthker + \bar{\alpha} \dv{g}_{i-1}^\discsmthker)|
\\\le &
\frac{C_i}{2}(\dv{g}_i^\discsmthker-\dv{g}_{i-1}^\discsmthker)^2
\end{align*}
where $C_i$ is the maximum of $|\hf''(u)|$ for $u$ in $[\dv{g}_{i-1}^\discsmthker,\dv{g}_{i}^\discsmthker].$

We now have
\begin{align*}
&\Bigl|\int_{x_{i-1}}^{x_i}
\bigl(h_{[\tff,\tgS]}
(\dv{\tgS}(x)) 
-
\hf(\dv{\tgS}(x))\bigr)
\text{d} \dv{\tgS}(x)\Bigr|
\\& \le
\frac{C_i}{2}(\dv{g}_i^\discsmthker-\dv{g}_{i-1}^\discsmthker)^3\,.
\end{align*}
Since $\sum_i (\dv{g}_i^\discsmthker-\dv{g}_{i-1}^\discsmthker) \le 1$
and
\[
\dv{g}_i^\discsmthker-\dv{g}_{i-1}^\discsmthker
\le \Delta \|\smthker\|_\infty
\]
we obtain by summing and changing variables
\begin{align*}
&\Bigl|\int_{0}^{1}
\bigl(h_{[\tff,\tgS]}
(u) 
-
\hf(u)\bigr)
\text{d} u\Bigr|
\\& \le
\frac{\|\hf''\|_\infty}{2} ( \|\smthker\|_\infty)^2 \Delta^2\,.
\end{align*}

A similar argument applies to $\hg$ and $h_{[\tfg,\tfS]},$
and Theorem \ref{thm:discreteFPsum} follows.

\section{Existence of Travelling Wave Solution: Final Case}\label{app:B}

In this section we prove Theorem \ref{thm:mainexist}  for the case where
$\jump{\hf}\cap \jump{\hginv} \neq \emptyset$ or
$\jump{\hg}\cap \jump{\hfinv} \neq \emptyset$
and $\ashift \neq 0.$
Without loss of generality we assume $\ashift>0.$
The main part of the proof is the construction of an approximating
sequence with special regularity properites.
The construction is encapsulated in the following lemma.

\begin{lemma}\label{lem:regularapprox}
Given $(\hf,\hg)$ that satisfy the strictly positive gap condition there exists
$(\hf^i,\hg^i) \rightarrow (\hf,\hg)$ such that 
$(\hf^i,\hg^i)$ satisfies the strictly positive gap condition for each $i$
and all of the following properties hold
\begin{itemize}
\item[A.] 
\(
\jump{\hfiinv} \subset \jump{\hfinv}\,,
\)
\(
\jump{\hgiinv} \subset \jump{\hginv}\,.
\)
\item[B.] 
\(
\jump{\hf^i} \cap \jump{\hginv} =\emptyset\,,
\)
\(
\jump{\hg^i} \cap \jump{\hfinv} =\emptyset\,.
\)
\item[C.]
If $u \in   \jump{\hginv}\cap\jump{\hf}$ then
$\hf^i(u)=\hf(u)$ for all $i$ large enough.

If $v \in   \jump{\hfinv}\cap\jump{\hg}$ then
$\hg^i(v)=\hg(v)$ for all $i$ large enough.
\item[D.]
If $v \in \jump{\hfinv}$ then 
\[
[\hfiinv(v-),\hfiinv(v+)] \rightarrow [\hfinv(v-),\hfinv(v+)].
\]
If $u \in \jump{\hginv}$ then 
\[
[\hgiinv(u-),\hgiinv(u+)] \rightarrow [\hginv(v-),\hginv(u+)].
\]
\item[E.]
If $v  \in \jump{\hfinv}$ then, setting $u=\hfinv(v+),$ we have
$\hf^i(u)=\hf(u)$ for all $i$ large enough. 

If $u  \in \jump{\hginv}$ then, setting $v=\hginv(u+),$ we have
$\hg^i(v)=\hg(v)$ for all $i$ large enough. 
\end{itemize}
\end{lemma}
\begin{IEEEproof}[Proof of Lemma \ref{lem:regularapprox}]
Note that if
\(
\jump{\hf} \cap \jump{\hginv} =\emptyset\,
\)
and
\(
\jump{\hg} \cap \jump{\hfinv} =\emptyset\,
\)
then we can simply set $(\hf^i,\hg^i) =(\hf,\hg).$
Thus, the lemma targets the case where this does not hold.

We will describe the construction of $\hf^i,$ the construction of $\hg^i$ is analogous.

Consider the countable\footnote{The case where $\jump{\hf} \cap \jump{\hginv}$
is finite can be handled similarly to the countably infinite case. To avoid notational overhead
we present the argument only for the infinite case.}  set 
\[
\jump{\hf} \cap \jump{\hginv} =\{ u_1,u_2,\ldots \}
\]
and also the set
\[
\{  \hfinv(v-),\hfinv(v+) : v \in \jump{\hfinv} \}
\backslash (\jump{\hf} \cap \jump{\hginv})
= 
\{ t_1,t_2,\ldots \}\,.
\]
For each $k=1,2,\ldots$ set
\[
d_k = \min \{ u_k,1-u_k;|u_k-u_j|,|u_k-t_j| : j < k \}
\]
and note that $d_k > 0$ (we cannot have $u_k=0$ or $u_k=1$ since $\hf$ is continuous at $0$ and $1$ by Lemma \ref{lem:zocontinuity}. )
For each $i=1,2,\ldots$ we define sequences $\eta_{i,k},$
$k=1,2,\ldots$ such that 
\[
0 < \eta_{i,k} < \frac{1}{2}\min \{ 3^{-ik},d_k \}
\] 
and such that 
\[
\{ u_k \pm \eta_{i,k} \} \cap \jump{ \hginv } = \emptyset\,.
\]
Note that $2\sum_k \eta_{i,k} \le \frac{1}{2^i}.$

For  each $k$ we define $H_k = \unitstep_{r_k}$ (which is a unit step function
except that we set $H_k(0)  =r_k$)
where  $r_k= \frac{\hf(u_k)-\hf(u_k-)}{\hf(u_k+)-\hf(u_k-)}.$
This function represents the jump in $\hf$ at $u_k.$
We will substitute for this a function continuous at $0$:
\[
S_{i,k}(x) = \begin{cases}
0 & x < 1- \eta_{i,k}\\
0   \vee  (x+r_k) \wedge 1 & |x| \le \eta_{i,k} \\
1 & x > 1+ \eta_{i,k}
\end{cases}
\]
where $0   \vee  z \wedge 1 = \min\{ \max \{ 0,z\} ,1\}.$
Define
\begin{align*}
&\hf^i(x) = \hf(x) \\&- \sum_k ( \hf(u_k+)-\hf(u_k-)) 
(H_k(x-u_k) - S_{i,k}(x-u_k))\,.
\end{align*}
Note that  $\sum_k ( \hf(u_k+)-\hf(u_k-)) \le 1$ and $|H_k(x) - S_{i,k}(x)|\le 1$
so the sum is well defined.
The function $\hf^i(x)$ can be expressed as the sum of two functions,
\[
h_1(x)= \hf(x) - \sum_k ( \hf(u_k+)-\hf(u_k-)) H_k(x-u_k) \,
\]
and
\[
h_{2,i}(x)= \sum_k ( \hf(u_k+)-\hf(u_k-)) S_{i,k}(x-u_k)\,,
\]
both of which are in $\exitfns,$ i.e., both of which are non-decreasing.
The function $h_1$ is continuous for all $u \in \jump{\hginv} \cap \jump{\hf}$
since $H_k(0+)-H_k(0-)=1$  and $H_k(u+)-H_k(u-)=0$ for $u \neq 0.$
If $u \in \jump{\hginv} \backslash \jump{\hf}$ then $\hf$ is continuous at $u$ and
therefore $h_1$ is continuous at $u.$
If follows that $\hf^i \in \exitfns$ and $\hf^i \xrightarrow{i\rightarrow\infty} \hf.$
We assume a similar definition of $\hg^i.$

We will now show that properties A through E hold for this sequence.
Each property has two essentially equivalent forms (through the symmetry of substitution of $f$ and $g$). In each case we will show the first form.

Consider part A. Let $v \in \jump{\hfiinv}.$  Since $\gSa$ is continuous, there is non-empty interval
$I =(u',u'')$ such that $\hf^i$ is evaluates to $v$ on $I.$
Since both $h_1$ and $h_{2,i}$ are non-decreasing it follows that both are constant
on $I.$  From the fact that $h_{2,i}$ is constant on $I$ we easily obtain that
$\sum_k ( \hf(u_k+)-\hf(u_k-)) H_{k}(u-u_k)\,$ is also constant on $I$ and 
we deduce that $\hf$ is constant on $I.$
Hence $v \in \jump{\hfinv}$
and part A is proved.

Consider part B. The function  $S_{i,k}(u-u_k)$ is continuous at $u$
unless $u = u_k \pm \eta_{i,k}$ and, 
by construction, $u_k \pm \eta_{i,k} \not\in \jump{\hginv}.$
Hence, $h_{2,i}(u)$ is continuous at all $u \in \jump{\hginv}.$
Since $h_1$ is continuous on $\jump{\hginv}$ and $\hf^i$ is continuous at all $u\in \jump{\hginv},$ part B is proved.

Consider part C. Let $u \in\jump{\hginv}\cap \jump{\hf}\,,$ i.e., $u=u_j$ for some $j.$
We prove part C  by showing that
$H_k(u_j-u_k) - S_{i,k}(u_j-u_k) =0$ for all $k$ for all $i$ large enough.
For $k=j$ we note $H_k(0) - S_{i,k}(0) =0$ by construction.
For $k<j$ we have $H_k(u_j-u_k) - S_{i,k}(u_j-u_k) =0$ for all $i$ such that
\(
\frac{2}{3^i}<\min_{k < j} \{ |u_k-u_j| \}.
\)
For $j<k$ we have $H_k(u_j-u_k) - S_{i,k}(u_j-u_k) =0$ by construction, i.e.,
by the requirement that $\eta_{i,k}  < d_k$ which implies $|u_j -  u_k| > \eta_{i,k}.$
This proves part C.

Consider part D.
Assume $v \in \jump{\hfinv},$ then  $\hfinv(v-)<\hfinv(v+).$
Since $(\hfinv(v-),\hfinv(v+)) \cap \jump{\hf} = \emptyset$ and $\eta_{i,k} < 2^{-i}$ it follows
that for $u \in (\hfinv(v-)+2^{-i},\hfinv(v+)-2^{-i})$ we have
$\hf^i(u) = \hf(u) = v$ and for 
$u \not\in (\hfinv(v-)-2^{-i},\hfinv(v+)+2^{-i})$
we have
$\hf^i(u) \neq v.$
Part D now follows.

Consider part E.
Let $v  \in \jump{\hfinv}$ and set $u=\hfinv(v+).$
If $u \in \jump{\hginv}\cap \jump{\hf}$ then $u=u_k$ for some $k$ and
property C implies property E.
Otherwise, we have $u = t_k$ for some $k.$
For $j\ge k$ we have $  |u_j-t_k| > \eta_{i,k}$ for all $i.$
For $j <  k$ we have  $\frac{2}{3^i} < \min_{j<k} \{|u_j-t_k|\}$
for all $i$ sufficiently large.
Hence property E holds.

Now we address the satisfaction of the strictly positive gap condition.
In general, we may need to further modify the constructed sequence and take a subsequence.
Define 
\begin{align*}
 \hf^\delta(u) = \unitstep_0(u-\delta)\wedge\hf(u) \vee \unitstep_1(u - (1-\delta)) \\
 \hg^\delta(v) =\unitstep_0(v-\delta)\wedge \hg(v) \vee \unitstep_1(v - (1-\delta))\,.
\end{align*}
We claim that for all $\delta$ sufficiently small
\(
(\hf^\delta,\hg^\delta)
\)
satisfies the strictly positive gap condition with $A>0$.
Since $\hf,\hg$ satisfies the strictly positive gap condition and
$A(\hf,\hg)>0,$ Lemma \ref{lem:Sstructure} there exists a minimal element 
$(u^*,v^*) in \intcross(\hf,\hg).$
By Lemma \ref{lem:zocontinuity} we have $(0,0) < (u^*,v^*) < (1,1)$
and
if $\delta$ is small enough then $\hf^\delta,\hg^\delta$ has $(u^*,v^*)$ as a non-trivial crossing point. Clearly we get
$A(\hf^\delta,\hg^\delta) >0$ for $\delta$ small enough.
For $\delta$ small enough we have
$(\delta,\hf(\delta)),(\hg(\delta),\delta) \in S(\hf,\hg)$ and
no new crossing point is introduced for $u \le \delta$ or $v\le\delta.$

Comparing potentials, we have
\begin{align*}
&\altPhi(\hf,\hg;u,v)-
\altPhi(\hf^\delta,\hg^\delta;u,v)\\
=&
\int_0^u (\hginv(x)-{(\hg^\delta)}^{-1}(x)) dx
+
\int_0^v (\hfinv(x)-{(\hf^\delta)}^{-1}(x)) dx
\end{align*}
so for $u\le \hginv(1-\delta)$ and $v\le \hfinv(1-\delta)$ we have
$\altPhi(\hf^\delta,\hg^\delta;u,v)>\altPhi(\hf,\hg;u,v).$
Similarly, for $u\ge\hginv(\delta)$ and $v\ge \hfinv(\delta)$ we have
\begin{align*}
&\altPhi(\hf^\delta,\hg^\delta;u,v) - A(\hf^\delta,\hg^\delta)\\
\ge&
\altPhi(\hf,\hg;u,v) - A(\hf,\hg)\\
>&0
\end{align*}
which establishes the claim. 

Let us define $\delta_j \rightarrow 0$ with 
$1-\delta_j, \delta_j \not\in \jump{\hginv} \cup \jump{\hfinv}$ so that, for each $j,$
\begin{align*}
\hf^j (u) =  \unitstep_0(u-\delta_j) \wedge \hf(u) \vee \unitstep_1(u - (1-\delta_j)) \\
\hg^j (v)= \unitstep_0(v-\delta_j) \wedge \hg(v) \vee \unitstep_1(v - (1-\delta_j)) 
\end{align*}
satisfies the strictly positive gap condition with $A >0.$  Now, for each $i$ we define the sequence
\begin{align*}
\hf^{i,j}(u) &=
\unitstep_0(u-\delta_j) \wedge \hf^i(u) \vee \unitstep_1(u - (1-\delta_j)) \\
\hg^{i,j}(v) &=
\unitstep_0(v-\delta_j) \wedge \hg^i(v) \vee \unitstep_1(v - (1-\delta_j)) \,.
\end{align*}
Then we have
\begin{align*}
\hf^{i,j} \xrightarrow{i \rightarrow \infty} \hf^j , \quad
\hg^{i,j} \xrightarrow{i \rightarrow \infty} \hg^j \,.
\end{align*}
Properties A and B still hold for all $i$ and $j.$

Clearly, for each $j,$ $(\hf^{i,j},\hg^{i,j})$ satisfies the strictly positive gap condition
with $A(\hf^{i,j},\hg^{i,j}) >0$ for all $i$ large enough.
Hence, for each $j$ we can find $i(j)$  such that $(\hf^{i,j},\hg^{i,j})$
satisfies the strictly positive gap condition for all $i \ge i(j).$
We can assume $i(j)$ is increasing in $j.$
Consider the diagonal sequence 
$(\hf^{i(j),j},\hg^{i(j),j})\, j=1,2,\ldots.$
Let us re-index this as 
$(\hf^{i},\hg^{i})\, i=1,2,\ldots$ with corresponding $\delta_i.$
We now show that properties C,D, and E continue to hold.

Property C holds since, by Lemma \ref{lem:zocontinuity}, $u \in \jump{\hf}$ implies $u \in (0,1)$
and $v \in \jump{\hg}$ implies $v \in (0,1).$
Now we show property D.
Assume $v \in \jump{\hfinv}.$ If $v \in (0,1)$ then
$ [\hfinv(v-),\hfinv(v+)] \subset (0,1)$ by Lemma \ref{lem:zocontinuity} and property D clearly holds.
If $v=0$ then $\hfinv(v-)=\hfiinv(v-)=0$ and 
$\hfinv(v+)  < 1$ by Lemma \ref{lem:zocontinuity}.
Since  $1-\delta_i \rightarrow 1$ we  have $\hfiinv(v+) \rightarrow \hfinv(v+).$ 
Similarly, if $v=1$ then $\hfinv(v+)=\hfiinv(v+)=1$ and 
$\hfinv(v-)  >0$ and we  have $\hfiinv(v-) \rightarrow \hfinv(v-).$ 
Thus, property D holds generally.

Finally we consider property E.
Let $v  \in \jump{\hfinv}$ and set $u=\hfinv(v+).$
Then $u>0$  and if $u<1$ then we clearly have
$\hf^i(u)=\hf(u)$ for all $i$ large enough.
If $u=1$ then $\hf(u)=1$
and $\hf^i(u)=1$ for all $i.$
Thus, property E holds.
\end{IEEEproof}

%

Given an interval $I$ let 
$\upperx{I}$ denote its right end point and
let
$\lowerx{I}$ denote its left end point.
For two closed intervals $I^1,I^2$ we say $I^1 \le I^2$ 
if $\upperx{I} \le \lowerx{I^2}.$
The interval $I^1+x$ denotes the interval $I^1$ translated by $x.$ 
For a non-empty interval $I$ and $\epsilon>0$ by $\neigh{I}{-\epsilon}$ we mean
$(\lowerx{I}+\epsilon,\upperx{I}-\epsilon).$

\begin{lemma}\label{lem:flatseparate}
Let $I^1,I^2 \in \flats{\fS}$ be distinct where $\ff\in \sptfns$ 
and $\smthker$ is regular.
Then $I^1 \le I^2$  implies $I^1+2W \le I^2.$
\end{lemma}
\begin{IEEEproof}
Since $I^1$ and $I^2$ are both maximal we have $\fS(I^1) < \fS(I^2).$
Since $\smthker$ is regular, we have $\ff(x)=\fS(I^1)$ for $x\in\neigh{I^1}{W}$ 
and $\ff(x)=\fS(I^2)$ for $x\in\neigh{I^2}{W}.$
It follows that $\neigh{I^1}{W}$ and $\neigh{I^2}{W}$ are disjoint.\end{IEEEproof}

Given regular $\smthker$ and shift $\ashift >0$ 
we say $I \in \intflats{\fS}$ is {\em linked to}
$I' \in \flats{\gS}$ if  $\lowerx{I'}<\upperx{I}+W +\ashift \le\upperx{I'}$
and
we say $I' \in \intflats{\gS}$ is {\em linked to}
$I'' \in \flats{\fS}$ if  $\lowerx{I''}<\upperx{I'}+W \le \upperx{I''}.$
If we have a sequence $I^1,I^2,\ldots$
such that $I^j$ is linked to $I^{j+1}$ then we call this a {\em chain}.
Note that by construction all intervals in a chain in either $\flats{\fS}$
or $\flats{\gS}$ must be distinct.
The chain {\em terminates} if the last element in the chain is not linked to 
another interval.


\begin{lemma}\label{lem:linkterminate}
Let $(\hf,\hg) \in \exitfns^2$ satisfy the strictly positive gap condition 
with $A(\hf,\hg)>0.$
Let $(f,g)\in\sptfns^2$ be $(0,1)$-interpolating and let $\smthker$ be regular.
Assume $\ff \veq \hf \circ \gSa$ and 
$\fg \veq \hg \circ \fS$
(hence $\ashift>0$),
then any chain in $\flats{\fS},\flats{\gS}$ terminates.
\end{lemma}
\begin{IEEEproof}
Let $(u^*,v^*)$ be the minimal element in $\intcross(\hf,\hg)$
as guaranteed by Lemma \ref{lem:Sstructure}.
There exists  finite $y$ such that $g(y) \ge u^*$ and 
$f(y) \ge v^*.$
By Lemma \ref{lem:phiflat} if $z \in I \in \{\intflats{\gS} \cup \intflats{\fS}\}$
then $\altPhi(\hf,\hg;g(z),f(z)) \in [0,A(\hf,\hg)]$ and we therefore have
$(g(z),f(z)) < (u^*,v^*)$ componentwise by Lemma \ref{lem:Sstructure}.
Thus, we obtain $z<y.$
It now follows from Lemma \ref{lem:flatseparate} that any chain of linked intervals terminates.
\end{IEEEproof}

\begin{lemma}\label{lem:convprop}
Let $(\hf,\hg) \in \exitfns^2$ satisfy the strictly positive gap condition 
with $A(\hf,\hg) > 0$ and let $\smthker$ be regular.
Let $(\hf^i,\hg^i) \rightarrow (\hf,\hg)$ be given as in Lemma \ref{lem:regularapprox}.
Assume there exist $(0,1)$ interpolating  $f,g$ 
such that
$\ff \veq \hf \circ \gSa$ and 
$\fg \veq \hg \circ \fS$
and $(0,1)$-interpolating sequences
$\ff_i \rightarrow \ff$ and $\fg_i\rightarrow \fg$ and $\ashift_i \rightarrow \ashift$ 
where
$\ff_i = \hf^i \circ \gSai_i$ and 
$\fg_i = \hg^i \circ \fS_i$ for each $i.$

Then $\ashift>0$ and 
the following properties hold for any $I\in\intflats{\fS}.$
\begin{itemize}
\item[A.]
If $I$ is not linked to an $I'\in\flats{\gS}$
then  for any $\epsilon >0$ we have 
$\fS_i(x) =\fS(I)$ for $x\in \neigh{I}{-\epsilon}$
for all $i$ large enough.
\item[B.]
If $I$ is linked to $I' \in \flats{\gS}$ and
for any $\delta>0$ we have
$\gS_i$ is a fixed constant, denoted $U,$ on $\neigh{I'}{-\delta}$ for all $i$ large enough,
then,
for any $\epsilon >0$ we have 
$\fS_i(x) =\fS(I)$ for $x\in \neigh{I}{-\epsilon}$
for all $i$ large enough.
\item[C.]
Assume that for any $\epsilon>0$ we have
$\fSi_i(x) = \fS(I)$ for $x \in \neigh{I}{-\epsilon}$ for all $i$ large enough.
Then we have $\fg(x) = \hg( \fS(I))$ for all $x \in (\lowerx{I},\upperx{I}).$
\end{itemize}
\end{lemma}
\begin{IEEEproof}
Let $I \in \intflats{\fS}$ and let $v=\fS(I).$
We have $\ff(x)=v$ for all $x \in \neigh{I}{W}$
and $v\in \jump{\hfinv}$  
by Lemma \ref{lem:rightincrease}.

Assume $I$ is not linked to any $I' \in\flats{\gS}.$
Then $\gSa$ is strictly increasing from the left at $\upperx{I}+W.$
By Lemma \ref{lem:rightincrease} $\gSa$ is strictly increasing to the
right at $\lowerx{I}-W.$
We conclude from this that
\[ (\gSa(\lowerx{I}-W),\gSa(\upperx{I}+W))
\subset
[\hf^{-1}(v-),\hf^{-1}(v+)]\,.
\]
Moreover, given any $\epsilon>0$ property D of Lemma \ref{lem:regularapprox}
and the uniform convergence of
$g_i^{\smthker,\ashift_i}$ to $\gSa$ now imply that
\begin{align*}
(g_i^{\smthker,\ashift_i}&(\lowerx{I}-W+\epsilon),g_i^{\smthker,\ashift_i}(\upperx{I}+W-\epsilon))
\\ &\subset
[\hfiinv(v-),\hfiinv(v+)]
\end{align*}
for all $i$ large enough.
We conclude from this that $\ff_i(x)=v$ for $x \in \neigh{I}{W-\epsilon}$
for all $i$ large enough which implies that $\fS_i(x) = v$ 
for $x \in \neigh{I}{-\epsilon}$
for all $i$ large enough, proving part A.

%

Consider part B.
Given the stated conditions it follows that 
$\gS=U$ on $I'$ and $\hf(U)=v.$ 
Since $I$ is maximal,
we have $U = \hfinv(v+),$ 
hence
\[
(\gSa(\lowerx{I}-W),U]
\subset
[\hfinv(v-),\hfinv(v+)]\,.
\]
We now apply property E of Lemma \ref{lem:regularapprox} to conclude that $\hf^i(U)  = v$ for all $i$ large enough.
Given $\epsilon>0$  we combine this
 with property D of Lemma \ref{lem:regularapprox} 
and Lemma \ref{lem:rightincrease} to obtain
\[
(g_i^{\smthker,\ashift_i}(\lowerx{I}-W+\epsilon),U]
\subset
[\hfiinv(v-),\hfiinv(v+)]
\]
for all $i$ large enough.
Let $\delta = \epsilon,$ then for all $i$ large enough
we have $\gS_i(x) = U$ for $x \in\neigh{I'}{-\epsilon}.$
We conclude that $\ff_i(x)=v$ for $x \in \neigh{I}{W-\epsilon}$
for all $i$ large enough which implies that $\fS_i(x) = v$ 
for $x \in \neigh{I}{-\epsilon}$
for all $i$ large enough, proving part B.

Consider part C. 
If $\hg$ is continuous at  $v$ then we must have
$g(x)=\hg(v)$ on $I.$
Assume now that $v  \in \jump{\hg}.$
We now have $v   \in \jump{\hfinv} \cap \jump{\hg}.$ 
Property $C$ of Lemma \ref{lem:regularapprox} now gives $\hg^i(v)=\hg(v)$ 
for all $i$ large enough.
This implies that for any $\epsilon>0$ we now have
$\fg_i(x)=\hg(v)$  for all $x \in \neigh{I}{-\epsilon}$
for all $i$ large enough.
Since $\ff_i \rightarrow \ff$ this proves part C.
\end{IEEEproof}

Lemma \ref{lem:convprop} essentially completes the proof of Theorem \ref{thm:mainexist} and we state the main
result as the following.
\begin{corollary}
Let $(\hf,\hg)$ satisfy the strictly positive gap condition. 
Then there exists $(0,1)$-interpolating $\ff,\fg$ such that 
$\ff = \hf \circ \gS$ and 
$\fg = \hg \circ \fSa.$
\end{corollary}
\begin{IEEEproof}
Lemma \ref{lem:pathology} covers the result unless
$\jump{\hf} \cap \jump{\hginv}=\emptyset$
or
$\jump{\hg} \cap \jump{\hfinv}=\emptyset$
and $A \neq 0.$ For this case we assume $A>0$ without loss of
generality and proceed as follows.

Let $(\hf^i,\hg^i) \rightarrow (\hf,\hg)$ be given as in Lemma \ref{lem:regularapprox}.
By Lemma \ref{lem:weakexistence} and Lemma \ref{lem:pathology}
there exists $(0,1)$-interpolating $\ff_i,\fg_i$  such that
$\ff_i = \hf^i \circ \gS_i$ and 
$\fg_i = \hg^i \circ \fS_i$ for each $i.$
Let $\ff$ and $\fg$  be  $(0,1)$-interpolating limits so that
$\ff \veq \hf \circ \gSa$ and 
$\fg \veq \hg \circ \fS$ as guaranteed by Lemma \ref{lem:limitexist}.

Let us first note that  if $1 \in \jump{\hfinv}$
then there exists a half-infinite interval $I \in \flats{\fS}$
with  $\fS(I)=1.$  
It follows easily from \ref{lem:regularapprox} and uniform convergence
of $g_i^{\smthker}$ to $\gS$ that for any $\epsilon> 0$ we have $f_i^{\smthker}(\lowerx{I}+\epsilon) = 1$
for all $i$ large enough.
A similar argument applies if $1 \in \jump{\hginv}.$

Lemma \ref{lem:linkterminate} states that any element in $I\in \intflats{\fS}$ must be part of a terminating
chain.
Parts A and B of Lemma \ref{lem:convprop} show 
(with suitable restatements for $I' \in\flats{\gS}$ and the inclusion
of the above $1$-valued case) 
that for any $\epsilon >  0$  we have $\fS_i(x)=\fS(I)$ for all $x\in \neigh{I}{-\epsilon}$ for all $i$ large enough.
Part C of  Lemma \ref{lem:convprop} then shows that $\fg(x) = \hg(\fS(x))$ for all
$x$ in the interior of $I.$
Since $\hf$ is continuous at $0$ and $1$ by Lemma \ref{lem:zocontinuity},
Lemma  \ref{lem:notequal} states that if  $\fg \not\equiv \hg \circ \fS$ then there exists
$I\in\intflats{\fS}$ such that $\fg(x) \not\equiv \hg(\fS(x))$ on a subset of positive measure in $I.$
Since this is not the case
we can now conclude that $\fg \equiv \hg \circ \gS.$
A similar argument shows that  $\ff \equiv \hf \circ \gSa.$
We can obtain equality by modifying $\ff$ and $\fg$ on a set of measure $0.$
\end{IEEEproof}
\section{Two Sided Termination with Positive Gap}\label{app:C}

In this section we prove Theorems
\ref{thm:twoterminatedexist} and
\ref{thm:discretetwoterminatedexist}.
The two results have much in common and we begin with some
constructions that apply to both.

We assume that $\smthker$ is regular and that
$(\hf,\hg)$ satisfies the strictly positive gap condition with $A(\hf,\hg) < 0.$
It  follows from Lemma \ref{lem:Sstructure} that we may choose $\delta>0$ sufficiently small so that
\begin{align}
&\altPhi(\hf,\hg;\cdot,\cdot)>0 \text{ on } \neigh{\intcross(\hf,\hg)}{\delta},\label{eqn:innneigh}\\
&\altPhi(\hf,\hg;\cdot,\cdot)>0 \text{ on } \neigh{(0,0)}{\delta}\backslash (0,0),\\
&\neigh{(1,1)}{\delta} \cap  \neigh{\intcross(\hf,\hg)}{\delta} = \emptyset.\label{eqn:ooneigh}
\end{align}


Consider the following parametric modification of $(\hf,\hg)$ 
\begin{align}
\begin{split}
\thf(u) &\defeq(\hf(u)-\eta)^+\,\\ \label{eqn:FirstMod}
\thg(v) &\defeq (\hg(v)-\eta)^+ \,.
\end{split}
\end{align}
(Here we have introduced the notation $h(u)^+\defeq h(u)\vee 0.$)
By Lemma \ref{lem:crosspointlimit} we have 
$\cross(\thf,\thg) \subset \neigh{\cross(\hf,\hg)}{\delta}$ for all $\eta$ sufficiently small.
In this case let $(u^*,v^*)$ the minimum point (coordinate-wise) of 
$\cross(\thf,\thg) \cap \neigh{(1,1)}{\delta}.$
As $\eta\rightarrow 0$ we deduce from \eqref{eqn:ooneigh} 
and Lemma \ref{lem:crosspointlimit}
that $(u^*,v^*) \rightarrow (1,1)$
and therefore
$\altPhi(\thf,\thg;u^*,v^*) \rightarrow A(\hf,\hg).$ 
Hence, given arbitrary $\epsilon > 0$  we have for all $\eta$ small enough that
$u^*,v^* > 1-\epsilon$ and $\altPhi(\thf,\thg;u^*,v^*) < 0.$
Since $\altPhi(\thf,\thg;\cdot,\cdot) \ge \altPhi(\hf,\hg;\cdot,\cdot)$ it then follows from
\ref{eqn:innneigh} and the minimality of $u^*,v^*$ that 
$\thf,\thg$ satisfies the strictly positive gap condition over $[0,u^*]\times[0,v^*].$
By Lemma \ref{lem:zocontinuity} we have 
we have $\thf(u^*-) = v^*$ and $\thg(v^*-) = u^*.$
If $\thf(u^*) > v^*$ or $\thg(v^*) > u^*$ then we can
can reduce them slightly by redefining
$\thf(u^*) = v^*$ and $\thg(v^*) = u^*.$
This ensures that $u^*,v^*$ is a fixed point of $\thf,\thg.$

We can now apply Theorem \ref{thm:mainexist} over $[0,u^*]\times[0,v^*]$
 to obtain
 ${\tmplF},{\tmplG} \in \Psi_{[-\infty,\infty]}$ 
interpolating between $(0,v^*)$ and $(0,u^*)$ respectively
and $\ashift < 0$ so that
setting $\ff^t(x) = {\tmplF}(x-\ashift t)$ and
$\fg^t(x) = {\tmplG}(x-\ashift t)$ satisfies
\eqref{eqn:gfrecursion} for $(\thf ,\thg ).$
Since $\hf$ and $\hg$ are continuous at $0$ and $\eta>0$ we have
$\thf(u)=0$ for some neighborhood of $u=0$ and
$\thg(v)=0$ for some neighborhood of $v=0.$
Hence ${\tmplF}(x) = 0$ on some maximal interval, we may take to be $[-\infty,0),$
and ${\tmplG}(x) = 0$ on some maximal interval $[-\infty,x_g).$

Note that we have
\begin{equation}\label{eqn:ADFGbnd}
\hg(\tmplF^{\smthker}(x))
\ge
\tmplG(x) + \eta \unitstep_0 (x-x_g)
\end{equation}
and
\begin{equation}\label{eqn:ADGFbnd}
\hf(\tmplG^{\smthker}(x+\ashift))
\ge
\tmplF(x) + \eta \unitstep_0 (x)\,.
\end{equation}

Applying Lemma \ref{lem:shiftupperbound} and Lemma \ref{lem:stposbound} we can
assert the existence of a bound $S<2W$ such that $-\ashift \le S$ for all $\eta$ sufficiently small.  
We assume $Z = Z(\epsilon)$ large enough so that
\begin{align}
{\tmplF}(\tfrac{1}{4}Z+\ashift-1/\|\smthker\|_\infty) & > v^*-\frac{\eta}{4},\label{eqn:topF}\\
{\tmplG}(\tfrac{1}{4}Z+\ashift-1/\|\smthker\|_\infty) & > u^*-\frac{\eta}{4} \label{eqn:topG}\,.
\end{align}
(We require the $1/\|\smthker\|_\infty$ term  for the discrete case where we will use
$\Delta \le 1/\|\smthker\|_\infty.$)

Let us define 
\begin{equation}\label{eqn:f0initial}
\ff^{0}(x) = \tmplF(x+\ashift) +  \eta \unitstep_0(x+\ashift)
\end{equation}
for $x \le \half Z$ and for $x > \half Z$
initialize symmetrically using $\ff^{0}(x) = \ff^{0}(Z-x).$
Clearly this is even about $\half Z$ and we have
$\ff^0(x) \le 1.$ 
For $x \in [\tfrac{1}{4}Z,\tfrac{1}{2}Z]$ we have
${\tmplF}(x+\ashift)  > v^*-\frac{\eta}{4}$  by \eqref{eqn:topF}
and for all $x$ we have ${\tmplF}(x) \le v^*.$  This gives for all $x$ the bound
\begin{equation}\label{eqn:f0initialbound}
\ff^{0}(x) \ge \tmplF(x+\ashift) +  \tfrac{3}{4} \eta \unitstep_0(x+\ashift)
-
\unitstep_1(x-\tfrac{3}{4}Z)
\end{equation}

\begin{IEEEproof}[Proof of Theorem \ref{thm:twoterminatedexist}]

We assume $Z$ large enough so that
\begin{align}
\tfrac{3}{4}\eta \Omega(0)
- \Omega(-\tfrac{1}{4}Z) & \ge 0 \label{eqn:lm15Obnd} \\
\tfrac{3}{4}\eta \Omega(-x_g+\ashift)
- \Omega(-\tfrac{1}{4}Z) & \ge 0 \label{eqn:lm15Obbnd} 
\end{align}

Let us initialize the system \eqref{eqn:gfrecursion} with 
\(
\ff^{0}(x) 
\)
as given in \eqref{eqn:f0initial}.
By \eqref{eqn:f0initialbound} we have
\[
\ff^{0,\smthker}(x) \ge \tmplF^{\smthker} (x+\ashift) +  \tfrac{3}{4} \eta \Omega(x+\ashift)
- \Omega(x-\tfrac{3}{4}Z)
\] 
and for $x \in [x_g-\ashift,\half Z]$ we have 
\(
\Omega(x+\ashift)
- \Omega(x-\tfrac{3}{4}Z) \ge 
\Omega(0)
- \Omega(-\tfrac{1}{4}Z)
\)
so by \eqref{eqn:lm15Obnd} we have
\[
\ff^{0,\smthker}(x) \ge \tmplF^{\smthker} (x+\ashift) \,
\] 
on this interval.

Consider now 
\(
g^{0}(x) = \hg(\ftS{0}(x)).
\)
We have for $x\in [x_g-\ashift,\half Z]$
\begin{align*}
g^{0}(x) &= \hg(\ftS{0}(x)) \\
& \ge \hg(\tmplF^{\smthker} (x+\ashift)) 
\\ & \stackrel{\eqref{eqn:ADFGbnd}}{\ge}
\tmplG(x+\ashift)+\eta\unitstep_0(x-x_g+\ashift)
\end{align*}
and we observe that since the right hand side is $0$ for
$x < x_g-\ashift$ the inequality holds for all $x \le \half Z.$

As in the derivation of \eqref{eqn:f0initialbound} we apply \eqref{eqn:topG}
to derive for all $x$ the bound
\begin{align*}
g^{0}(x) & \ge
\tmplG(x+\ashift)+\tfrac{3}{4}\eta\unitstep_0(x-x_g+\ashift) - \unitstep_1(x-\tfrac{3}{4}Z)
\end{align*}
and we obtain
\begin{align*}
g^{0,\smthker}(x) \ge \tmplG^{\smthker}(x+\ashift) + \tfrac{3}{4}\eta \Omega(x-x_g+\ashift) - \Omega(x-\tfrac{3}{4}Z)
\end{align*}
Which, by \eqref{eqn:lm15Obbnd}, gives
$g^{0,\smthker}(x) \ge \tmplG^{\smthker}(x+\ashift)$
for $x\in [0,\half Z].$

Now, define $\ff^1$ by
$\ff^1(x) = \hf(\gtS{0}(x))$ for $x\in[0,Z]$ and
$\ff^1(x) = 0$ otherwise.  For $x\in[0,\half Z]$ we have
\begin{align*}
f^{1}(x) &= \hf(\gtS{0}(x)) \\
& \ge \hf(\tmplG^{\smthker} (x+\ashift)) 
\\ & \stackrel{\eqref{eqn:ADGFbnd}}{\ge}
\tmplF(x)+\eta\unitstep_0(x)
\\ & \ge
f^{0}(x)
\end{align*}
This implies the existence of a fixed point lower bounded by
$f^0,g^0,$ which completes the proof since $f^0(Z/2),g^0(Z/2) > 1-\epsilon.$
\end{IEEEproof}

\begin{IEEEproof}[Proof of Theorem \ref{thm:discretetwoterminatedexist}]
The proof is similar to the proof of Lemma \ref{thm:twoterminatedexist}
but we require some stronger assumptions.
First, we assume that $Z=L\Delta$ for an integer $L.$
In addition we assume $\eta$ small enough so that 
$\altPhi(\thf,\thg;u^*,v^*)< -\Delta\|\smthker\|_\infty.$
Theorem \ref{thm:mainexist} now implies
$\ashift <-\Delta$ (actually we have $\ashift < -\Delta/(u^* v^*)$).
Finally, we assume $Z$ large enough so that
\begin{align}
\tfrac{3}{4}\eta \Omega(0)
- \Omega(-\tfrac{1}{4}Z+\half \Delta) & \ge 0 \label{eqn:lm16Obnd}
\\
\tfrac{3}{4}\eta \Omega(-x_g+\ashift-\Delta)
- \Omega(-\tfrac{1}{4}Z+ \Delta) & \ge 0 \label{eqn:lm16Obbnd}
\end{align}

Let us initialize the system \eqref{eqn:discretegfrecursion} with 
\(
\ff^{0}(x) 
\)
as given in \eqref{eqn:f0initial}.
By \eqref{eqn:f0initialbound} and Lemma \ref{lem:disccontbnd} we have
\begin{align*}
\ff^{0,\smthker}(x_i) \ge &\tmplF^{\smthker} (x_i+\ashift-\half\Delta) +  \tfrac{3}{4} \eta \Omega(x_i+\ashift-\half\Delta)\\
&- \Omega(x_i-\tfrac{3}{4}Z+\half\Delta)
\end{align*}
and for $x_i \in [x_g-\ashift+\half\Delta,\half Z]$ we have by \eqref{eqn:lm16Obnd}
\[
\ff^{0,\smthker}(x_i) \ge \tmplF^{\smthker} (x_i+\ashift-\half\Delta) \,.
\] 

Consider now 
\(
g^{0}(x_i) = \hg(\ftS{0}(x_i)).
\)
We have for $x_i\in [x_g-\ashift+\half\Delta,\half Z]$
\begin{align*}
g^{0}(x_i) &= \hg(\ftS{0}(x_i)) \\
& \ge \hg(\tmplF^{\smthker} (x_i+\ashift-\half\Delta)) 
\\ & \stackrel{\eqref{eqn:ADFGbnd}}{\ge}
\tmplG(x_i+\ashift-\half\Delta)+\eta\unitstep_0(x_i-x_g+\ashift-\half\Delta)
\end{align*}
and we observe that since the right hand side is $0$ for
$x_i < x_g-\ashift+\half\Delta$ the inequality holds for all $x_i \le \half Z.$

Again, as in the derivation of \eqref{eqn:f0initialbound} we apply \eqref{eqn:topG}
to derive for all $x$ the bound
\begin{align*}
g^{0}(x_i) \ge &
\tmplG(x_i+\ashift-\half\Delta)+\tfrac{3}{4}\eta\unitstep_0(x_i-x_g+\ashift-\half\Delta)
\\& - \unitstep_1(x_i-\tfrac{3}{4}Z)
\end{align*}
and, applying Lemma \ref{lem:disccontbnd}, we obtain
\begin{align*}
g^{0,\discsmthker}(x_i) \ge& \tmplG^{\smthker}(x_i+\ashift-\Delta) + \tfrac{3}{4}\eta \Omega(x_i-x_g+\ashift-\Delta)
\\& - \Omega(x_i-\tfrac{3}{4}Z+\Delta)
\end{align*}
Which, by \eqref{eqn:lm16Obbnd}, gives
$g^{0,\smthker}(x_i) \ge \tmplG^{\smthker}(x_i+\ashift-\Delta)$
for $x_i\in [0,\half Z].$

Now, define $\ff^1$ by
$\ff^1(x_i) = \hf(\gtS{0}(x_i))$ for $x_i\in[0,Z]$ and
$\ff^1(x_i) = 0$ otherwise.  For $x_i\in[0,\half Z]$ we have
\begin{align*}
f^{1}(x_i) &= \hf(\gtS{0}(x_i)) \\
& \ge \hf(\tmplG^{\smthker} (x_i+\ashift-\Delta)) 
\\ & \stackrel{\eqref{eqn:ADGFbnd}}{\ge}
\tmplF(x_i-\Delta)+\eta\unitstep_0(x_i-\Delta)
\\ & \ge
f^{0}(x_i)
\end{align*}
where the last inequality uses $\ashift <-\Delta.$
This implies the existence of a fixed point lower bounded by
$f^0,g^0,$ which completes the proof which completes the proof since $f^0(Z/2),g_0(Z/2) > 1-\epsilon.$
\end{IEEEproof}

\section{General Convergence Results}\label{app:E}

The existence of interpolating wave solutions often implies 
global convergence of the spatially coupled system.  The structure of $\cross(\hf,\hg)$ can, however, be complicated
enough to prevent direct application of the existence results for wave-like solutions.
Typically, the necessary conditions for existence of spatial fixed points are easier to apply.
Our technique to prove the general convergence results largely consists of applying those
conditions to a modified version of the spatial iterative system.
Monotonicity typcially implies convergence and necessary conditions on interpolating fixed points
provide the leverage needed to get the desired results.
We start with a Lemma that uses this approach in a canonical way.

\begin{lemma}\label{lem:liminfgap}
Let $(\hf,\hg)$ be given with $A(\hf,\hg)<0$ and 
$\altPhi(\hf,\hg;u,v) > A(\hf,\hg)$ for $(u,v) \neq (1,1).$
Consider the spatially continuous system \eqref{eqn:gfrecursion}.
If $\ff^0 \in \sptfns$ 
satisfies $\ff^0(\pinfty) = 1$ then for all $x\in\reals$ we have
\[
\lim_{t\rightarrow \infty} \ff^t(x) =1\,,\quad
\lim_{t\rightarrow \infty} \fg^t(x) =1
\]
\end{lemma}\begin{IEEEproof}
Given $\epsilon > 0$ we claim that we can find
a pair of EXIT functions $\thf,\thg$
and $1> u^*,v^* > 1-\epsilon$ such that 
$\thf \le \hf$ and $\thg \le \hg,$ and that, restricted to $[0,u^*]\times[0,v^*],$
the pair satisfies the strictly positive gap condition over $[0,u^*]\times[0,v^*]$ with
$\altPhi(\thf,\thg;u^*,v^*) < 0.$

Assume the claim, then by Theorem \ref{thm:mainexist} there exists $\tmplF,\tmplG \in \sptfns$ 
interpolating over $(0,u^*)$ and $(0,v^*)$ respectively 
and $\ashift \le -\altPhi(\thf,\thg;u^*,v^*)/\|\omega\|_\infty$
such that $\tmplF(x-\ashift t)$ and
$\tmplG(x-\ashift t)$ solves \eqref{eqn:gfrecursion} for
the pair $\thf,\thg.$

Since $\tmplF(\pinfty)=v^*<1$ and $\tmplF(x)=0$ for some finite $x,$ we see that 
for any $(0,1)$-interpolating function $f^0 \in \sptfns$ with $f^0(\pinfty) = 1$
we can assume (by applying an appropriate translation)  that $\tmplF \le f^0.$
Letting $f^t,g^t$ be sequence determined by \eqref{eqn:gfrecursion} for
the pair $\hf,\hg,$ we now have
$f^t(x) \ge \tmplF(x-\ashift t)$ and $g^t(x) \ge \tmplG(x-\ashift t).$
Hence $\liminf_{t\rightarrow\infty} f^t(x)\ge v(\eta)\ge 1-\epsilon$
and $\liminf_{t\rightarrow\infty} g^t(x)\ge u(\eta)\ge 1-\epsilon$
for all $x.$ Since $\epsilon$ is arbitrary this proves the lemma.

Now we prove the claim. 
We define $u^*,v^*$ slightly differently then in the last section.
Let us define
\[
\hf(u;\eta) =(\hf(u)-\eta)^+ 
\text{ and }
\hg(g;\eta) =(\hf(g)-\eta)^+\,.
\]
Let $m(\eta)$ 
be the minimum value of $\altPhi(\hf(\cdot;\eta), \hg(\cdot;\eta);u,v).$
The minimum is achieved at some coordinate-wise minimal point $(u^*,v^*),$ 
i.e. $\altPhi(\hf(\cdot;\eta), \hg(\cdot;\eta);u^*,v^*) = m(\eta)$
and $\altPhi(\hf(\cdot;\eta), \hg(\cdot;\eta);u,v) > m(\eta)$
for  $(u,v) \in \cross (\hf(\cdot;\eta), \hg(\cdot;\eta))$ and $(u,v) < (u^*,v^*).$
As $\eta\rightarrow 0$ we have 
$(u^*,v^*) \rightarrow (1,1)$ and
$\altPhi(\hf(\cdot;\eta),\hg(\cdot;\eta);u^*,v^*)\rightarrow A(\hf,\hg).$
Given $\epsilon>0$
we can choose $\eta$ small enough so that 
$u^*,v^* > 1-\epsilon$ and
$\altPhi(\hf(\cdot;\eta),\hg(\cdot;\eta);u^*,v^*)<0.$
Note that we have $u^*,v^* < 1$ since $\eta>0.$

Since $(u^*,v^*) \in \cross(\hf(\cdot;\eta),\hg(\cdot;\eta))$
is coordinate-wise minimal we have $v^* \le \hf(u^*-)$ and $u^* \le \hg(v^*-).$
(For example, if $v^* > \hf(u^*-)$ then since $v^*< \hf(u^*+)$
we obtain 
\(
\altPhi((\hf(\cdot)-\eta)^+, (\hg(\cdot)-\eta)^+;u^*,\hf(u^*-))
=
\altPhi((\hf(\cdot)-\eta)^+, (\hg(\cdot)-\eta)^+;u^*,v^*)
\)
which contradicts the corodinate-wise minimality of $(u^*,v^*).$)
We can therefore choose $\eta'>0$ sufficiently small so that
\begin{align*}
\hf(u) &\ge  \hf(u;\eta,\eta') \defeq \hf(u;\eta) \vee v^*\unitstep (u-(u^*-\eta'))\,\\
\hg(v) &\ge \hg(v;\eta,\eta') \defeq \hg(v;\eta) \vee u^*\unitstep (v-(v^*-\eta'))\,.
\end{align*}
It now easily follows that $\altPhi(\hf(\cdot;\eta,\eta'),\hg(\cdot;\eta,\eta');u,v)$ is minimized at $u^*,v^*$ and that there exists $\delta>0$ such that if
 $(u,v) \in \cross(\hf(\cdot;\eta,\eta'),\hg(\cdot;\eta,\eta')) \cap [0,u^*]\times[0,v^*]$
and  $(u,v)\neq(u^*,v^*)$ then
$\altPhi(\hf(\cdot;\eta,\eta'),\hg(\cdot;\eta,\eta');u,v)\ge \altPhi(\hf(\cdot;\eta,\eta'),\hg(\cdot;\eta,\eta');u^*,v^*)+\delta.$

Now consider
\begin{align*}
  \hf(u;z,\eta,\eta') &\defeq \unitstep_1 (u-z)\wedge \hf(u;\eta,\eta')
\\
 \hg(v;z,\eta,\eta') &\defeq \unitstep_1 (u-z)\wedge \hg(v;\eta,\eta')  \,.
\end{align*}
We can choose $z$ so that we have
\(
\altPhi(\hf(\cdot;z,\eta,\eta'),\hg(\cdot;z,\eta,\eta');u^*,v^*)<-\delta/2
\) 
and for all 
$(u,v) \in \cross(\hf(\cdot;z,\eta,\eta'),\hg(\cdot;z,\eta,\eta')) \cap [0,u^*)\times[0,v^*)$
we have
 $\altPhi(\hf(\cdot;z,\eta,\eta'),\hg(\cdot;z,\eta,\eta');u,v)>0.$  

We can now take $\thf(u) = \hf(\cdot;z,\eta,\eta')$ and
$\thg(v) = \hg(\cdot;z,\eta,\eta')$ and
if necessary we further reduce $\thf(u^*)$ and $\thg(v^*)$
so that they equal $v^*$ and $u^*$ respectively.
This proves the claim.
\end{IEEEproof}

The above proof can be easily adapted to the spatially discrete case.
\begin{lemma}\label{lem:discreteliminfgap}
Let $(\hf,\hg)$ be given with $A(\hf,\hg)<0$ and 
$\altPhi(\hf,\hg;u,v) > A(\hf,\hg)$ for $(u,v) \neq (1,1).$
Consider the spatially discrete system \eqref{eqn:discretegfrecursion}.
For any $\epsilon>0,$ if $\Delta$ is sufficiently small then
for all $x\in\reals$ we have
\[
\lim_{t\rightarrow \infty} \ff^t(x) \ge 1-\epsilon\,,\quad
\lim_{t\rightarrow \infty} \fg^t(x) \ge 1-\epsilon
\]
for any $\ff^0 \in \sptfns$ satisfying $\ff^0(\pinfty) = 1.$
\end{lemma}
\begin{IEEEproof}
We use the construction from the proof of Lemma \ref{lem:liminfgap}
and recall
the existence of $\tmplF,\tmplG \in \sptfns$ 
interpolating over $(0,v^*)$ and $(0,u^*)$ respecively
and $\ashift \le -(\delta/2)/\|\omega\|_\infty$
such that $f^t(x)=\tmplF(x-\ashift t)$ and
$g^t(x)=\tmplG(x-\ashift t)$ solves \eqref{eqn:gfrecursion} for
the pair $\thf,\thg.$
Assume $\Delta \le |\ashift|.$  

Given  $f^0$ satisfying $\ff^0(\pinfty) = 1$ we can find $y$ 
such that $f^0(x_i) \ge \tmplF(x_i-y)$ for all $x.$
We can apply Theorem \ref{thm:mainquantize} and the inequalities
$\hf \ge \thf$ and $\hg \ge \thg$
to obtain
$f^t(x_i) \ge \tmplF(x_i-y-(\ashift+\Delta)t)$ and
$g^t(x_i) \ge \tmplG(x_i-y-(\ashift+\Delta)t).$

The Lemma now follows.
\end{IEEEproof}

Recall that in the statement of Theorem \ref{thm:globalconv} 
we have $(0,0)\le (u',v') \le (u'',v'') \le (1,1)$ and $\altPhi$ is minimized on
$(u',v')$ and $(u'',v'')$ where it takes the value  $m(\hf,\hg).$
Furthermore, $(u',v')$ and $(u'',v'')$ are the extreme points where the minimum is attained.

\begin{IEEEproof}[Proof of Theorem \ref{thm:globalconv}]
We will prove the first statement in the Theorem,  i.e.,
\(
\liminf_{t\rightarrow \infty} f^t(x) \ge v'\,,
\)
the other cases being similar.

If $u' =0$ or $v'=0$ then $m=0$ and $(u',v')=(0,0)$ and the result is immediate.
Let us assume that $m<0$ and hence that $(u',v')>(0,0).$
Consider the system restricted to $[0,u']\times [0,v'].$
If $\hf(u')>v'$ then let us redefine $\hf(u')=v'$ and 
if $\hg(v')>u'$ then let us redefine $\hf(u')=u'.$
This makes $(u',v')$ a fixed point of the underlying system.
This reduction will not affect the remaining argument.
Let us reduce $\ff^0$ by saturating it at $v',$ i.e., replacing it with
$\ff^0 \wedge v'.$

We can now apply Lemma \ref{lem:liminfgap} to obtain $\ff^t(x) \rightarrow v'$
and $\fg^t(x) \rightarrow u'.$
Since $\ff^t$ and $\fg^t$ in the original system are only larger, the result follows.
\end{IEEEproof}

\begin{IEEEproof}[Proof of Theorem \ref{thm:discreteglobalconv}]
We can use Lemma \ref{lem:discreteliminfgap} to prove Theorem \ref{thm:discreteglobalconv}
in the same manner that  Lemma \ref{lem:liminfgap} is used to prove Theorem \ref{thm:globalconv}.
The argument is essentially the same so we omit it.
\end{IEEEproof}

We now consider the one-sided termination scenario.
\begin{IEEEproof}[Proof of Theorem \ref{thm:terminatedzero}]
The case where $\altPhi(\hf,\hg;u,v) > 0$ for $(u,v) \neq (0,0)$
follows easily from Theorem \ref{thm:globalconv}.  We assume now that
$\altPhi(\hf,\hg;u,v) = 0$ and   $\hf$ and $\hg$ are strictly positive on $(0,1].$

Define 
\[
\ff^0(x)=\unitstep_1(x)
\] 
We will show that $\ff^t\rightarrow 0,$ which implies the same for arbitrary initial conditions.
By monotonicity in $t,$ $\ff^t$ has a point-wise limit $\ff^\infty \in \sptfns$
and $\fg^t$ has a point-wise limit $\fg^\infty \in \sptfns$
By continuity we have $(\ff^\infty(\pinfty),\fg^\infty(\pinfty)) \in \cross(\hf,\hg).$

In general $h_{[\ff^\infty,\fg^{\smthker,\infty}]}$
is well defined on
$[0,\fg^{\infty}(\pinfty)]$ and
$h_{[\fg^\infty,\ff^{\smthker,\infty}]}$
is well defined on
$[0,\ff^{\infty}(\pinfty)]$ 
and we have
\begin{align}
\begin{split}\label{eqn:AineqB}
0 \le  &\altPhi(\hf,\hg;\fg^{\infty}(\pinfty),\ff^{\infty}(\pinfty))
\\ \le &
\altPhi(h_{[\ff^\infty,\fg^{\smthker,\infty}]},h_{[\fg^\infty,\ff^{\smthker,\infty}]};\fg^{\infty}(\pinfty),\ff^{\infty}(\pinfty))\,.
\\ = &0\,.
\end{split}
\end{align}

Assume that $\ff^\infty \neq 0.$
Let $z = \sup \{x:\ff^\infty(x)=0\}.$
We have $\ff^{\smthker,\infty}(x)>0$ on $\neigh{z}{W}$ and therefore 
$\fg^{\infty}(x)>0$ on $\neigh{z}{W}$ and
$\fg^{\smthker,\infty}(x)>0$ on $\neigh{z}{2W}.$
Hence $\ff^\infty(x) > 0$  for $x \in \neigh{z}{2W} \cap (0,\infty)$
but $\ff^\infty(x)=0$ for $x<z.$
This implies that $z=0$ and that $\ff^\infty(x)$ is discontinuous at $x=0.$
We now have 
\begin{align}
 &\altPhi(\hf,\hg;u,v)
<
\altPhi(h_{[\ff^\infty,\fg^{\smthker,\infty}]},h_{[\fg^\infty,\ff^{\smthker,\infty}]};u,v)
\end{align}
for $u > 0$ and we easily conclude that 
$(\fg^{\infty}(\pinfty),\ff^{\infty}(\pinfty)) = (0,0)$
from Lemma \ref{lem:FPequal} (part C).
\end{IEEEproof}

\begin{IEEEproof}[Proof of Theorem \ref{thm:discreteterminatedexistB}]
Theorem \ref{thm:discreteterminatedexistB} can be proved along lines similar to
Lemma \ref{lem:liminfgap} with some additional features introduced to handle the spatial discreteness.

Let us define $m(\eta),\hf(\cdot,\eta),\hg(\cdot,\eta)$ and $u^*,v^*$ as in the proof of Lemma \ref{lem:liminfgap}.
Given $\epsilon>0$ we may choose $\eta$ small enough so that
$\tilde{u},\tilde{v} \ge 1-\epsilon$ and 
$m(\eta)<0.$

Now define
\begin{align*}
\hf(u;z,\eta) &\defeq\unitstep_1(u-z) \wedge \hf(u;\eta) 
\\
\hg(v;z,\eta) &\defeq\unitstep_1(v-z) \wedge \hg(v;\eta) 
\end{align*}
We can choose $z(\eta) (>0)$ so that
\[
 \altPhi\bigl(\hf(\cdot;z,\eta),\hg(\cdot;z,\eta);u^*,v^* \bigr)=0
\]
If necessary we further reduce $\hf(u^*;z,\eta)$ and $\hg(v^*;z,\eta)$
so that they equal $v^*$ and $u^*$ respectively.
Then 
$(\hf(\cdot;z,\eta),\hg(\cdot;z,\eta))$
satisfies the strictly positive gap condition on
$[0,u^*]\times[0,v^*].$

By Theorem \ref{thm:mainexist} there exists
 $\tmplF$ and $\tmplG$ that form a fixed point for 
\eqref{eqn:gfrecursion}
with $(\tmplF(\minfty),\tmplG(\minfty))=(0,0)$ and
$(\tmplF(\pinfty),\tmplG(\pinfty))=(v^*,u^*).$

Let us translate the solution so that $0 =\sup_x \{\tmplF(x)=0\}$
and let us then define $x_g = \sup_x \{\tmplG(x)=0\}.$
It follows that $|x_g| < W.$
Let us choose $\Delta$ sufficiently small so that
the following holds:
\begin{equation}\label{eqn:9condB}
\eta\intsmthker(-|x_g|-\half\Delta)
\ge
\half\Delta \|\omega\|_\infty\,.
\end{equation}


Consider initializing \eqref{eqn:discretegfrecursion} with
\begin{align*}
f^{0}(x_i)&=
\tmplF(x_i) +\eta \unitstep_0(x_i)\,.
\end{align*} 
Applying Lemma \ref{lem:disccontbnd} this yields for $x_i\ge x_g$
\begin{align*}
f^{0,\discsmthker}(x_i)&\ge
\tmplF^\smthker (x_i-\half\Delta) + \eta\intsmthker(x_i-\half\Delta) 
\\
&\stackrel{\eqref{eqn:9condB}}{\ge}
\tmplF^\smthker (x_i-\half\Delta) +\half\Delta\|\omega\|_\infty\,
\\
&{\ge}
\tmplF^\smthker (x_i) \,.
\end{align*}
We now obtain for $x_i  \ge x_g$
\begin{align*}
g^{0}(x_i)&=\hg(f^{0,\discsmthker}(x_i))
\\
&\ge
\hg(\tmplF^{\smthker}(x_i))
\\
&\ge
\hg(\tmplF^{\smthker}(x_i);z,\eta)+\eta\unitstep_0(x_i-x_g)
\\
&\ge
\tmplG(x_i)+\eta\unitstep_0(x_i-x_g)
\end{align*}
and we observe that since $\tmplG(x_i)=0$ for
$x_i < x_g$ this bound  holds for all $x_i.$
We now have
\begin{align*}
g^{0,\discsmthker}(x_i)&\ge
\tmplG^\smthker (x_i-\half\Delta)
+\eta\intsmthker(x_i-x_g-\half\Delta)\,.
\end{align*}
 For  $x_i \ge 0$ we obtain
\begin{align*}
g^{0,\discsmthker}(x_i)
&\stackrel{\eqref{eqn:9condB}}{\ge}
\tmplG^\smthker (x_i-\half\Delta)
+\half\Delta \|\omega\|_\infty
\\ & \ge 
\tmplG^\smthker (x_i)\,.
\end{align*}
Thus we have 
\begin{align*}
\ff^{1}(x_i)&=\hf(\fg^{0,\discsmthker}(x_i))
\\&\ge \hf(\tmplG^{\smthker}(x_i))
\\&\ge \tmplF(x_i)+\eta\unitstep_0(x_i)
\\&= \ff^0(x_i)
\end{align*}
and the consequently increasing sequence establishes the existence a
fixed point that we denote $\ff^{\infty},\fg^{\infty}$.
We have $\ff^{\infty}(\pinfty),\fg^{\infty}(\pinfty) \ge 1-\epsilon + \eta.$
Letting $\Delta \rightarrow 0$ we can have $\epsilon,\eta \rightarrow 0.$
\end{IEEEproof}

\begin{IEEEproof}[Proof of Theorem \ref{thm:discretetwoterminatedexistGB}]
We assume $Z = L\Delta$ for an integer $L.$
The termination $\hf(x_i,\cdot)=0$ holds for $x_i<0$ and $x_i > Z.$
This means that symmetry holds about $\half Z.$
The proof follows that of Theorem \ref{thm:discreteterminatedexistB}
up to the point where requirements on $\Delta$ are given.
Continuing from there we
%
%
%
choose $\Delta$ small enough and $Z$ large enough 
so that
all of the following hold.
\begin{equation}\label{eqn:14condB}
\tfrac{3}{4}\eta\intsmthker(-|x_g|-\Delta)
- \intsmthker(-\tfrac{1}{4}Z+\half\Delta )
\ge
\half\Delta \|\omega\|_\infty
\end{equation}
\begin{equation}\label{eqn:14condA}
\tmplF(\tfrac{1}{4}Z) > v^*-\eta/4
\end{equation}
\begin{equation}\label{eqn:14condE}
\tmplG(\tfrac{1}{4}Z) > u^*-\eta/4
\end{equation}

Consider initializing for $x_i \le \half Z$ with
\[
f^{0}(x_i)=
\tmplF(x_i) +\eta \unitstep_0(x_i)\,.
\]
and for $x_i > \half Z$ initializing symmetrially with $\ff^{0}(x_i)=\ff^{0}(x_{L-i}).$
As in the derivation of \eqref{eqn:f0initialbound}, this, by \eqref{eqn:14condA}, implies for all $x$
\[
f^{0}(x_i)=
\tmplF(x_i) +\tfrac{3}{4}\eta \unitstep_0(x_i)
-\unitstep_1(x_i - \tfrac{3}{4}Z)\,.
\]
which gives by Lemma \ref{lem:disccontbnd},
\[
f^{0,\discsmthker}(x_i)\ge
\tmplF^{\smthker}(x_i-\half\Delta) +\tfrac{3}{4}\eta \intsmthker(x_i-\half\Delta)
-\intsmthker(x_i - \tfrac{3}{4}Z+\half\Delta)\,.
\]

This yields for $x_i\in[x_g-\half\Delta,\tfrac{1}{2} Z]$
\begin{align*}
f^{0,\discsmthker}(x_i)
&\ge
\tmplF^\smthker (x_i) + \tfrac{3}{4}\eta\intsmthker(x_g-\Delta) - \intsmthker(-\tfrac{1}{4} Z+\half\Delta )
\\
&\stackrel{\eqref{eqn:14condB}}{\ge}
\tmplF^\smthker (x_i) +\half\Delta\|\omega\|_\infty\,
\\
&{\ge}
\tmplF^\smthker (x_i+\half\Delta) \,.
\end{align*}

We now obtain for $x_i  \in [x_g-\half \Delta, \half Z]$
\begin{align*}
g^{0}(x_i)&=\hg(f^{0,\discsmthker}(x_i))
\\
&\ge
\hg(\tmplF(x_i+\half\Delta))
\\
&\ge
\tmplG(x_i+\half\Delta)+\eta\unitstep_0(x_i+\half\Delta-x_g)
\end{align*}
and we observe that since the right hand side is $0$ for
$x_i+\half\Delta < x_g$ this bound  holds for all $x_i \le \half Z.$
As in the derivation of \eqref{eqn:f0initialbound}, by \eqref{eqn:14condE} we now have
\[
g^{0}(x_i)\ge
\tmplG(x_i+\half\Delta)+\tfrac{3}{4}\eta\unitstep_0(x_i+\half\Delta-x_g)-\unitstep_1(x_i-\tfrac{3}{4}Z)
\]
which gives by Lemma \ref{lem:disccontbnd}
\[
g^{0,\discsmthker}(x_i)\ge
\tmplG^\smthker(x_i)+\tfrac{3}{4}\eta\Omega(x_i-x_g)-\Omega(x_i-\tfrac{3}{4}Z+\half\Delta)
\]
which by \eqref{eqn:14condB} yields for $x_i \in [0,\half Z],$
\[
g^{0,\discsmthker}(x_i)\ge
\tmplG^\smthker(x_i)\,.
\]

Thus we have for $x_i \in [0,\half Z],$
\begin{align*}
\ff^{1}(x_i)&=\hf(\fg^{0,\discsmthker}(x_i))
\\&\ge \hf(\tmplG^{\smthker}(x_i))
\\&\ge \tmplF(x_i)+\eta\unitstep_0(x_i)
\\&= \ff^0(x_i)
\end{align*}
and the consequently increasing sequence establishes the existence of the
desired fixed point.
\end{IEEEproof}

\bibliographystyle{IEEEtran}
\bibliography{lth,lthpub,cspub,oneD}


\end{document}